\documentclass[rmp,aps,twocolumn,showpacs]{revtex4}
\usepackage{graphicx}
\usepackage{color}
\usepackage{epsf}
\usepackage{amsmath}
\usepackage{amssymb}
\usepackage[english]{babel}
\usepackage{bm}
\usepackage{graphics}
\usepackage{color}
\usepackage{dcolumn}
\usepackage{sidecap}
\usepackage{enumitem}

\usepackage{epsfig}

\usepackage[hidelinks]{hyperref}
\usepackage{xcolor}
\usepackage{floatrow}

\hypersetup{
    colorlinks,
    linkcolor={black},
    citecolor={blue!40!black},
    urlcolor={black}
}

\newcommand {\cci} {\ensuremath{{\bf \chi}}}

\newcommand {\ff} {\ensuremath{{\bf f}}}

\newcommand {\cc} {\ensuremath{{\bf c}}}
\newcommand {\DD} {\ensuremath{{\bf D}}}

\newcommand{\bes}{ \begin{equation} \begin{split} }
\newcommand{\ees}{ \end{split} \end{equation} }

\usepackage[normalem]{ulem}

\newcommand{\ignore}[1]{}

\newcommand{\red}[1]{{ {#1}}}
\newcommand{\blue}[1]{{{#1}}}

\begin{document}

\title{Coupling functions: Universal insights into dynamical interaction mechanisms}
\author{Tomislav Stankovski$^{1,2}$}%
\author{Tiago Pereira$^{3,4}$}
\author{Peter V. E. McClintock$^2$}%
\author{Aneta Stefanovska$^2$}

\affiliation{$^1$ Faculty of Medicine, Ss Cyril and Methodius University, 50 Divizija 6, Skopje 1000, Macedonia}
\affiliation{$^2$ Department of Physics, Lancaster University, Lancaster, LA1 4YB, United Kingdom}
\affiliation{$^3$ Department of Mathematics, Imperial College London, London SW7 2AZ, United Kingdom}
\affiliation{$^4$ Institute of Mathematical and Computer Sciences, University of S\~{a}o Paulo, S\~{a}o Carlos 13566-590, Brazil}

\begin{abstract}
The dynamical systems found in Nature are rarely isolated. Instead they interact and influence each other. The coupling functions that connect them contain detailed information about the functional mechanisms underlying the interactions and prescribe the physical rule specifying how an interaction occurs. Here, we aim to present a coherent and comprehensive review encompassing the rapid progress made recently in the analysis, understanding and applications of coupling functions. The basic concepts and characteristics of coupling functions are presented through demonstrative examples of different domains, revealing the mechanisms and emphasizing their multivariate nature. The theory of coupling functions is discussed through gradually increasing complexity from strong and weak interactions to globally-coupled systems and networks. A variety of methods that have been developed for the detection and reconstruction of coupling functions from measured data is described. These methods are based on different statistical techniques for dynamical inference. Stemming from physics, such methods are being applied in diverse areas of science and technology, including chemistry, biology, physiology, neuroscience, social sciences, mechanics and secure communications. This breadth of application illustrates the universality of coupling functions for studying the interaction mechanisms of coupled dynamical systems.
\end{abstract}


\date{\today}

\maketitle
\tableofcontents

\section{Introduction}\label{sec1:Intro}

\red{
\subsection{Coupling functions, their nature and uses}\label{sec12:CF_Intro}
}

\noindent Interacting dynamical systems abound in science and technology, with examples ranging from physics and chemistry, through biology and population dynamics, to communications and climate \cite{Kuramoto:84,Winfree:80,Pikovsky:01,Strogatz:03b,Haken:83}.

The interactions are defined by two main aspects: structure and function. The structural links  determine the connections and communications between the systems, or the topology of a network. The functions are quite special from the dynamical systems viewpoint, as they define the laws by which the action and co-evolution of the systems are governed. The functional mechanisms can lead to a variety of qualitative changes in the systems. 
Depending on the coupling functions, the resultant dynamics can be quite intricate, manifesting a whole range of qualitatively different states, physical effects, phenomena and characteristics, including synchronization \cite{Pikovsky:01,Acebron:05,Lehnertz:98,Kapitaniak:12}, oscillation and amplitude death \cite{Saxena:12,Koseska:13}, birth of oscillations \cite{Pogromsky:99,Smale:76}, breathers \cite{Mackay:94}, coexisting phases \cite{Keller:92}, fractal dimensions \cite{Aguirre:09},  network dynamics \cite{Boccaletti:06,Arenas:08}, and coupling strength and directionality \cite{Rosenblum:01,Hlavackova:07,Marwan:07,Stefanovska:99a}. Knowledge of such coupling function mechanisms can be used to detect,  engineer or predict  certain physical effects, to solve some man-made problems and, in living systems, to reveal their state of health and to investigate changes due to disease.

\red{Coupling functions possess unique characteristics carrying implications that go beyond the collective dynamics (e.g.\ synchronization or oscillation death). In particular, the form of the coupling function can be used, not only to understand, but also to control and predict the interactions. Individual units can be relatively simple, but the nature of the coupling function can make their collective dynamics particular, enabling special behaviour. Additionally, there exist applications which depend just and only on the coupling functions, including examples of applications in social sciences and secure communication.}

Given these properties, it is hardly surprising that coupling functions have recently attracted considerable attention within the scientific community. They have mediated applications, not only in different subfields of physics, but also beyond physics, predicated by the development of powerful methods enabling the reconstruction of coupling functions from measured data.   The reconstruction within these methods is based on a variety of inference techniques, e.g.\ least squares and kernel smoothing fits \cite{Rosenblum:01,Kralemann:13b}, dynamical Bayesian inference \cite{Stankovski:12b}, maximum likelihood (multiple-shooting) methods \cite{Tokuda:07}, stochastic modeling \cite{Schwabedal:10} and the phase resetting \cite{Galan:05,Levnajic:11,Timme:07}.

\red{
Both the connectivity between systems, and the associated methods employed for revealing it, are often differentiated into structural, functional and effective connectivity \cite{Park:13,Friston:11}. \emph{Structural connectivity} is defined by the existence of a physical link, like anatomical synaptic links in the brain or a conducting wire between electronic systems. \emph{Functional connectivity} refers to the statistical dependences between systems, like for example correlation or coherence measures. \emph{Effective connectivity} is defined as the influence one system exerts over another, under a particular model of causal dynamics. Importantly in this context, the methods used for the reconstruction of coupling functions belong to the group of effective connectivity techniques i.e.\ they exploit a model of differential equations and allow for dynamical mechanisms -- like the coupling functions themselves -- to be inferred from data.
}

\begin{figure}[t]
\includegraphics[width=1\linewidth,angle=0]{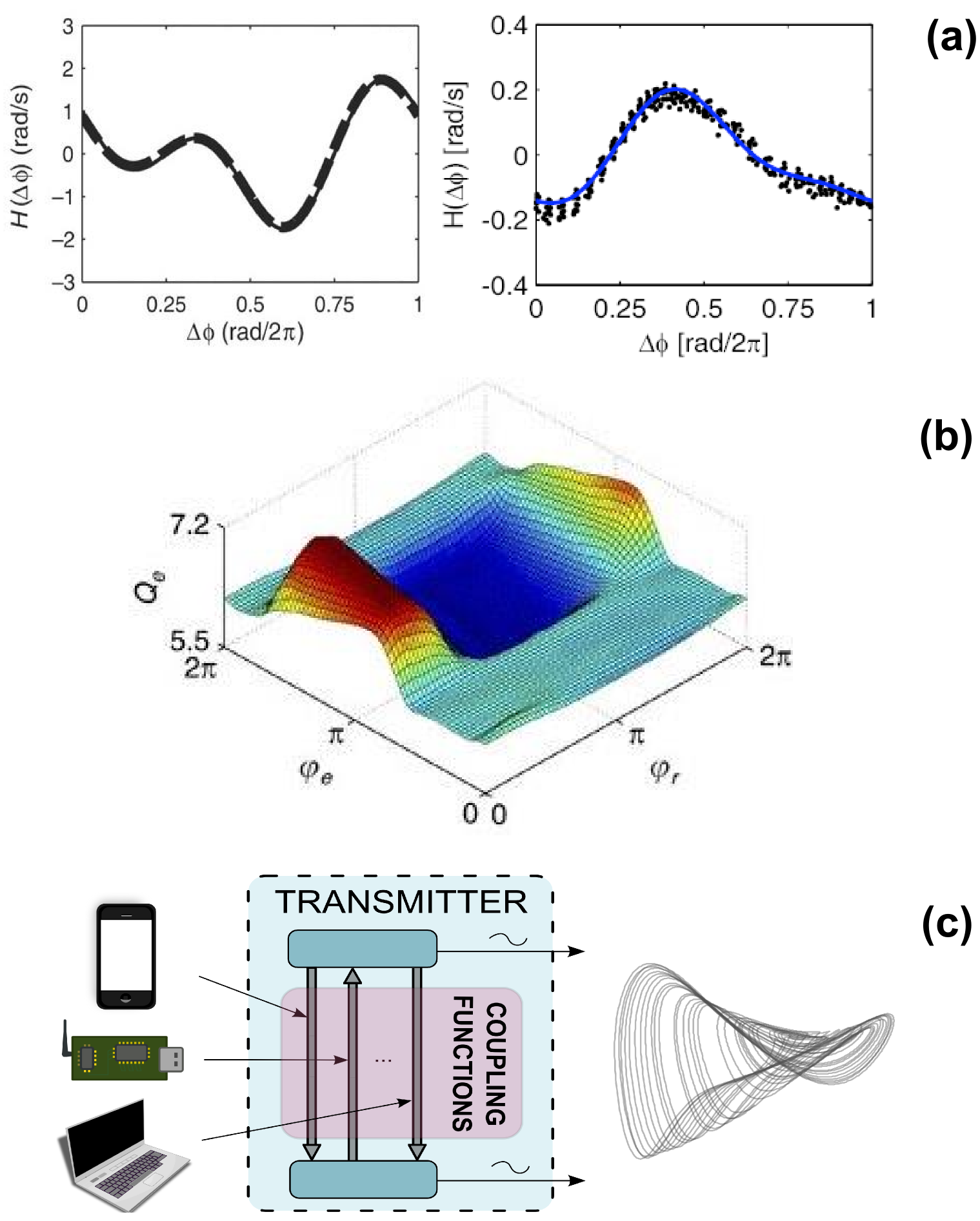}
\caption{\label{fig:app} (color online). Examples of coupling functions used in chemistry, cardiorespiratory physiology and secure communications, to demonstrate their diversity of applications. (a) Coupling functions used for controlling and engineering the interactions of two (left) and four (right) non-identical electrochemical oscillations.  (b) Human cardiorespiratory coupling function $Q_e$ reconstructed from  the phase dynamics the heart $\varphi_e$  and respiration $\varphi_r$ phases. (c) Schematic description of the coupling function encryption protocol. Multiple information signals are encrypted by modulating the parameters of linearly-independent coupling functions between (chaotic) dynamical systems at the transmitter. These applications are discussed in detail in Sec.\ \ref{sec64:Applc}.  Fig.\ \ref{fig:app}(a) is from \citet{Kiss:07}, (b) from \citet{Kralemann:13b} and (c) from \citet{Stankovski:14a}.}
\end{figure}

Coupling function methods have been applied widely (Fig.\ \ref{fig:app}), and to good effect: in \emph{chemistry}, for understanding, effecting, or predicting interactions between oscillatory electrochemical reactions \cite{Kiss:07,Miyazaki:06,Tokuda:07,Blaha:11,Kori:14}; in \emph{cardiorespiratory physiology} \cite{Kralemann:13b,Stankovski:12b,Iatsenko:13a} for reconstruction of the human cardiorespiratory coupling function and phase resetting curve, for assessing cardiorespiratory time-variability and for studying the evolution of the cardiorespiratory coupling functions with age; in \emph{neuroscience} for revealing the cross-frequency coupling functions between neural oscillations \cite{Stankovski:15a}; in {\it social sciences} for determining the function underlying the interactions between democracy and economic growth \cite{Ranganathan:14}; for \emph{mechanical} interactions between coupled metronomes \cite{Kralemann:08}; and in \emph{secure communications} where a new protocol was developed explicitly based on amplitude coupling functions \cite{Stankovski:14a}.

In parallel with their use to support experimental work, coupling functions are also at the centre of intense theoretical research \cite{Strogatz:00,Daido:96b,Crawford:95,Acebron:05}. Particular choices of coupling functions can allow for a multiplicity of singular synchronized states \cite{Komarov:13}. Coupling functions are responsible for the overall coherence in complex networks of non-identical oscillators \cite{Pereira:13,Ullner:16,Luccioli:10} and for the formation of waves and antiwaves in coupled neurons \cite{Urban:12}.  Coupling functions play important roles in the phenomena resulting from interaction such as synchronization \cite{Kuramoto:84,Daido:96b,Maia:15}, amplitude and oscillation death \cite{Aronson:90,Koseska:13,Zakharova:14,Schneider:15}, the low-dimensional dynamics of ensembles \cite{Ott:08,Watanabe:93}, and clustering in networks \cite{Ashwin:05a,Kori:14}. 
The findings of these theoretical works are fostering further the development of methods for coupling function reconstruction, paving the way to additional applications.

\red{
\subsection{Significance for interacting systems more generally}\label{sec12:General}
}

An interaction can result from a structural link through which causal information is exchanged between the system and one or more other systems  \cite{Kuramoto:84,Winfree:80,Pikovsky:01,Strogatz:03b,Haken:83}. \blue{Often it is not so much the nature of the individual parts and systems, but how they interact, that determines their collective behaviour. One example is circadian rhythms, which occur across different scales and organisms \cite{Dewoskin:14}.} The systems themselves can be diverse in nature -- for example, they can be either static or dynamical, including oscillatory, nonautonomous, chaotic, or stochastic characteristics \cite{Katok:97,Kloeden:11,Landa:13,Strogatz:01b,Suprunenko:13,Gardiner:04}. From the extensive set of possibilities, we focus in this review on dynamical systems, concentrating especially on nonlinear oscillators because of their particular interest and importance.

\vspace{0.3cm}

\subsubsection{Physical effects of interactions: Synchronization, amplitude and oscillation death} \label{sec21:PhysEffects}

An intriguing feature is that their mutual interactions can change the qualitative state of the systems. Thus they can cause transitions into or out of physical states such as synchronization, amplitude or oscillation death, or quasi-synchronized states in networks of oscillators.


The existence of a physical effect is, in essence, defined by the presence of a \emph{stable state} for the coupled systems. Their stability is often probed through a dimensionally-reduced dynamics, for example the dynamics of their phase difference or of the driven system only. By determining the stability of the reduced dynamics, one can derive useful conclusions about the collective behaviour. In such cases, the coupling functions describe how the stable state is reached and the detailed conditions for the coupled systems to gain or lose stability. In data analysis, the existence of the physical effects is often assessed through measures that quantify -- either directly or indirectly -- the resultant statistical properties of the state that remains stable under interaction.

The physical effects often converge to a manifold, such as a limit cycle. Even after that, however, coupled dynamical systems can still exhibit their own individual dynamics, making them especially interesting objects for study.

Arguably, \emph{synchronization} is the most studied of all such physical effects. \red{It is defined as an adjustment of the rhythms of the oscillators, caused by their weak interaction \cite{Pikovsky:01}.} Synchronization is the underlying qualitative state that results from many cooperative interactions in nature. Examples include cardiorespiratory synchronization \cite{Schaefer:98,Stefanovska:00a,Kenner:76}, brain seizures \cite{Lehnertz:98}, neuromuscular activity \cite{Tass:98},  chemistry \cite{Kiss:07,Miyazaki:06}, the flashing of fireflies \cite{Mirollo:90,Buck:68} and ecological synchronization \cite{Blasius:99}. Depending on the domain, the observable properties and the underlying phenomena, several different definitions and types of synchronization have been studied. These include phase synchronization, generalized synchronization, frequency synchronization, complete (identical) synchronization, lag synchronization and anomalous synchronization \cite{Kuramoto:84,Brown:00,Rulkov:95,Kocarev:96,Pecora:90,Blasius:03,Pikovsky:01,Rosenblum:96,Ermentrout:81,Arnhold:99,Eroglu:17}.


Another important group of physical phenomena attributable to interactions are those associated with oscillation and amplitude deaths \cite{Bar:85,Mirollo:90,Prasad:05,SuarezVargas:09,Zakharova:13,Schneider:15,Koseska:13}. \red{Oscillation death is defined as a complete cessation of oscillation caused by the interactions, when an inhomogeneous steady state is reached. Similarly, in amplitude death, due to the interactions a homogeneous steady state is reached and the oscillations disappear.} The mechanisms leading to these two oscillation quenching phenomena are mediated by different coupling functions and conditions of interaction, including strong coupling \cite{Mirollo:90,Zhai:04}, conjugate coupling \cite{Karnatak:07},  nonlinear coupling \cite{Prasad:10}, repulsive links \cite{Hens:13} and environmental coupling \cite{Resmi:11}. These phenomena are mediated, not only by the phase dynamics of the interacting oscillators, but also by their amplitude dynamics, where the shear amplitude terms and the nonisochronicity play significant roles. Coupling functions define the mechanism through which the interaction causes the disappearance of the oscillations.

There is a large body of earlier work in which physical effects, qualitative states, or quantitative characteristics of the interactions were studied, where coupling functions constituted an integral part of the underlying interaction model, regardless of whether or not the term was used explicitly. Physical effects are very important and they are closely connected with the coupling functions. In such investigations, however, the coupling functions themselves were often not assessed, or considered as entities in their own right. In simple words, such investigations posed the question of \emph{whether} physical effects occur; while for the coupling function investigations the question is rather \emph{how} they occur.  Our emphasis will therefore be on coupling functions as entities, on the exploration and assessment of different coupling functions, and on the consequences of the interactions.

\subsubsection{Coupling strength and directionality}\label{sec22:CplDirc}

The coupling strength gives a quantitative measure of \red{the information flow between the coupled systems. In an information-theoretic context, this is defined as the transfer of information between variables in a given process.} In a theoretical treatment the coupling strength is clearly the scaling parameter of the coupling functions. There is great interest in being able to evaluate the coupling strength, for which many effective methods have been designed  \cite{Smirnov:09,Chicharro:09,Staniek:08,Palus:03a,Bahraminasab:08,Jamsek:10,Sun:15,Faes:11,Marwan:07,Mormann:00,Rosenblum:01}.
The dominant direction of influence, i.e.\ the direction of the stronger coupling, corresponds to the directionality of the interactions. Earlier, it was impossible to detect the absolute value of the coupling strength, and a number of methods exist for detection only of the directionality through measurements of the relative magnitudes of the interactions -- for example, when detecting mutual information \cite{Smirnov:09,Staniek:08,Palus:03a}, but not the physical coupling strength. The assessment of the strength of the coupling and its predominant direction can be used to establish if certain interactions exist at all. In this way, one can determine whether some apparent interactions are in fact genuine, and whether the systems under study are \red{truly} connected or not.

When the coupling function results from a number of functional components, its net strength is usually evaluated as the Euclidian norm of the individual components' coupling strengths. Grouping the separate components, for example the Fourier components of periodic phase dynamics, one can evaluate the coupling strengths of the functional groups of interest. The latter could include the coupling strength from either one system or the other, or from both of them. Thus one can detect the strengths of the self, direct and common coupling components, or of the phase response curve \cite{Kralemann:11,Faes:15,Iatsenko:13a}. In a very similar way, these ideas can be generalized for multivariate coupling in networks of interacting systems.

It is worth noting that, when inferring couplings even from completely uncoupled or very weakly-coupled systems, the methods will usually detect non-zero coupling strengths. This results mainly from the statistical properties of the signals. Therefore, one needs to be able to ascertain whether the detected coupling strengths are genuine, or spurious, just resulting from the inference method. To overcome this difficulty, one can apply surrogate testing \cite{Schreiber:00b,Kreuz:04,Andrzejak:03,Palus:98a} which generates independent, uncoupled, signals that have the same statistical properties as the original signals. The apparent coupling strength evaluated for the surrogate signals should then reflect a ``zero-level'' of apparent coupling for the uncoupled signals. By comparison, one can then assess whether the detected couplings are likely to be genuine. This surrogate testing process is also important for coupling function detection -- one first needs to establish whether a coupling relation is genuine and then, if so, to try to infer the form of the coupling function.

\subsubsection{Coupling functions in general interactions}\label{sec23:GenNonOscilatory}

The present review is focused mainly on coupling functions between interacting dynamical systems, and especially between oscillatory systems, because most studies to date have been developed in that context. However, interactions have also been studied in a broader sense for non-oscillatory, non-dynamical, systems, spread over many different fields, including for example quantum plasma interactions \cite{Marklund:06,Shukla:11}, solid state physics \cite{Higuchi:03,Farid:97,Zhang:13}, interactions in semiconductor superlattices \cite{Bonilla:05}, Josephson junction interactions \cite{Golubov:04}, laser diagnostics \cite{Stepowski:92}, interactions in nuclear physics \cite{Mitchell:10,Guelfi:07}, geophysics \cite{Murayama:82}, space science \cite{Feldstein:92,Lifton:05}, cosmology \cite{Faraoni:06,Baldi:11}, biochemistry \cite{Khramov:07}, plant science \cite{Doidy:12}, oxygenation and pulmonary circulation \cite{Ward:08}, cerebral neuroscience \cite{Liao:13}, immunology \cite{Robertson:16}, biomolecular systems \cite{Christen:08,Stamenovic:09,Dong:14}, gap junctions \cite{Wei:04}  and protein interactions  \cite{Jones:96,Teasdale:96,Okamoto:09,Gaballo:02}. In many such cases, the interactions are different in nature. They are often structural, and not effective connections in the dynamics; or the corresponding coupling functions may not have been studied in this context before. Even though we do not discuss such systems directly in this review, many of the concepts and ideas that we introduce in connection with dynamical systems can also be useful for the investigation of interactions more generally.

\section{Basic Concept of Coupling Functions}\label{sec3:Basic}
\red{
\subsection{Principle meaning}\label{sec31:Defins}

\subsubsection{Generic form of coupled systems} \label{sec311:Generic}
}
\red{
The main problem of interest is to understand the dynamics of coupled systems from their building blocks.  We start from the isolated dynamics:
\[
\dot{x} = f(x;\mu),
\]
where $f: \mathbb{R}^m \times \mathbb{R}^n \rightarrow \mathbb{R}^n$ is a differentiable vector field with $\mu$ being the set of parameters. For sake of simplicity, whenever there is no risk of confusion, we will omit the parameters.  Over the last fifty years, developments in the theory of dynamical systems have illuminated the dynamics of isolated systems. For instance, we understand their bifurcations, including those that generate periodic orbits as well as those giving rise to chaotic motion. Hence we understand the dynamics of isolated systems in some detail.

In contrast, our main interest here is to understand the dynamics of the coupled equations:
\begin{eqnarray}
\dot{x} &=& f_1(x) + g_1(x,y) \label{couplingG} \\
\dot{y} &=& f_2(y) + g_2(x,y),
\end{eqnarray}
where $f_{1,2}$ are vector fields describing the isolated dynamics (perhaps with different dimensions) and $g_{1,2}$ are the coupling functions. The latter are our main objects of interest. We will assume that they are at least twice differentiable.

Note that we could also study this problem from an abstract point of view by representing the equations as:
\begin{eqnarray}
\dot{x} &=& q_1(x,y) \label{couplingQ} \\
\dot{y} &=& q_2(x,y),
\end{eqnarray}
where the functions $q_{1,2}$ incorporate both the isolated dynamics and the coupling functions. This notation for inclusion of coupling functions, with no additive splitting between the interactions and the isolated dynamics, can sometimes be quite useful \cite{Aronson:90,Pereira:14}. Examples include coupled cell networks \cite{Ashwin:05a}, or the provision of full Fourier expansions \cite{Rosenblum:01,Kiss:05} when inferring coupling functions from data.
}
\red{
\subsubsection{Coupling function definition} \label{sec312:Generic}
}

\begin{figure*}
\floatbox[{\capbeside\thisfloatsetup{capbesideposition={right,top},capbesidewidth=3.2cm}}]{figure}[\FBwidth]
{\caption{(color online). The state of synchronization described through phase difference dynamics, $\dot \psi$ versus $\psi$. Depending on the existence of stable equilibria, the oscillators can be synchronized (a),(c) or unsynchronized (b). Stable points are shown with white circles, while unstable with black circles.  Adapted from \citet{Kuramoto:84}. }\label{fig:CF_kuramoto}}
{\includegraphics[width=0.78\textwidth,angle=0]{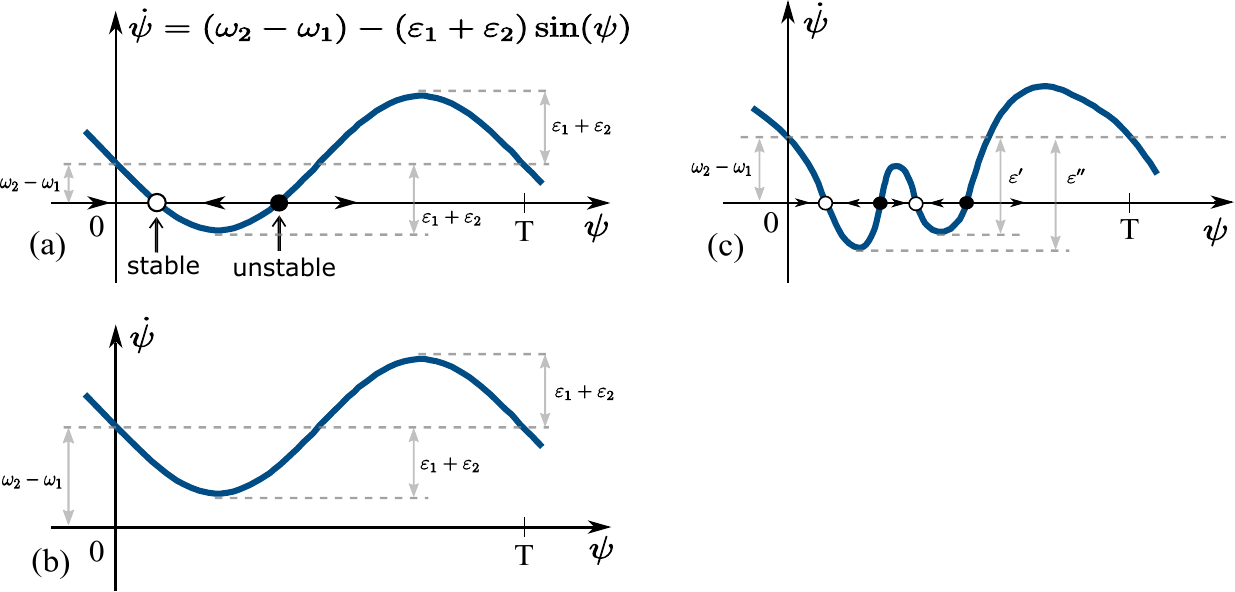}}
\end{figure*}

\red{
\emph{Coupling functions describe the physical rule specifying how the interactions occur}. Being directly connected with the functional dependences,  coupling functions focus not so much on \emph{whether} there are interactions, but more on \emph{how} they appear and develop. For instance,} the magnitude of the phase coupling function affects directly the oscillatory frequency and describes how the oscillations are being accelerated or decelerated by the influence of the other oscillator. Similarly, if one considers the amplitude dynamics of interacting dynamical systems, the magnitude of the coupling function will prescribe how the amplitude is increased or decreased by the interaction.

\red{
A coupling function can be described in terms of its {\it strength} and {\it form}. While the strength is a relatively well-studied quantity, this is not true of the form. It is the functional form that has provided a new dimension and perspective, probing directly the mechanisms of interaction.  In other words, the \emph{mechanism is defined by the functional form which, in turn, specifies the rule and process through which the input values are translated into output values} i.e.\ in terms of one system (System A) it prescribes how the input influence from another system (System B) gets translated into consequences in the output of System A. }
In this way the coupling function can describe the qualitative transitions between distinct states of the systems e.g.\ routes into and out of synchronization. 
Decomposition of a coupling function provides a description of the functional contributions from each separate subsystem within the coupling relationship. Hence, the use of coupling functions amounts to much more than just a way of investigating correlations and statistical effects: it reveals the mechanisms underlying the functionality of the interactions.

\red{
\subsubsection{Example of coupling function and synchronization} \label{sec313:ExSync}
}

To illustrate the fundamental role of coupling functions in synchronization, we consider a simple example of two coupled phase oscillators \cite{Kuramoto:84}:
\begin{equation}\label{eq:phs}
\begin{split}
\dot \phi_1 &=\omega_1+\varepsilon_1\sin(\phi_2-\phi_1)\\
\dot \phi_2 &=\omega_2+\varepsilon_2\sin(\phi_1-\phi_2),
\end{split}
\end{equation}
where $\phi_1,\phi_2$ are the phase variables of the oscillators, $\omega_1,\omega_2$ are their natural frequencies, $\varepsilon_1,\varepsilon_2$ are the coupling strength parameters, and the coupling functions of interest are both taken to be sinusoidal. (For further details including, in particular, the choice of the coupling functions, see also section \ref{sec4:Theory}). Further, we consider coupling that depends only on the phase difference $\psi=\phi_2-\phi_1$. In this case, from $\dot \psi = \dot \phi_2 - \dot \phi_1$ and Eqs.\ (\ref{eq:phs}) we can express the interaction in terms of $\psi$ as:
\begin{equation}\label{eq:psi}
\begin{split}
\dot \psi &=  \Delta_\omega+\varepsilon q(\psi)=(\omega_2-\omega_1)-(\varepsilon_1+\varepsilon_2)\sin(\psi).
\end{split}
\end{equation}
Synchronization will then occur if the phase difference $\psi$ is bounded, i.e.\ if  Eq.\ (\ref{eq:psi}) has at least one stable-unstable pair of solutions \cite{Kuramoto:84}. Depending on the form of the coupling function, in this case the sine form $q(\psi)= \sin(\psi)$, and on the specific parameter values, a solution may exist. For the coupling function given by Eq.\ (\ref{eq:psi}) one can determine that the condition for synchronization to occur is $|\varepsilon_1+\varepsilon_2|\geq |\omega_2-\omega_1|$.

Fig.\ \ref{fig:CF_kuramoto} illustrates schematically the connection between the coupling function and synchronization. An example of a synchronized state is sketched in Fig.\ \ref{fig:CF_kuramoto}(a). The resultant coupling strength $ \varepsilon=(\varepsilon_1+\varepsilon_2)$ has larger values of the frequency difference $\Delta_\omega=\omega_2-\omega_1$ at certain points within the oscillation cycle. As the condition $\dot \psi=0$ is fulfilled, there is a pair of stable and unstable equilibria, and synchronization exists between the oscillators. Fig.\ \ref{fig:CF_kuramoto}(b) shows the same functional form, but the oscillators are not synchronized because the frequency difference is larger than the resultant coupling strength. By comparing Figs.\ \ref{fig:CF_kuramoto}(a) and (b) one can note that while the form of the curve defined by the coupling function is the same in each case, the curve can be shifted up or down by choice of the frequency and coupling strength parameters. \red{For certain critical parameters, the system undergoes a saddle-node bifurcation, leading to a stable synchronization.}

The coupling functions of real systems are often more complex than the simple sine function presented in Fig.\ \ref{fig:CF_kuramoto}(a) and (b). For example, Fig.\ \ref{fig:CF_kuramoto}(c) also shows a synchronized state, but with an arbitrary form of coupling function that has two pairs of stable-unstable points. 
As a result, there could be two critical coupling strengths ($\varepsilon'$ and $\varepsilon''$) and either one, or both, of them can be larger than the frequency difference $\omega_2-\omega_1$, leading to stable equilibria and fulfilling the synchronization condition. This complex situation could cause bistability (as will be presented below in relation to chemical experiments Sec.\ \ref{sec641:MetChem}). Thus comparison of Fig.\ \ref{fig:CF_kuramoto}(a) and (c) illustrates the fact that, within the synchronization state, there can be different mechanisms defined by different forms of coupling function.

\subsection{History}\label{sec32:History}

The concepts of coupling functions, and of interactions more generally, had emerged as early as the first studies of the physical effects of interactions, such as the synchronization and oscillation death phenomena. In the seventeenth century, Christiaan Huygens observed and described the interaction phenomenon exhibited by two mechanical clocks \cite{Huygens:7386}. He noticed that their pendula, which beat differently when the clocks were attached to a rigid wall, would synchronize themselves when the clocks were attached to a thin beam. He realised that the cause of the synchronization was the very small motion of the beam, and that its oscillations communicated some kind of motion to the clocks. In this way, Huygens described the physical notion of the coupling -- the small motion of the beam which mediated the mutual motion (information flow) between the clocks that were fixed to it.

In the nineteenth century, John William Strutt, Lord Rayleigh, documented the first comprehensive theory of sound \cite{Rayleigh:96}. He observed and described the interaction of two organ pipes with holes distributed in a row. His peculiar observation was that for some cases the pipes could almost reduce one another to silence. He was thus observing the oscillation death phenomenon, as exemplified by the quenching of sound waves.


Theoretical investigations of oscillatory interactions emerged soon after the discovery of the triode generator in 1920 and the ensuing great interest in periodically alternating electrical currents. Appleton and van der Pol considered coupling in electronic systems and attributed it to the effect of synchronizing a generator with a weak external force \cite{Appleton:22,vanderpol:27}. Other theoretical works on coupled nonlinear systems included studies of the synchronization of mechanically unbalanced vibrators and rotors \cite{Blekhman:53}, and the theory of general nonlinear oscillatory systems \cite{Malkin:56}. Further theoretical studies of coupled dynamical systems, explained phenomena ranging from biology, to laser physics, to chemistry \cite{Winfree:67,Kuramoto:75,Glass:79,Haken:75,Wiener:63}. Two of these earlier theoretical works \cite{Winfree:67,Kuramoto:75} have particular importance and impact for the theory of coupling functions.

In his seminal work \citet{Winfree:67} studied biological oscillations and population dynamics of limit-cycle oscillators theoretically. 
Notably, he considered the phase dynamics of interacting oscillators, where the coupling function was a product of two periodic functions of the form:
\begin{equation}\label{eq:Winfree}
\begin{split}
q_1(\phi_1,\phi_2)=Z(\phi_1)I(\phi_2).
\end{split}
\end{equation}
Here, $I(\phi_2)$ is the influence function through which the second oscillator affects the first, while the sensitivity function $Z(\phi_1)$ describes how the first observed oscillator responds to the influence of the second one. (This was subsequently generalized for the whole population in terms of a mean field \cite{Winfree:67,Winfree:80}). Thus, the influence and sensitivity functions $I(\phi_2)$, $Z(\phi_1)$, as integral components of the coupling function, described the physical meaning of the separate roles within the interaction between the two oscillators. The special case $I(\phi_2)=1+\cos(\phi_2)$ and $Z(\phi_1)=\sin(\phi_1)$ has often been used \cite{Ariaratnam:01,Winfree:80}.

Arguably, the most studied framework of coupled oscillators is the Kuramoto model. It was originally introduced in 1975 through a short conference paper \cite{Kuramoto:75}, followed by a more comprehensive description in an epoch-making book \cite{Kuramoto:84}. Today this model is the cornerstone for many studies and applications \cite{Acebron:05,Strogatz:00}, including neuroscience \cite{Breakspear:10,Cumin:07,Cabral:14}, Josephson-junction arrays \cite{Wiesenfeld:96,Wiesenfeld:98,Filatrella:00}, power grids \cite{Dorfler:12,Filatrella:08}, glassy states \cite{Iatsenko:14a} and laser arrays \cite{Vladimirov:03}. The model reduces the full oscillatory dynamics of the oscillators to their phase dynamics, i.e.\ to so-called phase oscillators, and it studies synchronization phenomena in a large population of such oscillators \cite{Kuramoto:84}. By setting out a mean-field description for the interactions, the model provides an exact analytic solution.

At a recent conference celebrating ``40 years of the Kuramoto Model", held at the Max Planck Institute for the Physics of Complex Systems, Dresden, Germany, Yoshiki Kuramoto presented his own views of how the model was developed, and described its path from initial ignorance on the part of the scientific community to dawning recognition followed by general acceptance: a video message is available \cite{Kuramoto_Video}. Kuramoto  devoted particular attention to the \emph{coupling function} of his model, noting that:
\begin{quote}
In the year of 1974, I first came across Art Winfree's famous paper [\cite{Winfree:67}] \ldots I was instantly fascinated by the first few paragraphs of the introductory section of the paper, and especially my interest was stimulated when he spoke of the analogy between synchronization transitions and phase transitions of ferroelectrics, [\ldots]. [There was a] problem that mutual coupling between two magnets (spins) and mutual coupling of oscillators are quite different. For magnetic spins the interaction energy is given by a scalar product of a two spin vectors, which means that in a particular case of planar spins the coupling function is given by a sinusoidal function of phase difference. In contrast, Winfree's coupling function for two oscillators is given by a product of two periodic functions, [\ldots], and it seemed that this product form coupling was a main obstacle to mathematical analysis. [\ldots] I knew that product form coupling is more natural and realistic, but I preferred the sinusoidal form of coupling because my interest was in finding out a solvable model.
\end{quote}

\red{Kuramoto studied complex equations describing oscillatory chemical reactions \cite{Kuramoto:75b}. In building his model, he considered  phase dynamics and all-to-all diffusive coupling rather than local coupling, took the mean-field limit, introduced a random frequency distribution, and assumed that a limit-cycle orbit is strongly attractive \cite{Kuramoto:75}.} As already mentioned, Kuramoto's coupling function was a sinusoidal function of the phase difference:
\begin{equation}\label{eq:Kuramoto}
\begin{split}
q_1(\phi_1,\phi_2)=\sin(\phi_2-\phi_1).
\end{split}
\end{equation}
The use of the phase difference reduces the dimensionality of the two phases and provides a means whereby the synchronization state can be determined analytically in a more convenient way (see also Fig.\ \ref{fig:CF_kuramoto}).


The inference of coupling functions from data appeared much later than the theoretical models. The development of these methods was mostly dictated by the increasing accessibility and power of the available computers. One of the first methods for the extraction of coupling functions from data was effectively associated with detection of the directionality of coupling \cite{Rosenblum:01}. Although directionality was the main focus, the method also included the reconstruction of functions that closely resemble coupling functions. Several other methods for coupling function extraction followed, including those by \citet{Kiss:05}, \citet{Miyazaki:06}, \citet{Tokuda:07}, \citet{Kralemann:08}, and \citet{Stankovski:12b}, and it remains a highly active field of research.

\red{
\subsection{Different domains and usage}\label{sec33:PhaseCF}
}

\subsubsection{Phase coupling functions}\label{sec331:PhaseCF}


A widely used approach for the study of the coupling functions between interacting oscillators is through their phase dynamics \cite{Kuramoto:84,Winfree:67,Pikovsky:01,Ermentrout:86}. If the system has a stable limit-cycle, one can apply \emph{phase reduction procedures} (see Sec.\ \ref{sec:weakCpl} for further theoretical details) which systematically approximate the high-dimensional dynamical equation of a perturbed limit cycle oscillator with a one-dimensional reduced-phase equation, with just a \emph{single} phase variable $\phi$ representing the oscillator state \cite{Nakao:15}. In uncoupled or weakly-coupled contexts, the phases are associated with zero Lyapunov stability, which means that they are susceptible to tiny perturbations. In this case, one loses the amplitude dynamics, but gains simplicity in terms of the single dimension phase dynamics, which is often sufficient to treat certain effects of the interactions, e.g.\ phase synchronization. \red{Thus phase connectivity is defined by the connection and influence between such phase systems.}

To present the basic physics underlying a coupling function in the phase domain, we consider an elementary example of two phase oscillators that are unidirectionally phase-coupled:
\begin{equation}
\begin{split}
\dot {\phi_1}=&\omega_1\\
\dot {\phi_2}=&\omega_2 +q_2(\phi_1,\phi_2)=\omega_2 +\cos(\phi_1+\pi /2.5).
\label{eq:exe}
\end{split}
\end{equation}
Our aim is to describe the effect of the coupling function $q_2(\phi_1,\phi_2)$ through which the first oscillator influences the second one. From the expression for $\dot {\phi_2}$ in Eq.\ (\ref{eq:exe}) one can appreciate the fundamental role of the coupling function: $q_2(\phi_1,\phi_2)$ is added to the frequency $\omega_2$. Thus \emph{changes in the magnitude of $q_2(\phi_1,\phi_2)$ will contribute to the overall change of the frequency of the second oscillator.} Hence, depending on the value of $q_2(\phi_1,\phi_2)$, the second oscillator will either accelerate or decelerate relative to its uncoupled motion.

\begin{figure}
{\caption{(color online). Schematic illustration of a phase dynamics coupling function. The first oscillator $x_1$  influences the second oscillator $x_2$ unidirectionally, as indicated by the directional diagram on the left of the figure. (a) Amplitude signal $x_1(t)$ during one cycle of period $T_1$. (b) Coupling function $q_2(\phi_1,\phi_2)$ in $\{ \phi_1,\phi_2 \}$ space. (c) $\phi_2$-averaged projection of the coupling function $q_2(\phi_1,\phi_2)$. (d) Amplitude signal of the second driven oscillator $x_2(t)$, during one cycle of the first oscillator. From  \citet{Stankovski:15a}.}\label{fig:CF_ph}}
{\includegraphics[width=1\textwidth,angle=0]{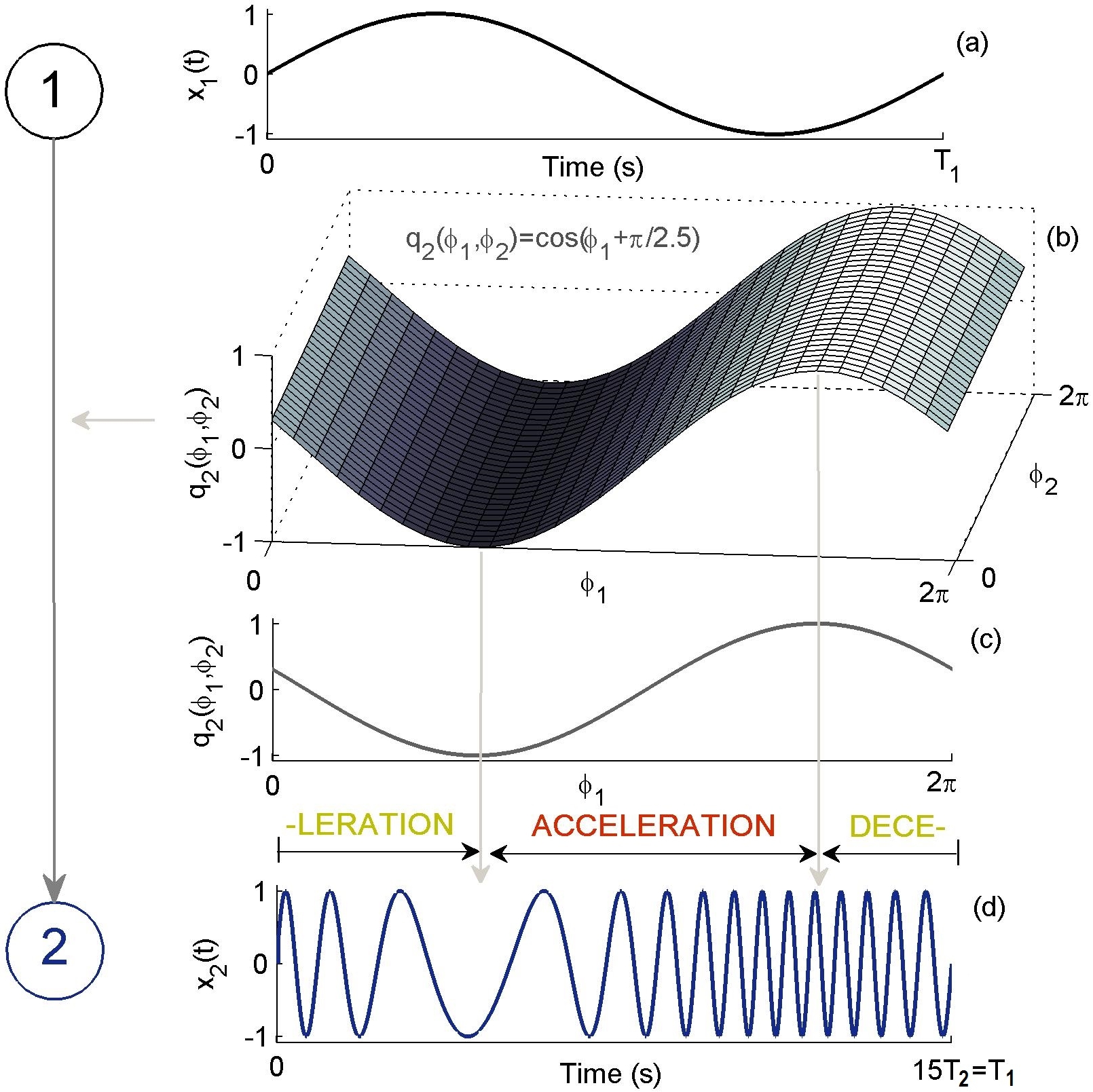}}
\end{figure}

The description of the phase coupling function is illustrated schematically in Fig.\ \ref{fig:CF_ph}. Because in real situations one measures the amplitude state of signals, we explain how the amplitude signals (Fig.\ \ref{fig:CF_ph}(a) and (d)) are affected depending on the specific phase coupling function (Fig.\ \ref{fig:CF_ph}(b) and (c)). In all plots, time is scaled relative to the period $T_1$ of the amplitude of the signal originating from the first oscillator $x_1(t)$ (e.g.\ $x_1(t)=\sin(\phi_1)$). For convenient visualisation of the effects we set the second oscillator to be fifteen times slower than the first oscillator: $\omega_2/\omega_1=15$. The particular coupling function $q_2(\phi_1,\phi_2)=\cos(\phi_1+\pi /2.5)$ presented on a $2\pi \times 2\pi$  grid (Fig.\ \ref{fig:CF_ph}(b)) resembles a shifted cosine wave, which changes only along the $\phi_1$-axis, \blue{like a direct coupling component}.  Because all the changes occur along the $\phi_1$-axis, and for easier comparison, we also present in Fig.\ \ref{fig:CF_ph}(c) a $\phi_2$-averaged projection of $q_2(\phi_1,\phi_2)$.

Finally, Fig.\ \ref{fig:CF_ph}(d) shows how the second oscillator $x_2(t)$ is affected by the first oscillator in time in relation to the phase of the coupling function: when the coupling function $q_2(\phi_1,\phi_2)$ is increasing, the second oscillator $x_2(t)$ accelerates; similarly, when $q_2(\phi_1,\phi_2)$ decreases, $x_2(t)$ decelerates. Thus the form of the coupling function $q_2(\phi_1,\phi_2)$ shows in detail the mechanism through which the dynamics and the oscillations of the second oscillator are affected: in this case they were alternately accelerated or decelerated by the influence of the  first oscillator.

\begin{figure} [t]
{\caption{(color online). Two characteristic coupling functions in the phase domain. (a) The coupling function $q(\phi_1,\phi_2)$ is of sinusoidal form for the phase difference, as used in the Kuramoto model. (b) The coupling function $q(\phi_1,\phi_2)$ is a product of the influence and sensitivity functions, as used in the Winfree model.}\label{fig:CF_ph2}}
{\includegraphics[width=1\textwidth,angle=0]{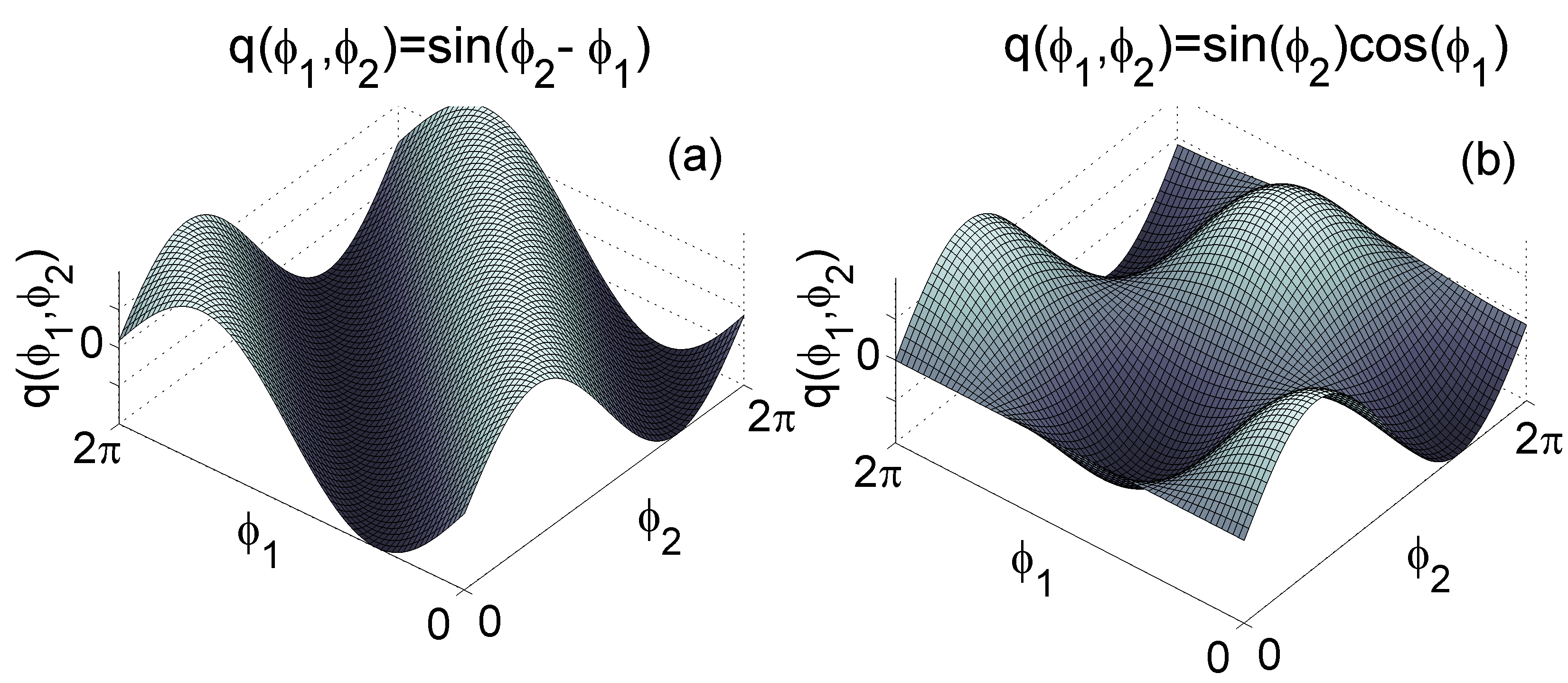}}
\end{figure}

Of course, coupling functions can in general be much more complex than the simple example presented ($\cos(\phi_1+\pi /2.5)$). This form of phase coupling function with a direct contribution (predominantly) only from the other oscillator is often found \blue{as a coupling component} in real applications, as will be discussed below. Other characteristic phase coupling functions of that kind could include the coupling functions from the Kuramoto model (Eq.\ (\ref{eq:Kuramoto})) and the Winfree model (Eq.\ (\ref{eq:Winfree})), as shown in Fig.\ \ref{fig:CF_ph2}. The sinusoidal function of the phase difference from the Kuramoto model exhibits a diagonal form in Fig.\ \ref{fig:CF_ph2}(a), while the influence-sensitivity product function of Winfree model is given by a more complex form spread differently along the two-dimensional space in Fig.\ \ref{fig:CF_ph2}(b). Although these two functions differ from those in the previous example (Fig.\ \ref{fig:CF_ph}), the procedure used for their interpretation is the same.

\subsubsection{Amplitude coupling functions}\label{sec332:AmplitudeCF}

Arguably, it is more natural to study amplitude dynamics than phase dynamics, as the former is directly observable while the phase needs to be derived. 
Real systems often suffer from the ``curse of dimensionality'' \cite{Keogh:11} in that not all of the features of a possible (hidden) higher-dimensional space are necessarily observable through the low-dimensional space of the measurements. Frequently, a delay embedding theorem \cite{Takens:81} is used to reconstruct the multi-dimensional dynamical system from data. In real application with non-autonomous and non-stationary dynamics, however the theorem often does not give the desired result \cite{Clemson:14b}. Nevertheless, amplitude state interactions also have a wide range of applications both in theory and methods, especially in the cases of chaotic systems, strong couplings, delayed systems, and large nonlinearities, 
including cases where complete synchronization \cite{Cuomo:93,Stankovski:14a,Kocarev:95} and \emph{generalized synchronization} \cite{Rulkov:95,Kocarev:96,Stam:02,Abarbanel:93,Arnhold:99} has been assessed through observation of  amplitude state space variables.

Amplitude coupling functions affect the interacting dynamics by increasing or decreasing the state variables. 
\red{Thus amplitude connectivity is defined by the connection and influence between the amplitude dynamics of the systems.} The form of the amplitude coupling function can often be a polynomial function or diffusive difference between the states.

To present the basics of amplitude coupling functions, we discuss a simple example of two interacting Poincar\'e limit-cycle oscillators. In the autonomous case, each of them  is given by the polar (radial $r$ and angular $\phi$) coordinates as: $\dot r=r(1-r)$ and $\dot \phi = \omega$. In this way, a Poincar\'e oscillator is given by a circular limit-cycle and monotonically growing (isochronous) phase defined by the frequency parameter. In our example, we transform the polar variables to Cartesian (state space) coordinates $x=r \cos(\phi)$, $y=r \sin(\phi)$, and we set unidirectional coupling, such that the first (autonomous) oscillator:
\begin{equation}
\begin{split}
\dot x_1&= \big (1-\sqrt{x_1^2+y_1^2} \big) x_1  -\omega_1\, y_1,\\
\dot y_1&= \big (1-\sqrt{x_1^2+y_1^2} \big) y_1  +\omega_1\, x_1,
\label{equ:num_osc1}
\end{split}
\end{equation}
is influencing the $x_2$ state of the second oscillator through the quadratic coupling function $q_2(x_1,y_1,x_2,y_2)=x_1^2$:
\begin{equation}
\begin{split}
\dot x_2&= \big (1-\sqrt{x_2^2+y_2^2} \big)  x_2  -\omega_2\, y_2  + \varepsilon x_1^2,\\
\dot y_2&= \big (1-\sqrt{x_2^2+y_2^2} \big)  y_2  +\omega_2\, x_2.
\label{equ:num_osc2}
\end{split}
\end{equation}
For simpler visual presentation we choose the first oscillator to be twenty times faster than the second one, i.e.\ their frequencies are in the ratio $\omega_2/\omega_1=20$, and we set a relatively high coupling strength $\varepsilon=5$.

\begin{figure}
{\caption{(color online). Schematic illustration of an amplitude dynamics coupling function. The first oscillator Eqs.\ (\ref{equ:num_osc1})   is influencing the second oscillator Eqs.\ (\ref{equ:num_osc2}) unidirectionally, as indicated by the directional diagram on the left of the figure. (a) Amplitude state signal $x_1(t)$ during one cycle of period $T_1$. (b) Coupling function $q_2(x_1,x_2)$ in $\{ x_1,x_2 \}$ space during one period of each of the oscillations. (c) $x_2$-averaged projection of the coupling function $q_2(x_1,x_2)$. (d) Amplitude signal of the second (driven) oscillator $x_2(t)$, during one cycle of the first oscillator.}\label{fig:CF_amp}}
{\includegraphics[width=1\textwidth,angle=0]{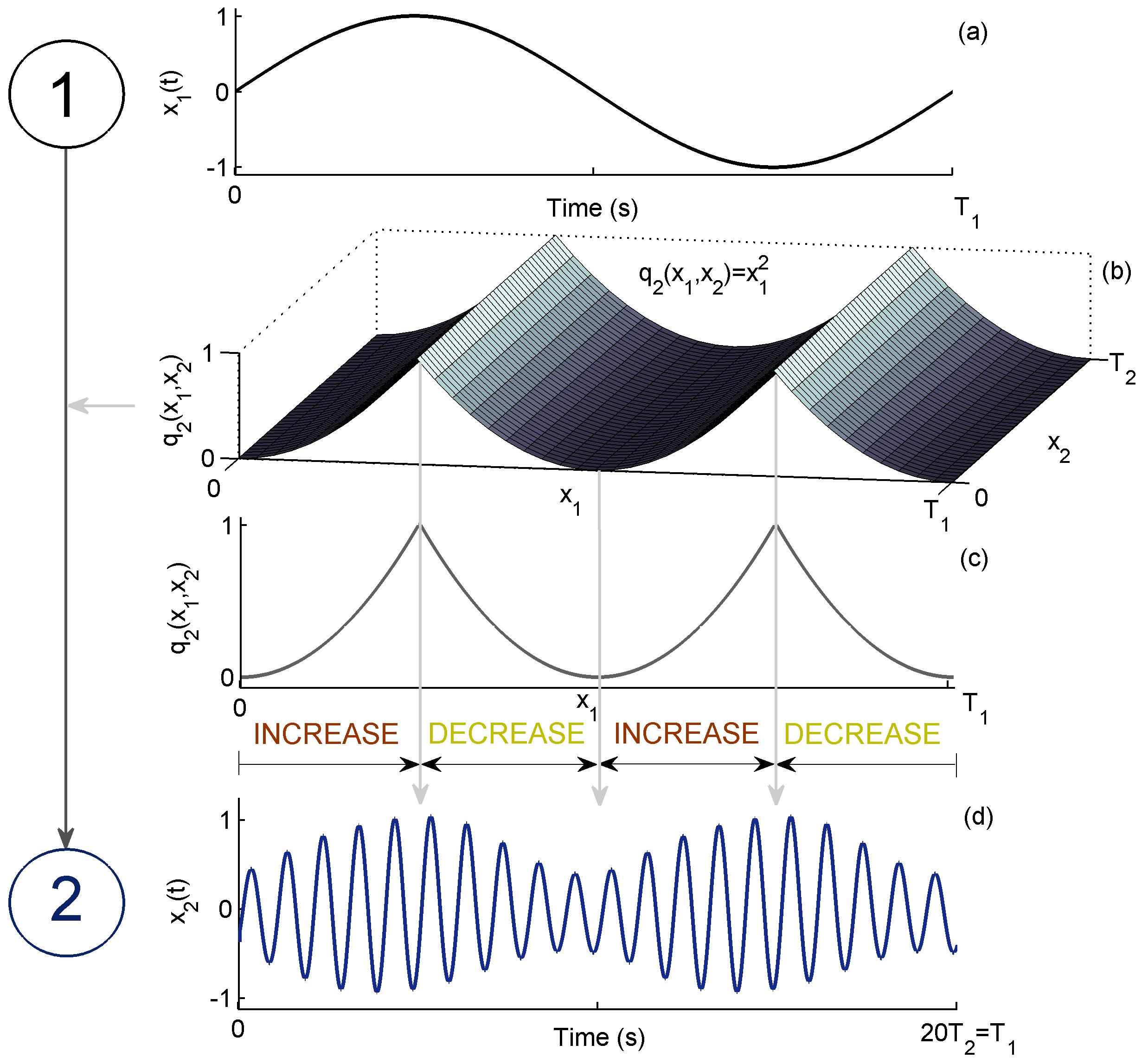}}
\end{figure}

\begin{figure*}
\floatbox[{\capbeside\thisfloatsetup{capbesideposition={right,center},capbesidewidth=6.75cm}}]{figure}[\FBwidth]
{\caption{(color online). Inference of multivariate interactions. True (structural) configurations (left), and the reconstructed phase model (right). Middle: the table shows the corresponding inferred coupling strengths. Note the multivariate triplet link -- the arrows from the centres of the diagrams. From \citet{Kralemann:11}. }\label{fig:CF_multi1}}
{\includegraphics[width=0.58\textwidth,angle=0]{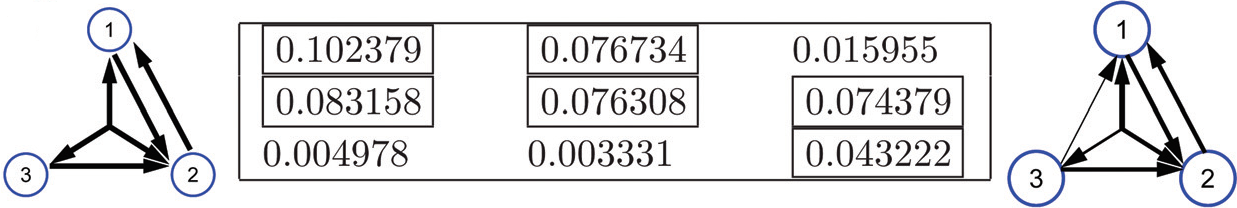}}
\end{figure*}

The description of the amplitude coupling function is illustrated schematically in Fig.\ \ref{fig:CF_amp}. 
In theory, the coupling function $q_2(x_1,y_1,x_2,y_2)$ has four variables, but for better visual illustration, and because the dependence is only on $x_1$, we show it only in respect to the two variables $x_1$ and $x_2$ i.e.\ $q_2(x_1,x_2)$. 
The form of the coupling function is quadratic, and it changes only along the $x_1$-axis, as shown in Figs.\ \ref{fig:CF_amp}(b) and (c). Finally, Fig.\ \ref{fig:CF_amp}(d) shows how the second oscillator $x_2(t)$ is affected by the first oscillator in time via the coupling function: when the quadratic coupling function $q_2(x_1,x_2)$ is increasing, the amplitude of the second oscillator $x_2(t)$ increases; similarly, when $q_2(x_1,x_2)$ decreases, $x_2(t)$ decreases as well. 

The particular example chosen for presentation used a quadratic function $x_1^2$; other examples include a direct linear coupling function e.g.\ $x_1$, or a diffusive coupling e.g.\ $x_2-x_1$ \cite{Aronson:90,Kocarev:96,Mirollo:90}. There are a number of methods which have inferred models that include amplitude coupling functions inherently \cite{Voss:04,Smelyanskiy:05a,Friston:02} or have pre-estimated most probable models \cite{Berger:96}, but without including explicit assessment of the coupling functions. 
Due to the multi-dimensionality and the lack of a general property in a dynamical system (like for example the periodicity in phase dynamics), there are countless possibilities for generalization of the coupling function. In a sense, this lack of general models is a deficiency in relation to the wider treatment of amplitude coupling functions. There are open questions here and much room for further work on generalising such models, in terms both of theory and methods, taking into account the amplitude properties of subgroups of dynamical systems, including for example the chaotic, oscillatory, or reaction-diffusion nature of the systems.

\subsubsection{ Multivariate coupling functions}\label{sec333:MultiCF}

Thus far, we have been discussing pairwise coupling functions between two systems. In general, when interactions occur between more than two dynamical systems, in a network (Sec.\ \ref{sec47:Thr_Net}), there may be multivariate coupling functions with more than two input variables. For example, a multivariate phase coupling function could be $q_1(\phi_1,\phi_2,\phi_3)$, which is a triplet function of influence in the dynamics of the first phase oscillator caused by a common dependence on three other phase oscillators. Such joint functional dependences can appear as clusters of subnetworks within a network \cite{Albert:02}.

Multivariate interactions have been the subject of much attention recently, especially in developing methods for detecting the couplings \cite{Baselli:97,Nawrath:10,Frenzel:07,Palus:07a,Kralemann:11,Duggento:12,Faes:11}. This is particularly relevant in networks, where one can miss part of the interactions if only pairwise links are inferred, or a spurious pairwise link can be inferred as being independent when they are actually part of a multivariate joint function. In terms of networks and graph theory, the multivariate coupling functions relate to \emph{hypergraph}, which is \red{defined as} a generalization of a graph where an edge (or connection) can connect any number of nodes (or vertices) \cite{Karypis:00,Zass:08,Weighill:15}.

Multivariate coupling functions have been studied by inference of small-scale networks where the structural coupling can differ from the inferred effective coupling \cite{Kralemann:11}. The authors considered a network of three van der Pol oscillators where, in addition to pairwise couplings, there was also a joint multivariate cross-coupling function, for example of the form $\varepsilon x_2 x_3$ in the dynamics of the first oscillator $\ddot x_1$. Due to the latter coupling, the effective phase coupling function is of a multivariate triplet nature.  By extracting the phases and applying an inference method, the effective phase coupling was reconstructed, as illustrated by the example in Fig.\ \ref{fig:CF_multi1}. Comparing the true (Fig.\ \ref{fig:CF_multi1} left) and the inferred  effective (Fig.\ \ref{fig:CF_multi1} right) diagrams, one can see that an additional pairwise link from the third to the first oscillator has been inferred. If the pairwise inference alone was being investigated one might conclude, wrongly, that this direct pairwise coupling was genuine and the only link -- whereas in reality it is just an indirect effect from the actual joint multivariate coupling. In this way, the inference of multivariate coupling functions can provide a deeper insight into the connections in the network.

A corollary is the detection of triplet synchronization \cite{Kralemann:13,Jia:15}. This is a synchronization phenomenon which has an explicit multivariate coupling function of the form $q_1(\phi_1,\phi_2,\phi_3)$ and which is tested in respect of the condition $|m\phi_1+n\phi_2+l\phi_3| \leq {\rm const}$, for $n,m,l$ negative or positive. It is shown that the state of triplet synchronization can exist, even though each pair of systems remains asynchronous.

\begin{figure}[b!]
{\caption{(color online). Multivariate triplet coupling functions between neural oscillations. The phase coupling function $q_\gamma(\phi_\theta,\phi_\alpha)$ shows the influence that $\theta$ and $\alpha$  jointly insert on the $\gamma$ cortical oscillations. From \citet{Stankovski:15a}. }\label{fig:CF_multi2}}
{\includegraphics[width=0.89\textwidth,angle=0]{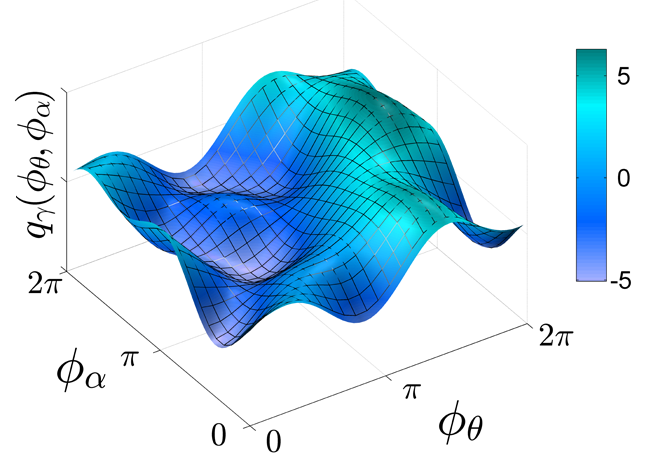}}
\end{figure}

The brain mediates many oscillations and interactions on different levels \cite{Park:13}. \red{Interactions between oscillations in different frequency bands are referred to as cross-frequency coupling in neuroscience \cite{Jensen:07}.} Recently, neural cross-frequency coupling functions were extracted from multivariate networks \cite{Stankovski:15a} (see also Sec.\ \ref{sec643:MetNeuro}). The network interactions between the five brainwave oscillations $\delta$, $\theta$, $\alpha$, $\beta$ and $\gamma$ were analysed by reconstruction of the multivariate phase dynamics, including the inference of triplet and quadruplet coupling functions. Fig.\ \ref{fig:CF_multi2} shows a triplet coupling function of how the $\theta$ and $\alpha$ influence $\gamma$ brain oscillations. It was found that the influence from theta oscillations is greater than from alpha, and that there is significant acceleration of gamma oscillations when the theta phase cycle  changes from $\pi$ to $2\pi$.


Very recently, \citet{Bick:16b} have shown theoretically that symmetrically-coupled phase oscillators with multivariate (or non-pairwise) coupling functions can yield rich dynamics, including the emergence of chaos. This was observed even for as few as $N = 4$ oscillators. In contrast to the Kuramoto-Sakaguchi equations, the additional multivariate coupling functions mean that one can find attracting chaos for a range of normal-form parameter values. 
Similarly, it was found that even the standard Kuramoto model can be chaotic with a finite number of oscillators \cite{Popovych:05a}.


\red{
\subsubsection{ Generality of coupling functions}\label{sec334:CFGeneral}

\red{
The coupling function is well-defined from a theoretical perspective. That is, once we have the model (as in Eqs.\ \ref{couplingG}), the coupling function is unique and fixed. The solutions of the equations also depend continuously on the coupling function.  Small changes in the coupling function will cause only small changes in the solutions over finite time intervals. If solutions are attracted to some set exponentially and uniformly fast, then small changes in the coupling do not affect the stability of the system.

When we want to infer the coupling function from data we can face a number of challenges in obtaining a unique result (Sec.\ \ref{sec5:Methods}). Typically, we measure only projections of the coupling function, which might in itself lead to non-uniqueness of the estimate.  That is, we project the function (which is infinite-dimensional) onto a finite-dimensional vector space. In doing so, we could lose some information and, generically, it is not possible to estimate the function uniquely (even without taking account of noise and perturbations). Furthermore, the final  form of the estimated function will depend on the choice and number of base functions. For example, the choice of Fourier series or general orthogonal polynomials as base functions can affect slightly the final estimate of the coupling function. \emph{The choice of  which base functions to be used is infinite.} Even though many aspects of coupling functions (like the number of arguments, decomposition under an appropriate model, analysis of coupling function components, prediction with coupling functions, etc.), can be applied with great generality, the coupling functions themselves cannot be determined uniquely.}


\begin{table*}[]
\centering
\caption{\red{Different examples of coupling functions $q$. These pairwise coupling functions (CFs) are considered in relation to the system: $\dot x=f(x)+q(x,y)$.  }  }
\label{tab:CFs}
\begin{tabular}{l l l r}
  \hline
Type of CF  &  Model & Meaning & Reference \\
\hline
\hline
Direct     & $q(x,y) = q(y)$ & unidirectional influence & \cite{Aronson:90} \\
Diffusive & $q(x,y) = q(y-x)$ & dependence on state difference & \cite{Kuramoto:84} \\
Reactive & $q(x,y) = (\varepsilon + i \beta)q(x - y)$ & complex coupling strength & \cite{Cross:06} \\
Conjugate &  $q(x,y) = q(x-Py)$ & $P$ permutes the variables &  \cite{Karnatak:07} \\
Chemical synapse & $q(x,y) = g(x)S(y)$ & $S$ is a sigmoidal &  \cite{Cosenza:01} \\
 Environmental  & $q(x,y)  \approx \varepsilon \int_0^t e^{-\kappa (t-s)}(x(s) + y(s))ds \text{  }\text{  }$  &  given by a differential equation & \cite{Resmi:11} \\
  \hline
\end{tabular}
\end{table*}

In the literature, authors often speak of the commonly-used coupling functions including, but not limited to, those listed in Table \ref{tab:CFs}.  
Note that reactive and diffusive coupling have functionally the same form, the difference being that the reactive case includes complex amplitudes. This results in a phase difference between the coupling and the dynamics. Also in the literature, a diffusive coupling function $q(y-x)$ satisfying a local condition $q^{\prime}(0) <0$ is called dissipative coupling \cite{Rulkov:92}. This condition resembles Fick's law as the coupling forces the coupled system to converge towards the same state. When $q^{\prime}(0) >0$ the coupling is called repulsive \cite{Hens:13}. Chemical synapses are an important form of coupling where the influences of $x$ and $y$ appear together as a product. There are also other interesting forms of coupling such as the geometric mean and further generalizations \cite{Petereit:17,Prasad:10}. In environmental coupling, the function is given by the solution of a differential equation. In this case one can consider $\dot y = -\kappa y + \varepsilon (x(t) + y(t) )$ for $\kappa >0$, so that the variables are considered as external fields driving the equation. Its solution $y(t) = y(t;x,y)$ is taken as the coupling function $q(x,y)$ and, for $t\gg 1$, is given in the table. The generality of coupling functions, and the fact that the form can come from an unbounded set of functions, were used to construct the encryption key in a secure communications protocol \cite{Stankovski:14a} (see Sec.\ \ref{sec645:Comm}).

%

}

\subsection{ Coupling functions revealing mechanisms}\label{sec34:CFMechanism}

The functional form is a qualitative property that defines the mechanism and acts as an additional dimension to complement the quantitative characteristics such as the coupling strength, directionality, frequency parameter and limit-cycle shape parameters. 
By definition, the mechanism involves some kind of function or process leading to a change in the affected system. Its significance is that it may lead to qualitative transitions and induce or reduce physical effects, including synchronization, instability,  amplitude death, or oscillation death.

\begin{figure}
{\caption{ Coupling function determined from the phase dynamics of two interacting chemical Belousov-Zhabotinsky oscillators. The coupling function is reconstructed in terms of the phase difference $\psi=\phi_2-\phi_1$. Points obtained from reactors 1 and 2 are
plotted with open circles and triangles, respectively. The full curves represent smooth interpolations. From \citet{Miyazaki:06}. }\label{fig:BZ_chemical}}
{\includegraphics[width=0.8\textwidth,angle=0]{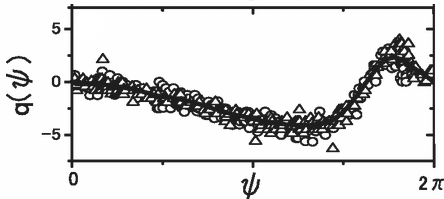}}
\end{figure}

But why is the mechanism important, and how it can be used? The first and foremost use of the coupling function mechanism is to illuminate the nature of the interactions themselves. For example, the coupling function of the Belousov-Zhabotinsky chemical oscillator has been reconstructed \cite{Miyazaki:06} with the help of a method for the inference of phase dynamics.  Fig.\ \ref{fig:BZ_chemical} shows such a coupling function, demonstrating a form that is very far from a sinusoidal function: a curve that gradually decreases in the region of a small $\psi$ and abruptly increases at a larger $\psi$,  with its minimum and maximum at around $5/4\pi$ and $7/4\pi$, respectively.

\red{Another important set of examples is the class of coupling functions and phase response curves used in neuroscience. In neuronal interactions, some variables are very spike-like i.e.\ they resemble delta functions. Consequently, neuronal coupling functions (which are convolution of phase response curves and perturbation functions) then depend only, or mainly, on the phase response curves. So the interaction mechanism is defined by the phase response curves: quite a lot of work has been done in this direction \cite{Schultheiss:11,Tateno:07,Gouwens:10,Ermentrout:96}; see also Sec.\ \ref{sec42:PRC}. For example, \citet{Tateno:07} and \citet{Gouwens:10} reconstructed experimentally the phase response curves for different types of interneurons in rat cortex, in order to better understand the mechanisms of neural synchronization.}

The mechanism of a coupling function depends on the differing contributions from individual oscillators. Changes in form may depend predominantly on only one of the phases (along one-axis), or they may depend on both phases, often resulting in a complicated and intuitively unclear dependance. The mechanism specified by the form of the coupling function can be used to distinguish the individual functional contributions to a coupling. One can decompose the net coupling function into components describing the self,  direct and indirect couplings \cite{Iatsenko:13a}. The self-coupling describes the inner dynamics of an oscillator which results from the interactions and has little physical meaning. Direct-coupling describes the influence of the direct (unidirectional) driving that one oscillator exerts on the other.  The last component, indirect-coupling, often called common-coupling, depends on the shared contributions of the two oscillators e.g.\ the diffusive coupling given with the phase difference terms. This functional coupling decomposition can be further generalized for multivariate coupling functions, where for example, a direct coupling from two oscillators to a third one can be determined \cite{Stankovski:15a}.

After learning the details of the reconstructed coupling function, one can use this knowledge to study or detect the physical effects of the interactions. In this way, the synchronous behavior of the two coupled Belousov-Zhabotinsky reactors can be explained in terms of the coupling function as illustrated by the examples given in Fig.\ \ref{fig:BZ_chemical} \cite{Miyazaki:06} and in Sec.\ \ref{sec641:MetChem}. Furthermore, the mechanisms and form of the coupling functions can be used to engineer and construct a particular complex dynamical structure, including sequential patterns and desynchronization of electrochemical oscillations \cite{Kiss:07}. Even more importantly, one can use knowledge about the mechanism of the reconstructed coupling function to predict transitions of the physical effects -- an important property described in detail for synchronization in the following section.

\subsection{ Synchronization prediction with coupling functions}\label{sec35:SyncPrediction}

Synchronization is a widespread phenomenon whose occurrence and disappearance can be of great importance. For example,  epileptic seizures in the brain are associated with excessive synchronization between a large number of neurons, so there is a need to control synchronization to provide a means of stopping or preventing seizures \cite{Schindler:07}; while in power grids the maintenance of synchronization is of crucial importance \cite{Rubido:15}. Therefore, one often needs to be able to control and predict the onset and disappearance of synchronization.

A seminal work on coupling functions by \citet{Kiss:05} uses the inferred knowledge of the \emph{coupling function to predict} characteristic synchronization phenomena in electrochemical oscillators. In particular, the authors demonstrated  the power of phase coupling functions, obtained from direct experiments on a single oscillator, to predict the dependence of synchronization characteristics such as order-disorder transitions on system parameters, both in small sets and in large populations of interacting electrochemical oscillators.

\begin{figure}
{\caption{Experimental coupling function from electrochemical oscillators, used for the prediction of synchronization. (a)-(c) Coupling function $q(\psi)$
evaluated in respect of the phase difference  $\psi=\phi_2-\phi_1$ shown on the left panel and its odd part $q_{-}(\Delta\phi)$ shown on the right panel -- for the case of (a) smooth oscillator, and (b) and (c) for relaxation oscillator with slightly different parameters. $H(\Delta\phi)$ on the plots is equivalent to the $q(\psi)$ notation used in the current review. From \citet{Kiss:05}. }\label{fig:predKiss1}}
{\includegraphics[width=0.95\textwidth,angle=0]{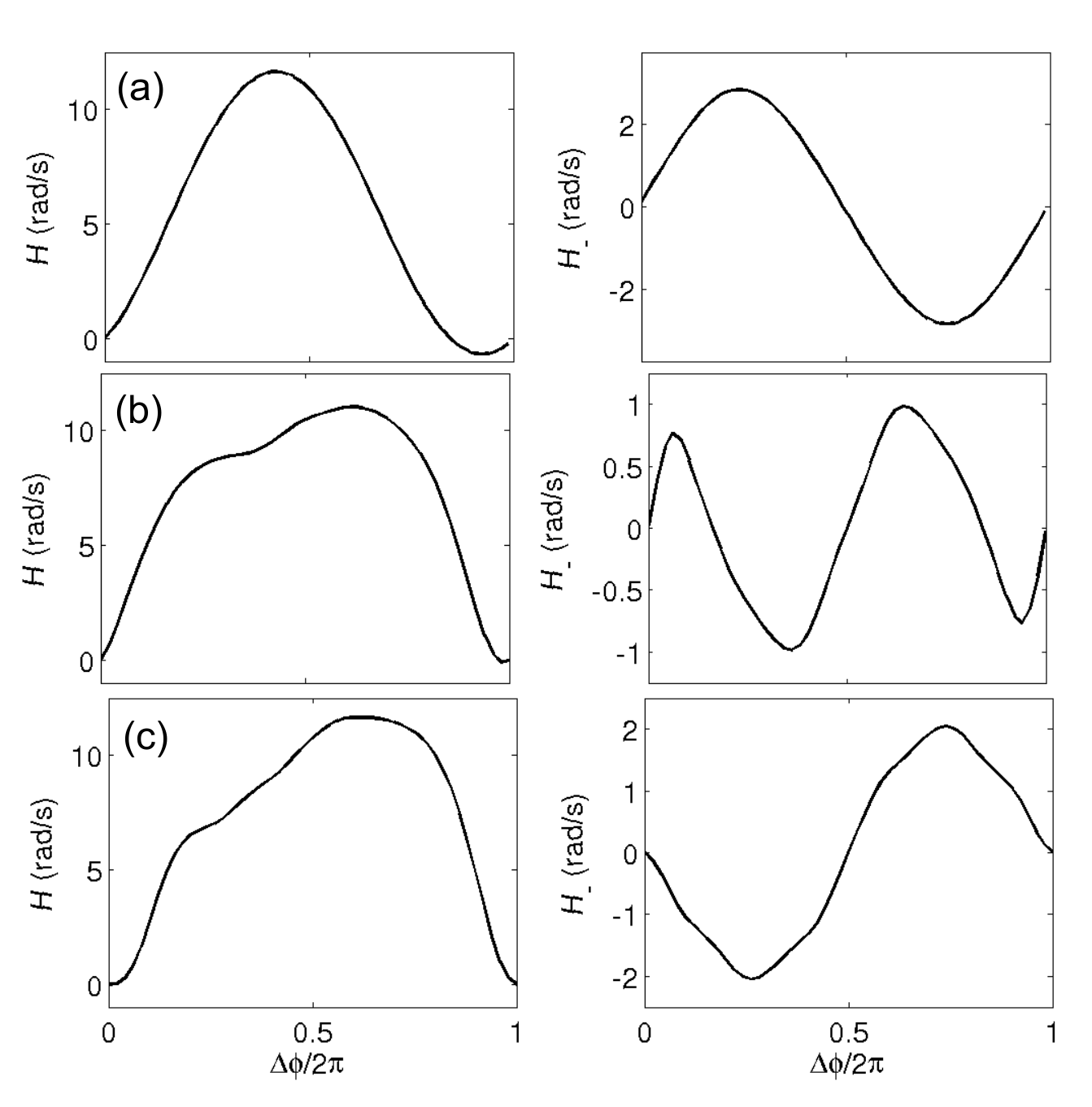}}
\end{figure}

The authors investigated the parametric dependence of mutual entrainment using an electrochemical reaction system, the electrodissolution of nickel in sulfuric acid (see also Sec.\ \ref{sec641:MetChem} for further applications on chemical coupling functions). A single nickel electrodissolution oscillator can have two main characteristic waveforms of periodic oscillation -- the smooth type and the relaxation oscillation type. The phase response curve is of the smooth type and is nearly sinusoidal, while being more asymmetric for the relaxation oscillations.

The coupling functions are calculated using the phase response curve obtained from  experimental data for the variable through which the oscillators are coupled. The coupling functions $q(\psi)$ of two coupled oscillators are reconstructed for three characteristic cases, as shown in Fig.\ \ref{fig:predKiss1}(a)-(c), left panels. The right panels in Fig.\ \ref{fig:predKiss1} show the corresponding odd (antisymmetric) part of the coupling functions $q_{-}(\psi)=[q(\psi)-q(-\psi)]/2$, which is important for determination of the synchronization. The coupling functions $q(\psi)$ Fig.\ \ref{fig:predKiss1} (a)-(c) have predominantly positive values, so the interactions contribute to the acceleration of the affected oscillators. The first coupling function Fig.\ \ref{fig:predKiss1}(a) for smooth oscillations has a sinusoidal $q_{-}(\psi)$ which can lead to in-phase synchronization at the phase difference of $\psi^*=0$. The third case of relaxation oscillations Fig.\ \ref{fig:predKiss1}(c)
has an inverted sinusoidal form $q_{-}(\psi)$, leading to stable anti-phase synchronization at $\psi^*=\pi$. The most peculiar case is the second one Fig.\ \ref{fig:predKiss1}(b) of relaxation oscillations, where the odd coupling function $q_{-}(\psi)$ takes the form of a second harmonic ($q_{-}(\psi)\approx \sin(2 \psi)$) and both the in-phase ($\psi^*=0$) and anti-phase ($\psi^*=\pi$) entrainments are stable, in which case the actual state attained will depend on the initial conditions.

\begin{figure}
{\caption{  Mutual entrainment and stable single-cluster (left panel) and two-cluster (right panel) states of a population of 64 \emph{globally-coupled} electrochemical relaxation oscillators under the same experimental conditions. The two-cluster state was obtained from the one-cluster state by a small perturbation acting as a different initial condition for the population. From \citet{Kiss:05}. }\label{fig:predKiss2}}
{\includegraphics[width=1\textwidth,angle=0]{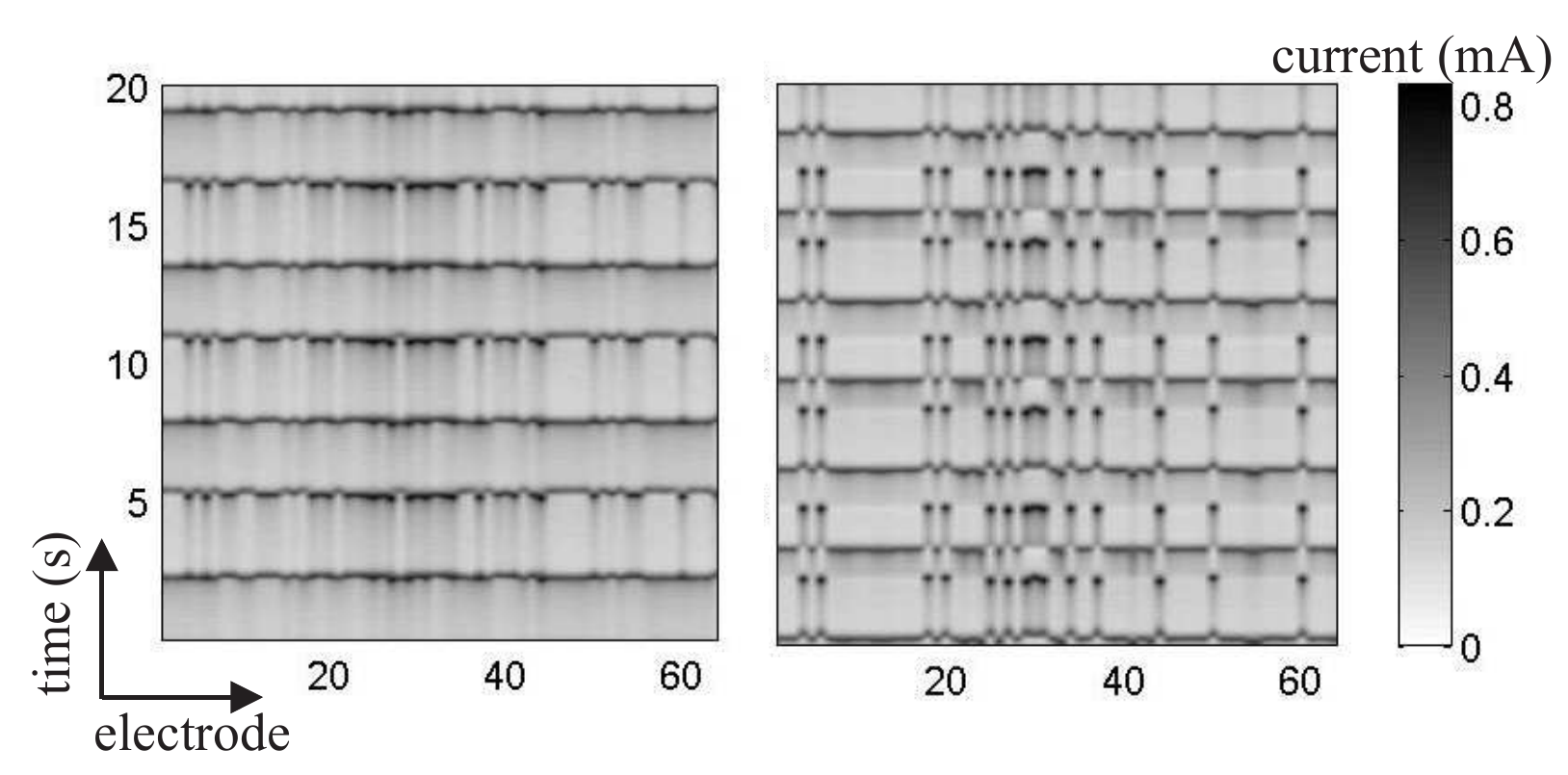}}
\end{figure}

Next, the knowledge obtained from experiments with a single oscillator was applied to predict the onset of synchronization in experiments with 64 globally coupled oscillators. The experiments confirmed that for smooth oscillators the interactions converge to a single cluster, and for relaxational oscillators they converge to a two-cluster synchronized state. Experiments in a parameter region between these states, in which bistability is predicted, are shown in Fig.\ \ref{fig:predKiss2}. A small perturbation of the stable one-cluster state (left panel of Fig.\ \ref{fig:predKiss2}) yields a stable two-cluster state (right panel of Fig.\ \ref{fig:predKiss2}). Therefore, all the synchronization behavior seen in the experiments was in agreement with prior predictions based on the coupling functions.

\red{In a separate line of work, synchronization was also predicted in neuroscience: interaction mechanisms involving individual neurons, usually in terms of phase-response curves (PRCs) or spike-time response-curves (STRCs), were used to understand and predict the synchronous behavior of networks of neurons \cite{Acker:03,Netoff:05,Schultheiss:11}. For example, \citet{Netoff:05}  studied experimentally the spike-time response-curves of individual neuronal cells. Results from these single-cell experiments were then used to predict the multi-cell network behaviors, which were found to be compatible with previous model-based predictions of how specific membrane mechanisms give rise to the empirically measured synchronization behavior.}


\subsection{ Unifying nomenclature}\label{sec37:Nomenclature}

Over the course of time, physicists have used a range of different terminology for coupling functions. For example, some publications refer to them as interaction functions and some as coupling functions.  
This inconsistency needs to be overcome by adopting a common nomenclature for the future.


\red{
The terms interaction function and coupling function have both been used to describe the physical and mathematical links between interacting dynamical systems. Of these, coupling function has been used about twice as often in the literature, including the most recent. The term coupling is closer to describing a connection between two systems, while the term interaction is more general. 
Coupling implies causality, whereas interaction does not necessarily do so. Often correlation and coherence are considered as signatures of interactions, while they do not necessarily imply the existence of couplings. We therefore propose that the terminology be unified, and the term \emph{coupling function} be used henceforth to characterise the link between two dynamical systems whose interaction is also causal.
}


\section{ Theory}\label{sec4:Theory}

In physics one is likely to examine stable static configurations whereas, in dynamical interaction between oscillators, solutions will converge to a subspace. For example, if two oscillators are in complete synchronization the subspace is called the {\it synchronization manifold} and corresponds to the case where the oscillators are in the same state for all time \cite{Pecora:90,Fujisaka:83}. So, within the subspace, the oscillators have their own dynamics and finer information on the coupling function is needed. 

The analytical techniques and methods needed to analyze the dynamics will depend on whether the coupling strength is strong or weak.   Roughly speaking, in the strong coupling regime, we will have to tackle the fully-coupled oscillators whereas in the weak coupling we can reduce the analysis to lower-dimensional equations.

\subsection{Strong interaction} \label{sec41:Thr_Amp}

To illustrate the main ideas and challenges of treating the case of strong interaction, while keeping technicalities to a minimum, we will first discuss the case of two coupled oscillators. These examples contain the main ideas and reveal the role of the coupling function and how it guides the system towards synchronization.

\subsubsection{Two coupled oscillators}

We start by illustrating the variety of dynamical phenomena that can be encountered and the role played by the coupling function in the strong coupling regime.

\begin{center}
\emph{Diffusion driven oscillations}
\end{center}

When two systems interact they may display oscillations solely because of the interaction. This is the nature of the problem posed by \citet{Smale:76} based on Turing's idea of morphogenesis \cite{Turing:52}. We consider two identical systems which, when isolated, each exhibit a globally asymptotically stable equilibrium, but which oscillate when diffusively coupled. This phenomenon is called {\it diffusion driven oscillation}.

Assume that the system
\begin{equation}\label{Eq_f}
\dot x = f(x),
\end{equation}
where $f: \mathbb{R}^n \rightarrow \mathbb{R}^n$ is a differentiable vector field with a globally stable attraction with point -- all trajectories will converge to this point. Now consider two of such systems coupled diffusively
\begin{eqnarray}\label{diff_coupled}
\dot{x}_1 &=& {f}(x_1) + \varepsilon H ({x}_2 - {x}_1)  \\
\dot{x}_2 &=& {f}(x_2) + \varepsilon H ({x}_1- {x}_2). \nonumber
\end{eqnarray}
The problem proposed by Smale was to find (if possible) a coupling function (positive definite matrix) $H$ such that the diffusively coupled system undergoes a Hopf bifurcation. Loosely speaking, one may think of two cells that by themselves are inert but which, when they interact diffusively, become alive in a dynamical sense and start to oscillate.

Interestingly, the dimension of the uncoupled systems comes into play.
Smale constructed an example in four dimensions. \citet{Pogromsky:99} constructed examples in three dimensions and also showed that, under suitable conditions, the minimum dimension for diffusive coupling to result in oscillation is $n=3$.  The following example illustrates the main ideas. Consider
\begin{equation}\label{S1}
f(x) = Ax (1+|x|^2) \mbox{   with   }
A =
\left(
\begin{array}{ccc}
1 & -1 & 1\\
1 & 0 &0\\
-4 & 2 & -3
\end{array}
\right),
\end{equation}
where $|x|^2 = x^T x$. Note that all the eigenvalues of $A$ have negative real parts. So the origin of the system Eq.\ (\ref{S1}) is exponentially attracting.
%

Consider the coupling function to be the identity
\begin{eqnarray}
\dot x_1 =  f(x_1) + \varepsilon (x_2 - x_1) \nonumber \\
\dot x_2 =  f(x_2)  + \varepsilon (x_1 - x_2). \nonumber
\end{eqnarray}
For $\varepsilon = 0$ the origin is globally attracting; the uniform attraction persists when $\varepsilon$ is very small, and so the origin is still globally attracting. However, for large values of the coupling
$\varepsilon > 0.6512$ the coupled systems exhibit oscillatory solutions (the origin has undergone a Hopf bifurcation).

{\it Generalizations:} In this example the coupling function was the identity.  \citet{Pogromsky:99} discussed further coupling functions, such as coupling functions of rank two that generate diffusion-driven oscillators. Further oscillations in originally passive systems have been reported in spatially extended systems  \cite{Gomez-Marin:07}. In diffusively coupled membranes, collective oscillation in a group of nonoscillatory cells can also occur as a result of spatially inhomogeneous activation factor \cite{Ma:09}. These ideas of diffusion leading to chemical differentiation have also been observed experimentally and generalized by including heterogeneity in the model \cite{Tompkins:14}.

\begin{center}
\emph{Oscillation death}
\end{center}

We now consider the opposite problem: Systems which when isolated exhibit oscillatory behaviour but which, when coupled diffusively, cease to oscillate and where the solutions converge to an equilibrium point.

As mentioned above in Sec.\ \ref{sec1:Intro}B, this phenomenon is called {\it oscillation death} \red{\cite{Bar:85,Koseska:13,Mirollo:90,Ermentrout:90}}.
To illustrate the essential features we consider a normal form of the Hopf bifurcation
$$
\dot x_j = f_j(x_j)
$$
where
$$
f_j(x) = \omega_j A x + (1 - |x|^2) x, \, \, \, \mbox{  with }
A =
\left(
\begin{array}{cc}
0 & -1 \\
1 & 0
\end{array}
\right).
$$
So, each isolated system has a limit cycle of amplitude $|x|^2 = 1$ and a frequency of $\omega_j$. Note that the origin $x=0$ is an  unstable equilibrium point. In oscillation death when the systems are coupled, the origin may become stable.

Focusing on diffusive coupling, again, the question concerns the nature of the coupling function. \citet{Aronson:90} remarked that the simplest coupling function to have the desired properties is the identity with strength $\varepsilon$. The equations have the same form as Eq. (\ref{diff_coupled}) with $H$ being the identity.

The effect can be better understood in terms of phase and amplitude variables. Let $r_1$, $r_2$ be the amplitudes and $\phi_1, \phi_2$ the phases of $x_1$ and $x_2$, respectively. We consider $r_1=r_2=r$ which captures the main causes of the effect, as well as the phase difference $\psi = \phi_1 - \phi_2$. Then the equations in these variables can be well approximated as
\begin{eqnarray}
\dot r &=& r ( 1 - \varepsilon - r^2) + \varepsilon r \cos \psi \label{r} \\
\dot \psi &=& \Delta_{\omega}  - 2 \varepsilon \sin \psi.
\end{eqnarray}
The conditions for oscillation death are a stable fixed point at $r=0$ along with a stable fixed point for the phase dynamics. The above equations provide the main mechanism for oscillation death.
\red{
First, we can determine the stable fixed point for the phase dynamics, as illustrated in Fig.\ \ref{fig:CF_kuramoto}. There is a fixed point $\psi^*$ if $\varepsilon > \Delta_{\omega}/2$ and $\sin \psi^* = \Delta_{\omega}/(2\varepsilon)$. We will assume that
$\Delta_{\omega}>2$ which implies that, when the fixed point $\psi^*$ exists, $\varepsilon > 1$.

Next, we analyse the stability of the fixed point $r^*=0$. This is determined by the linear part of Eq.\ \ref{r}. Hence, the condition for stability is
$$
1-\varepsilon +\varepsilon \cos \psi^* <0.
$$
Using the equation for the fixed point we have $\cos \psi^* = \sqrt{1 - [\Delta_{\omega}/(2\varepsilon)]^2}$. Replacing this in the stability condition we obtain $\varepsilon <  (4  + \Delta_{\omega}^2)/8$. The analysis reveals that the system will exhibit oscillation death if the coupling is neither too weak nor too strong. Because we are assuming that the mismatch is large enough, $\Delta_{\omega} >2$, then there are minimum and maximum coupling strengths for oscillation death
$$
 1< \varepsilon < \frac{4  + \Delta_{\omega}^2}{4}.
 $$
Within this range, there are no stable limit cycles: the only attracting point is the origin, and so the oscillations are dead.

The full equation is tackled in  \citet{Aronson:90}. The main principle is that the eigenvalues of the coupling function modify the original eigenvalues of the system and change their stability. It is possible to generalize these claims to coupling functions that are far from the identity \cite{Koseska:13}. The system may converge, not only to a single fixed point, but to many \cite{Koseska:13a}.
}
\begin{center}
\emph{Synchronization}
\end{center}

One of the main roles of coupling functions is to facilitate collective dynamics. Consider the diffusively coupled oscillators described by Eq.\ (\ref{diff_coupled}). We say that the diagonal
$$
{x}_1(t)={x}_2(t)
$$
is the complete synchronization manifold \cite{Brown:00}. Note that the synchronization manifold is an invariant subspace of the equations of motion for all values of the coupling strength. Indeed, when the oscillators synchronize the coupling term vanishes.
So, they will be synchronized for all future time. The main question is whether the synchronization manifold is attractive, that is, if the oscillators are not precisely synchronized will they converge towards synchronization? Similarly, if they are synchronized, and one perturbs the synchronization, will they return to synchronization?

\begin{figure}[h]
\centerline{\hbox{\psfig{file=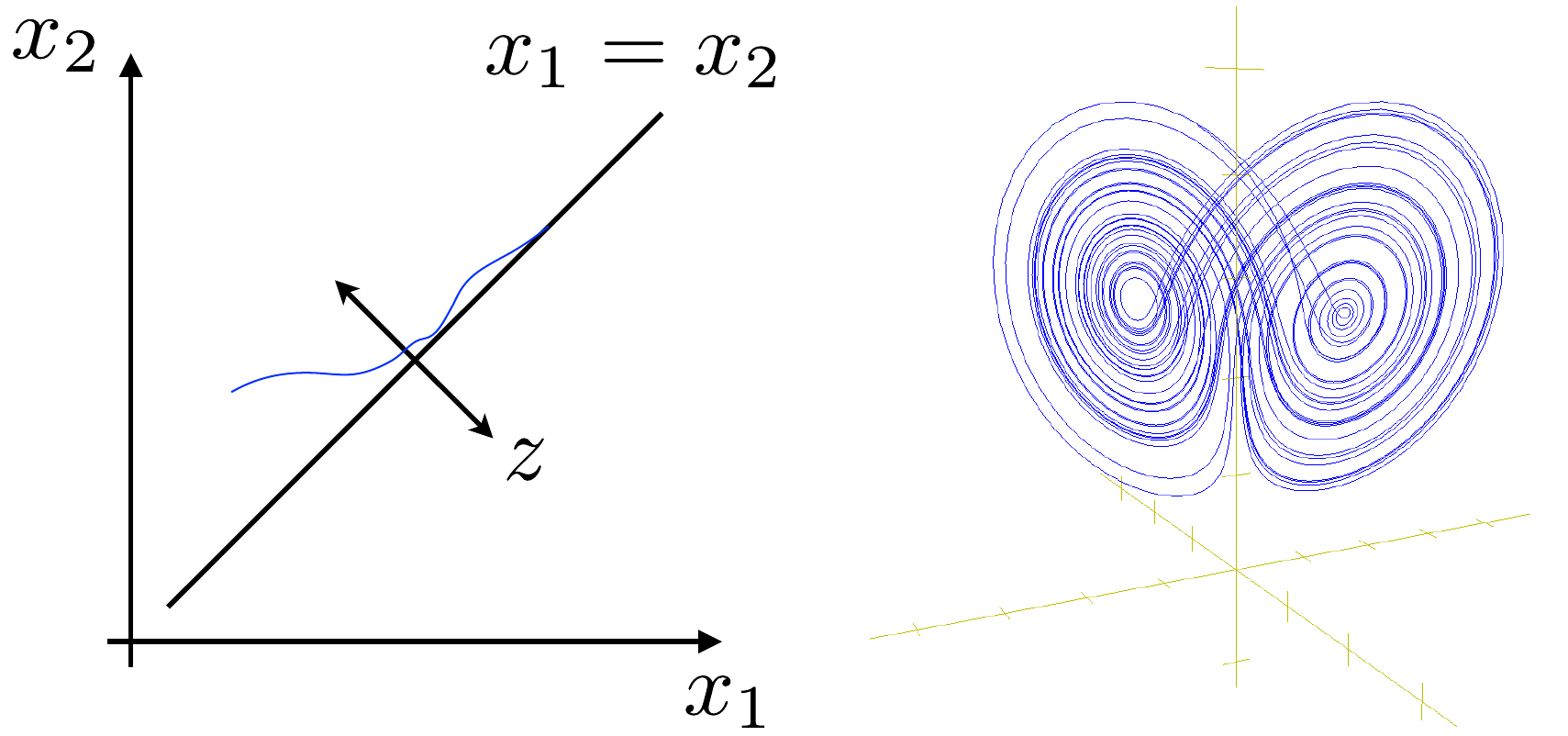,width=8.0cm}}}
\caption{(color online)
Illustration of the coordinates parallel $y$ and transverse $z$ to synchronization. In the left panel we also show a trajectory converging to the synchronization subspace implying that $z \rightarrow 0$. Once the coupled systems reach synchronization, their amplitudes will evolve together in time, but the evolution can be chaotic as illustrated in the right panel. The dynamics along the synchronization subspace is the Lorenz attractor.}
\label{coordi}
\end{figure}

Let us first consider the case where the coupling is the identity $H(x) = x$, and discuss the key mechanism for synchronization.  Note that there are natural coordinates to analyze  synchronization
$$
y = \frac{1}{2}(x_1 + x_2) \,\,\, \mbox{ and } z = \frac{1}{2}(x_1 - x_2).
$$
These coordinates have a natural meaning. If the system synchronizes,  $z \rightarrow 0$ and $y \rightarrow s$ with $\dot s = f(s)$.
Hence, we refer to $y$ as the coordinate parallel to the synchronization subspace $x_1=x_2$, and to $z$ as the coordinate transverse to the synchronization subspace, as illustrated in Fig.\ \ref{coordi}.

\noindent

The synchronization analysis follows two steps: (i) Obtaining a governing equation for the modes $z$ transverse to the synchronization subspace; and (ii) using the coupling function to damp instabilities in the transverse modes. \\

\noindent
{\bf (i) Obtain an equation for {\itshape z}}.
Let us assume that the initial disturbance of $z$ is small. Then we can obtain a linear equation for $z$ by {\it neglecting the high order terms proportional to $|z |^2$}. Noting that $\dot z = (\dot x_1 -  \dot x_2)/2$, using Eqs.\ (\ref{diff_coupled}) for $x_1$ and $x_2$, and expanding $f$ in a Taylor series we obtain
\begin{equation}\label{z}
\dot z = J(t) z - 2 \varepsilon z,
\end{equation}
where $J(t) = Df(x_2(t))$ is the Jacobian evaluated along a solution of $x_2$.  \\

\noindent
{\bf (ii) Coupling function to provide damping}. The term $-2 \varepsilon z$ coming from the coupling now plays the role of a damping term. So we expect that the coupling will win the competition with $J$ and will force the solutions of $z$ to decay exponentially fast to zero.  To see this, we observe that the first term
\begin{equation}\label{u}
\dot u = J(t) u
\end{equation}
depends on the dynamics of $x_2$ alone.
Typically,
$
\| u (t) \| \propto e^{\lambda t}
$
for $\lambda>0$. Now, to obtain a bound on the solution of Eq.\ (\ref{z}), we consider the ansatz
$
z = u e^{- 2 \varepsilon t},
$
and notice that differentiating $z$ we obtain Eq. (\ref{z}). Hence
$$
\| z (t) \| \propto e^{(\lambda - 2 \varepsilon) t},
$$
from the growth behaviour of the disturbance $z$ we can also obtain the critical coupling strength to observe synchronization.

{\bf Critical Coupling}: From this estimate, we can also obtain the critical coupling such that the solutions $z$ decay to zero. For coupling strengths
$$
\varepsilon > \frac{\lambda}{2},
$$
the oscillators synchronize.

{\it Meaning of $\lambda>0$}: This corresponds to chaotic behaviour in the synchronization manifold. If $z\rightarrow0$ then, $J(t)$ will be the Jacobian along a solution of $\dot s = f(s).$ So $\lambda$ depends on the dynamics on the synchronization manifold. If $\lambda> 0$ and the solutions are bounded, the dynamics of the synchronized system is chaotic. Roughly speaking, $\lambda>0$ means that two nearby trajectories will diverge exponentially fast for small times and, because the solutions are bounded, they will subsequently come close together again. So, the coupled systems can synchronize even if the dynamics of the synchronized system is chaotic, as shown in Fig. \ref{coordi} for the chaotic Lorenz attractor. The number $\lambda$ is the maximum Lyapunov exponent of the synchronization subspace.

There are intrinsic challenges associated with the analysis, and more when we attempt to generalize these ideas and also because of the nonlinearities that we neglected during the analysis.

\begin{enumerate}
\item General coupling functions: From a mathematical perspective the argument above worked because the identity commutes with all matrices.  For other coupling functions, the argument above cannot be applied, and we encounter three possible scenarios:

i)  The coupling function does not damp instabilities and the system never synchronizes \cite{Pecora:98,Boccaletti:02}.

ii) The coupling function damps out instabilities  only for a finite range of coupling strengths.
$$
\varepsilon_c^1 < \varepsilon < \varepsilon_c^2.
$$
For instance, this is the case for the R\"ossler system with coupling only in the first variable \cite{Huang:09}.

iii) The  coupling function damps instabilities and there is a single critical coupling $\varepsilon_c$. This is the case, when the coupling function \red{eigenvalues have positive real parts.} \cite{Pereira:14}.

\item Local versus global results: In the above argument we have expanded the vector field in a Taylor series and obtained a linear equation to describe how the systems synchronize. This means that any claim on synchronization is local. It is still an open question how to obtain global results.

\item Nonlinear effects: We have neglected the nonlinear terms (the Taylor remainders), which can make synchronization unstable. Many researchers have observed this phenomenon through the bubbling transition \cite{Ashwin:94,Viana:05,Venkataramani:96}, intermittent loss of synchronization \cite{Gauthier:96,Yanchuk:01}, and the riddling basin \cite{Heagy:94,Ashwin:05a}.
\end{enumerate}

\noindent To highlight the role of the coupling function and illustrate the above challenges, we will show how to obtain global results depending on the coupling function and discuss how local and global results are related.

{\bf Global argument:} Assume that $H$ is a Hermitian positive definite matrix.  The main idea is to turn the problem upside down. That is, we see the vector field as perturbing the coupling function. So, consider the system with only the coupling function and use the transverse coordinates
\begin{equation}\label{zH}
\dot z = - 2 \varepsilon H z.
\end{equation}
Since $H$ is positive definite we obtain
$
- z^T H z \le - 2  c \varepsilon |z|^2,
$
where $c = c(H)$ is the smallest eigenvalue of $H$. The global stability of the system can be obtained by constructing a Lyapunov function $V$. The system will be stable if $V$ is positive and its derivative $\dot V$ is negative. This system admits a quadratic Lyapunov function
$
V(z) = \frac{1}{2}z^T z.
$
Indeed, taking the derivative
$$
\dot V(z) = z^T \dot z \le - 2 c \varepsilon |z|^2.
$$
Hence all solutions of Eq.\ (\ref{zH}) will converge to zero exponentially fast. Next, consider the coupled system
$$
\dot z = - 2 \varepsilon  H z + J(t,z),
$$
where by the mean value theorem we obtain
\begin{eqnarray}
J(t,z) &=& f(x_1(t) + z(t)) - f(x_1(t)) \\ \nonumber
&=& \int_0^1 Df(x_1(t) + s z(t)) z(t) ds. \nonumber
\end{eqnarray}
Because we did not Taylor-expand the vector fields, the equation is globally valid.
Assuming that the Jacobian is bounded by a constant $M_f>0$, we obtain  $| J(t,z)| \le M_f |z|$.

Computing again the Lyapunov function for the coupled system (including the vector fields) we obtain
\begin{eqnarray}
\dot V(z) &\le& - ( 2 c \varepsilon - M_f)  |z|^2.
\end{eqnarray}
The system will synchronize if $\dot V$ is negative. So synchronization is attained if
$$
\varepsilon_c > \frac{M_f}{2 c}.
$$
Again the critical coupling has the same form as before. The coupling function came into play via the constant $c$, and instead of $\lambda$ we have $M_f$. Typically, $M_f$ is much larger than $\lambda$. So, {\it global bounds are not sharp}.  This conservative bound guarantees that the coupling function can damp all possible instabilities transverse to the synchronization manifold. Moreover, they are persistent under perturbation.

{\bf Local results:}
First, we Taylor-expand the system to obtain
\begin{equation}\label{zH2}
\dot z = J(t) z - 2 \varepsilon H z
\end{equation}
in just the same form as before. Note however that the trick we used previously, by defining $\dot u = J u $, is no longer applicable.
Indeed, we use the ansatz
$
z  = u e^{- 2 \varepsilon H t},
$
to obtain
$
\dot z = -2\varepsilon H z + e^{-2 \varepsilon H t} J(t) u,
$ and, since $H$ and $J(t)$ do not commute,
$$
e^{-2 \varepsilon H t} J(t) u \not =  J(t) e^{-2 \varepsilon H t}  u = J(t) z,
$$
the ansatz cannot be used. Thus we need a better way forward. So in the same way as we calculated the expansion rate $\lambda$ for $J(t)$, we calculate the expansion rate for Eq.\ (\ref{zH2}). Such Lyapunov exponents are very important in a variety of contexts. For us, it suffices to know that there are various ways to compute them \cite{Dieci:02,Pikovsky:16}. We calculate the Lyapunov exponent for each value of the coupling strength $\varepsilon$ to  obtain a function
$$
\varepsilon \mapsto \Lambda(\varepsilon).
$$
This function is called the {\it Master Stability Function} (MSF).
We will extract the synchronization properties from $\Lambda(\varepsilon)$.
As we already discussed, the solutions of Eq.\ (\ref{zH}) will behave as
$$
|z(t)| \propto Ce^{\Lambda(\varepsilon) t}.
$$
Now note that $\Lambda(0) = \lambda>0$ (the expansion rate of the uncoupled equation), since we considered the case of chaotic oscillators. Because of our assumptions, we know that there is a $\varepsilon_c$ such that
$$
\varepsilon > \varepsilon_c^{\tt msf} \Rightarrow \Lambda(\varepsilon) <0
$$
and for which $\Lambda(\varepsilon)$ will become negative, and $z$ will converge to zero. 

{\bf Global versus local results.} In the global analysis, the critical coupling depends on $M_f$, which is an upper bound for the Jacobian. This approach is rigorous and guarantees that all solution will synchronize. In the local analysis, we linearized the dynamics about the synchronization manifold and computed the Lyapunov exponent
associated with the transverse coordinate $z$. The critical coupling was then obtained by analysing the sign of the Lyapunov exponent. Generically, $\varepsilon_c^{\tt msf} \ll M_f/2c$. The main reasoning is as follows. The Lyapunov exponents measure the mean instability whereas, in the global argument, we consider the worst possible instability.  So the local method allows us to obtain a sharp estimate for the onset of synchronization.

{\bf The pitfalls of the local results.} The main challenge of the local method lies in the intricacies of the theory of Lyapunov exponents \cite{Barreira:02,Pikovsky:16}. These can be discontinuous functions of the vector field. In other words, the nonlinear terms we threw away as Taylor remainders can make the  Lyapunov exponent jump from negative to positive. Moreover, in the local case we cannot guarantee that all trajectories will be uniformly attracted to the synchronization manifold. In fact, for some initial conditions trajectories are attracted to the synchronization manifold, whereas nearby initial conditions are not. This phenomenon is called riddling \cite{Heagy:94}.

\subsubsection{Comparison between approaches}

As discussed above, there is a dichotomy between global versus local results, and sharp bounds for critical coupling. These issues depend on the coupling function. Some coupling functions allow one to employ a given technique and thereby obtain global or local results.

First, we compare the two main techniques used in the literature, that is, Lyapunov functions (LFs) and the master stability function (MSF). For a generic coupling function, the LFs are unknown; but Lyapunov exponents can be estimated efficiently by numerical methods \cite{Dieci:02,Froyland:13,Ginelli:07}.

Given additional information on the coupling, we can further compare the techniques.  Note that the coupling function $H$ can be nonlinear. In this case, we consider the Jacobian
$$
\Gamma = DH(0).
$$
Moreover, we say that $\Gamma$ belongs to the Lyapunov class if there are positive matrices $Q$ and $P$ such that
$$
\Gamma ^T P  + P \Gamma = -Q.
$$
Whenever the matrix $\Gamma$ is in the Lyapunov class we can construct the Lyapunov function algorithmically.

\begin{table}[t]
\centering
\begin{tabular}{lcccc} 
Coupling Function & Class & Technique & Global & Persistence \\
\hline
\hline	
$H$ +ve definite & Lyapunov & LF &  Yes & Yes \\
$DH$ +ve definite & Lyapunov &  LF & No & Yes  \\
$H$ differentiable & generic & LF & $-$ & $-$ \\
$H$ differentiable & generic & MSF & No & No
\end{tabular}
   \caption{Comparison between classes of coupling function and the techniques to obtain synchronization. Dashes indicate that typically we are unable to construct the Lyapunov function in such cases.}
   \label{Comp}
\end{table}

Table \ref{Comp} reveals that the MSF method is very versatile. Although it may not encompass nonlinear perturbations it provides a framework to tackle a generic class of coupling functions \cite{Huang:09}. In the theory of chaotic synchronization, therefore, this has been the preferred approach. However, it should be used with caution.

%
%
%
\subsection{Weak regime} \label{sec:weakCpl}

\red{In the weak-coupling regime, the coupling strength is by definition insufficient to affect the amplitudes significantly; however the coupling can still cause the phases to adapt and adjust their dynamics \cite{Kuramoto:84}.}  Many of the phenomena observed in nature relate to the weak coupling regime.

Mathematical descriptions of coupled oscillators in terms of their phases offer two advantages: first, it reduces the dimension of the problem, and secondly, it can reveal principles of collective dynamics and other phenomena.

The theory for the weak coupling regime is well-developed. In the seventies and early eighties \citet{Winfree:67,Winfree:80} and \citet{Kuramoto:75,Kuramoto:84} developed the idea of asymptotic phase and phase reduction.  Also starting from the seventies, the mathematical theory for such phase reduction was brought to completion in terms of normally hyperbolic invariant manifolds \cite{Wiggins:13,Eldering:13,Hirsch:77}. \red{Since then, the phase reduction theory \cite{Nakao:15} has been significantly extended and generalized, for inclusion of phase reduction in the case of strongly perturbed oscillations, for stochastic treatment of interacting oscillators subject to noise of different kinds, for oscillating neuronal populations, and for spatiotemporal oscillations in reaction-diffusion systems  \cite{Nakao:14,Goldobin:10,Teramae:09,Yoshimura:08,Kurebayashi:13,Brown:04}}. The main ingredient in this approach is an attracting periodic orbit.

\subsubsection{Stable Periodic Orbit and its phase}

If the system in question has an exponentially stable periodic orbit, the theory guarantees the existence of the reduction and provides a method to obtain it. Thanks to the works of \citet{Izhikevich:07,Hoppensteadt:12,Ermentrout:96,Ermentrout:08,Rinzel:98,Ermentrout:10}
we now have a phase description for certain classes of neurons and we understand its limitations \cite{Smeal:10}. The strategy is as follows. We assume that the system
\begin{equation}\label{unc}
\dot x = f(x),
\end{equation}
where $f: \mathbb{R}^n \rightarrow \mathbb{R}^n$, has a uniformly exponentially attracting periodic orbit $\gamma$ with period $T$, that is, $\gamma(t+T) = \gamma(t)$.
\red{
The orbit is exponentially stable if the trajectories of the system approach it exponentially fast and the rate
of convergence does not depend on the initial time or on initial conditions (for points sufficiently close to the orbit).

We can parameterize the orbit by its phase $\phi$, $\gamma(\phi + 2 \pi ) = \gamma(\phi)$. We can also re-parameterize time such that that phase $\phi$ increases uniformly along the orbit $\gamma$. That is, the phase is uniform frequency equal to unity. By the chain rule we then have
\[
\dot{\phi} = 1 = \nabla_{\gamma} \phi \cdot  f(\gamma)
\]
The key idea here is that weak coupling can adjust the rhythm of the phase dynamics. The goal is to obtain the phase reduction solely on the basis of information about the isolated system (the orbit $\gamma$). To this end we need to extend the phase $\varphi$ to a neighborhood of the orbit. The main ingredient necessary for the reduction of the problem to its phase dynamics is the concept of asymptotic phase \cite{Winfree:67,Winfree:80}, which will provide us with the coupling function.

{\bf Asymptotic phase:} Right now, the phase $\phi$ is defined only along the orbit $\gamma$. Our first step is to extend $\phi$ to a neighbourhood of $\gamma$. Since the periodic orbit is exponentially and uniformly attracting, it will attract an open neighbourhood of $\gamma$. We call this set the basin of attraction of the periodic orbit. Note that every initial point ${x}_0$ in the basin of attraction of the orbit will converge to the orbit. Hence, we have a $\phi (x_0)$ such that
$$
\displaystyle \lim_{t \to \infty}|{x}(t,{x}_0)-{\gamma} (t+\phi (x_0))|= 0,
$$
where ${x}(t,{x}_0)$ is the solution of the system with initial condition ${x}_0$. For each initial point in the basin of attraction of $\gamma$ we can assign a unique point in the orbit $\theta$. This $\phi \in [0, 2\pi]$ is called the {\it asymptotic phase}.
}

\begin{figure}[h]
\centerline{\hbox{\psfig{file=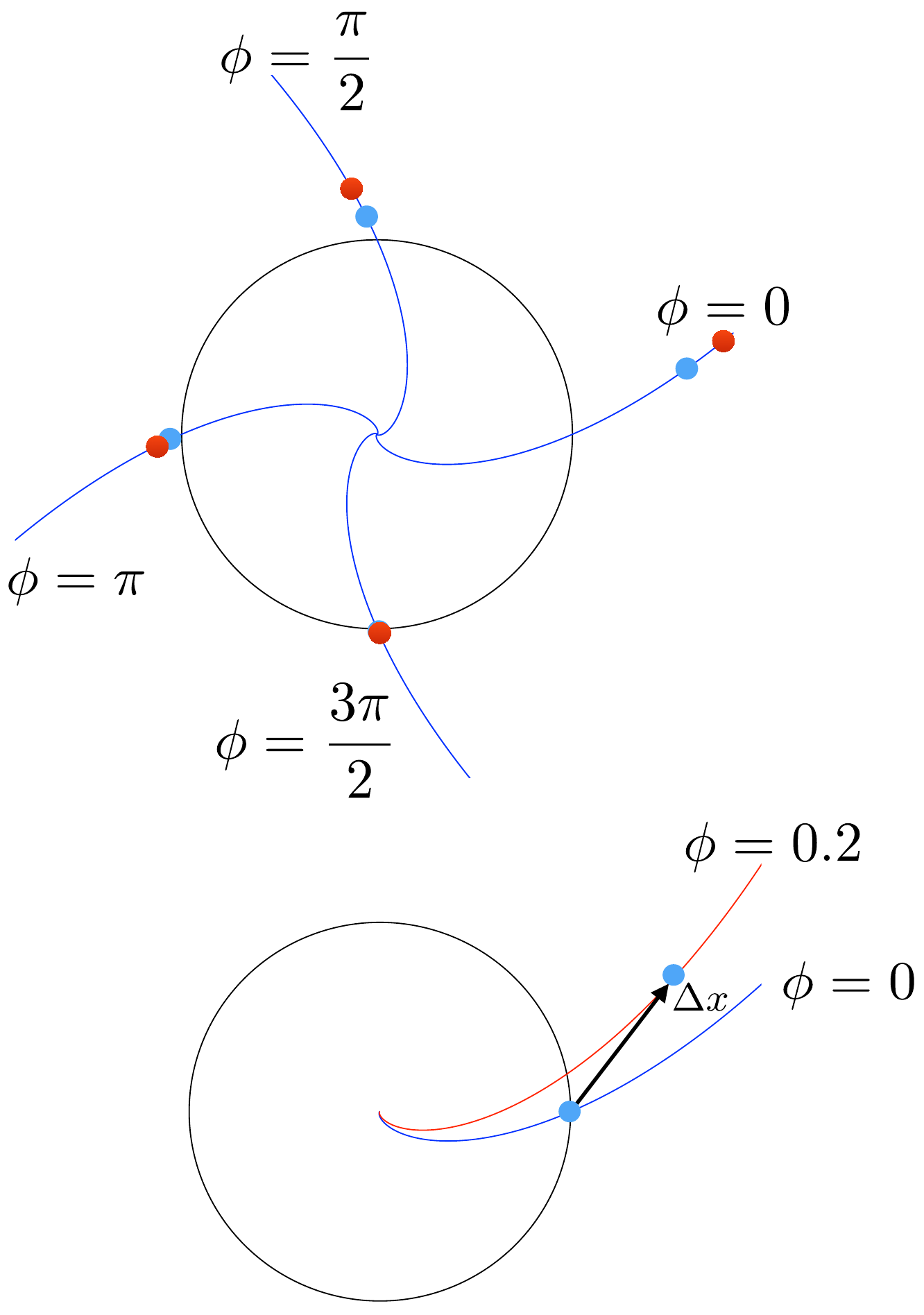,width=6.5cm}}}
\caption{(color online)
Periodic orbits \blue{are shown as filled (black) circles and isochrons as (blue)} lines. Every point in an isochron has the same value of asymptotic phase. Moreover, the distance between two points in the same isochron tends to zero exponentially fast, as illustrated by the \blue{red (dark) and blue (light)} points. In the lower figure, we show the effect on the phase dynamics of a small perturbation. The point is initially at phase zero. The perturbation $\Delta x$ moves the system from \blue{its} initial point to another isochron, thereby advancing the phase. The periodic orbit $\gamma$ and the isochrons are for Eq.\ \ref{exiso}.}
\label{Figphase}
\end{figure}
\noindent

\red{
{\it  Isochron.} For each value of phase $\phi$ in the orbit $\gamma$ we have a curve passing through this phase values. And along this curve every initial will have the same asymptotic phase. This set is called isochron.  That is, the isochron is a level set of $\phi ({x})$. So points in the isochron have the same value of phase and will move at the same speed. See Fig.\ \ref{Figphase} where points in the same isochron approach the orbit along the same isochron. The set of points where the isochron cannot be defined is called phaseless. Once we find the isochron we can perform the phase reduction.
}

\subsubsection{Coupling function and phase reduction}

Consider Eq.\ (\ref{unc}) with a stable periodic orbit $\gamma$ being perturbed
$$
\dot x = f(x) + \varepsilon I(\vartheta,x),
$$
where $\vartheta = \omega t$ is the phase of the external influence, and $I$ the influence is a periodic on $\vartheta$.  The weak coupling implies $\varepsilon \ll 1$.

One of the cornerstones of the theory of {\it invariant manifolds} is to guarantee that, when the system is perturbed and the coupling strength is weak $\varepsilon\ll 1$, there is a new attracting periodic orbit $\tilde \gamma$ close to the orbit $\gamma$, the difference between the orbits being of order $\varepsilon$.  Moreover, $\tilde \gamma$ is exponentially attractive and the isochrons also persist. So, while the amplitudes are basically unaffected, the dynamics of the phases change greatly.

\red{
With the help of the asymptotic phase, we define the phase in a neighborhood of the orbit $\gamma$. This neighborhood contains the new orbit $\tilde \gamma$ as we consider small $\varepsilon > 0$. So, calculating the phase along $\tilde \gamma$,  by the chain rule we obtain
$$
\dot{\phi} = \nabla_{\tilde\gamma} \phi \cdot [f(\tilde \gamma) + \varepsilon I (\vartheta, \tilde \gamma)].
$$
But by construction $\nabla \phi \cdot f = 1$ in a neighborhood of $\gamma$. Because $\tilde \gamma$ is $O(\varepsilon)$ distant from $\gamma$ we can expand both $f$ and $I$ and evaluate them along $\gamma$ at the expense of a perturbation of order $\varepsilon$. It is standard to denote $Z = \nabla_{\gamma} \phi$. In this setting we have
\[
\dot{\phi} = 1  + \varepsilon Z (\phi) \cdot I (\vartheta, \gamma(\phi)) + O(\varepsilon^2),
\]
and we have successfully reduced the problem to the phase of the unperturbed orbit $\gamma$. In general terms, we study problems of the following type
\begin{equation}\label{phase_red}
\dot \phi = 1 + \varepsilon q (\phi,\vartheta).
\end{equation}
The main insight was to obtain the coupling function in terms of how the phase of the  unperturbed orbit behaves near the orbit $\gamma$. We performed the following steps:
}

\begin{description}
\item[(1) Phase sensitivity $Z$ of the  unperturbed system.] Once we have the asymptotic phase, we can use it as the new phase variable $\phi$, extending the definition of phase along the orbit to a neighborhood of the orbit. From the phase, we can in turn compute the phase sensitivity function
$$
Z = \nabla \phi,
$$
where the gradient is evaluated along the orbit $\gamma$.

\item[(2) Obtain coupling function by $q = Z \cdot I$.] For this step, we need to take the inner product of $Z$ with the perturbation $p$. When studying collective phenomena $q$ will contain fast variables and slow variables. Typically, only the slow variable are of interest, so we will {\it average $q$ over the fast variables}.
\end{description}

{\bf Meaning of $q$}. In this approach we have a strong underlying assumption: that
{\it the phase responds linearly to perturbations}. That is, the coupling function is linear in the perturbations. If the phase is perturbed by $I_1$ and $I_2$ the net effect will be the sum of $I_1$ and $I_2$. Notice, that the linearity is only in terms of the perturbations. The equation itself is nonlinear in the phase variable $\varphi$. The linearity with respect to perturbations is because we have discarded all nonlinear terms and terms of order $\varepsilon^2$ (by computing $Z$ along the unperturbed orbit). We will discuss these issues in an example below. This linearity will facilitate the study of networks and large ensembles of oscillators.

These two steps will provide the phase description for weakly coupled oscillators. Using these steps, it is possible to explain the collective behaviour of neurons \cite{Ermentrout:96}, and circadian dynamics \cite{Winfree:80}, among other processes \cite{Kuramoto:84,Ermentrout:10}.

\subsubsection{Synchronization with external forcing}

\label{SyncEF}

We will illustrate and discuss how the above ideas can be applied to study the problem of synchronization with external forcing. Consider the system
 \begin{equation}\label{exiso}
 \dot{x}= {f}({x})
 = x + |x|^2 A x, \mbox{  with }
A = \left(
\begin{array}{cc}
 1 & -1\\
 1 & 1
\end{array}\right).
\end{equation}
By inspection, it is clear that $x=0$ is an unstable point and the system has an attracting periodic orbit $\gamma$ of radius 1. This can be better seen by changing to polar coordinates
$
x_1(r,\phi)=r\sin(\varphi)$ and  $x_2(r,\varphi)=r\cos(\varphi)$
using which, we obtain
\begin{equation}\label{poliso}
\begin{array}{lr}
\dot{r}=r(1-r^2), \,\, & \dot{\varphi}=r^2.
\end{array}
\end{equation}
The orbit $\gamma$ corresponds to  $r=1$ and is shown in Fig.\ \ref{Figphase}.

{\bf Asymptotic phase:} The phase $\varphi$ as defined in the orbit $\gamma$ has a constant
frequency equal to unity. Along the orbit ${\gamma}$, we therefore have $\dot{\varphi}=1$ (by inspection of the equations). For points outside the orbit, however, this is no longer true. The asymptotic phase $\phi$ will fix this issue because  the points then move at the same speed as the corresponding points in the orbit, so that $\dot{\phi}=1$ for points outside the orbit.

Because of the symmetry ($r$ does not depend on $\varphi$) we can use the ansatz
$$\phi(r,\varphi)=\varphi +\zeta(r),$$
where we aim to find the function $\zeta$. Differentiating we obtain
$\dot \phi=\dot \varphi+\frac{d\zeta}{dr}\frac{dr}{dt}
$ and, using the isochron's properties together with the equations for $r$ and $\varphi$, we obtain
$\dot{\zeta}= 1/r$ so $\zeta= \log r+ C.$ Since we want to extend the phase continuously from the orbit, if ${x}\in{\gamma}$ then $\phi(x)=\varphi(x)$. We  choose the constant $C=0.$ Therefore,
$$
\phi(r,\varphi)=\varphi+\log r,
$$
and we can define the isochron with asymptotic phase $\phi (r,\varphi)=c.$ In Fig.\ \ref{Figphase}(top) we show four level sets of the asymptotic phase corresponding to $\phi=0,\pi/2,\pi,$ and $3\pi/2$.

We can use the asymptotic phase to obtain a coordinate that decouples the phase dynamics from the other coordinates. Note that, by defining a new coordinate
$
\phi =\varphi-\eta(r),
$
we obtain
\begin{equation}\label{vf}
\dot{\phi}= \nabla \phi \cdot f = 1,
\end{equation}
which is valid, not only along the orbit $\gamma$ via Eq. \ref{poliso}, but also in a neighborhood of the orbit.
In the first equality we just stressed the identity between the frequency  (applying the chain rule) and the gradient of $\phi$.

We can now readily take the gradient (in polar coordinates), yielding
$
\nabla \phi= e_1 ( \sin(\phi-\log(r))+\cos(\phi-\log(r))/ r.
$
Along the unperturbed orbit ${\gamma}$ we have $Z(\phi) = \nabla_{\gamma}\phi$ so that the phase sensitivity functions are
$$
Z(\phi)= e_1 \sin(\phi+\pi/4).
$$
Next we obtain the coupling function.

{\bf Obtaining the coupling function of external forcing.} Now we consider the system being forced at frequency:
 \begin{equation}\label{expert}
 \dot{{x}}= {f}({x}) + \varepsilon {I}(\vartheta),
\end{equation}
where  $\vartheta = \omega t$.
We obtain the coupling function through the isochron. We now justify in detail why we have discarded the corrections in $\varepsilon^2$.

We compute the equation for the phase dynamics. Note that, by chain rule
$\dot{\phi} = \nabla\phi\cdot{\dot{x}} = \nabla\phi\cdot(f(x) +\varepsilon{I})$. Using Eq. (\ref{vf})
and evaluating the gradient along the orbit $\tilde \gamma$, we obtain
$$
\dot{\phi} = 1+\varepsilon \nabla_{\tilde \gamma}\phi \cdot{I}.
$$
For small $\varepsilon$, we know that the difference between
$\nabla_{\gamma} \phi$
and
$
\nabla_{\tilde{\gamma}} \phi
$
is of order $\varepsilon$, so we can replace the gradient along the perturbed orbit and unperturbed orbit with corrections of order $\varepsilon^2$ (because $\varepsilon$ is already multiplying the function).
Hence,
$$
\dot{\phi}=1+\varepsilon q(\phi,\vartheta)+{O}(\varepsilon^2).
$$

{\bf Synchronization and coupling function.}  The main idea is that the coupling function $q$ can help in adjusting the frequency of the system to the frequency $\omega$ of the forcing.
As we discussed above, we will neglect the terms ${O}(\varepsilon^2)$. Introducing the phase difference
$$
\psi = \phi -\vartheta
$$
and considering  $1-\omega = \Delta_{\omega}$ we obtain
$$
\dot{\psi}= \Delta_{\omega} +\varepsilon Z(\psi+\vartheta) \cdot I(\vartheta ).
$$
If $\Delta_{\omega}$ is of order $\varepsilon$ then the dynamics of $\psi$ will be  {\it slow} in comparison with the dynamics of $\theta$. Roughly speaking, for each cycle of $\psi$ we have $1/\varepsilon$ cycles of $\theta$.
Because, the dynamics of $\vartheta$ is faster than that of $\psi$, we use the \emph{averaging} method to obtain the coupling function
$$
q(\psi)=\frac{1}{T}\displaystyle{\int_0^T Z(\psi+\vartheta) \cdot I(\vartheta)d\vartheta},
$$
\noindent
where $T=2\pi$ is the period of $p$ as a function of $\vartheta$.
Note that, for our result, $Z$ is sinusoidal so that by integrating over $\vartheta$ while keeping $\psi$ fixed, we obtain $q(\psi) = A \sin (\psi + \beta)$. Hence we obtain the dynamics in terms of the phase difference
\begin{equation}\label{exmpAdler}
 \frac{d\psi}{dt}=\Delta_{\omega}+\varepsilon q(\psi),
\end{equation}
which is exactly the equation shown in Fig.\ \ref{fig:CF_kuramoto}.

We are now ready to study collective phenomena between the driving  and the system. For instance, the system will phase-lock with the driving when $\Delta + \varepsilon q(\psi^*) = 0$. In this case, the oscillators will have the same frequency. Moreover, because
$q$ is a periodic function, the fixed point $\psi^*$ will exist only when
$ | \Delta/\varepsilon| \le \max q$.

%
%
%

{\bf Higher order {\itshape n:m} phase locking}. Our assumption is that $\Delta = O(\varepsilon)$, so that Eq. (\ref{exmpAdler}) for the phase difference $\psi = \phi - \vartheta$ is a slow variable. It may happen that $\psi_{mn} = m \varphi - n \vartheta$ gives rise to a slow variable. In such cases, we perform the same analysis for $\psi_{mn}$ and further information on the higher order phase locking can be obtained \red{\cite{Ermentrout:81}}.

\subsubsection{Phase response curve} \label{sec:PRC_theory}

The phase sensitivity function $Z$ plays a major role in this analysis. It also has many names: \red{infinitesimal phase response curve (iPRC)},  linear response function,  infinitesimal phase resetting curve. It is deeply related to the so-called \red{phase response curve or phase resetting curve (PRC)}.  For an oscillator to be able to adjust its rhythm and synchronize, it must respond differently to the perturbations at different phases $\phi$. So the phase can advance or retard to adjust its rhythm to the external forcing. The PRC is a natural way of displaying the response of oscillators to perturbations and thereby to gain insight into the collective dynamics.

The main idea of the PRC is as follows: If we perform a small and short perturbation of the orbit, the phase may complete its cycle before expectation (in the absence of perturbations), or it may be delayed.  The unperturbed period of the orbit $\gamma$ is $T_0$. Every point on the orbit can be uniquely described by the phase $\phi$. A small perturbation applied at a phase $\phi_s$ can cause the phase to complete its full cycle at a time $T_1$.  The normalized phase difference between the cycles is
$$
PRC  = \frac{T_0-T_1}{T_0}.
$$
Note that the PRC  depends on the phase $\phi_s$ at which the small perturbation was applied, that is, $PRC  = PRC (\phi_s)$. This is the so-called phase response curve.

In the theory of weakly coupled oscillators, we use the concept of an infinitesimal PRC (iPRC). It is equivalent to the gradient of the phase $Z$, and it is defined as the PRC normalized by the amplitude of the perturbation $A$
$$
Z = \lim_{A\rightarrow 0} \frac{PRC}{A}.
$$
Indeed, the isochrons and the PRC are closely related. If a  point moving along the orbit $\gamma$ is instantaneous and the perturbation is small, the point will land on an isochron, which tells us the new phase $\phi$ of the point once it comes back to the orbit. Further considerations and the relationship of the PRC to experiments are given in Sec.\ \ref{sec42:PRC}.

In neuroscience, pulse-coupled oscillators are an important class of models. Here the interactions happen in instantaneous pulses of communication. The collective dynamics of such models are of great interest \cite{Mirollo:90}. The relationship between pulse-coupled oscillators and the phase reduction presented above has recently been elucidated by \citet{Politi:15} who showed that the models are equivalent.

\subsubsection{Examples of the phase sensitivity function}

Because of the works of \citet{Winfree:67,Winfree:80}, \citet{Kuramoto:84},  \citet{Ermentrout:96,Stiefel:08},  we now have a good understanding of the phase sensitivity $Z$ for many classes of systems such as heartbeats, circadian rhythms, and  in some neurons (with stable repetitive firing).

\begin{figure}
\centerline{\hbox{\psfig{file=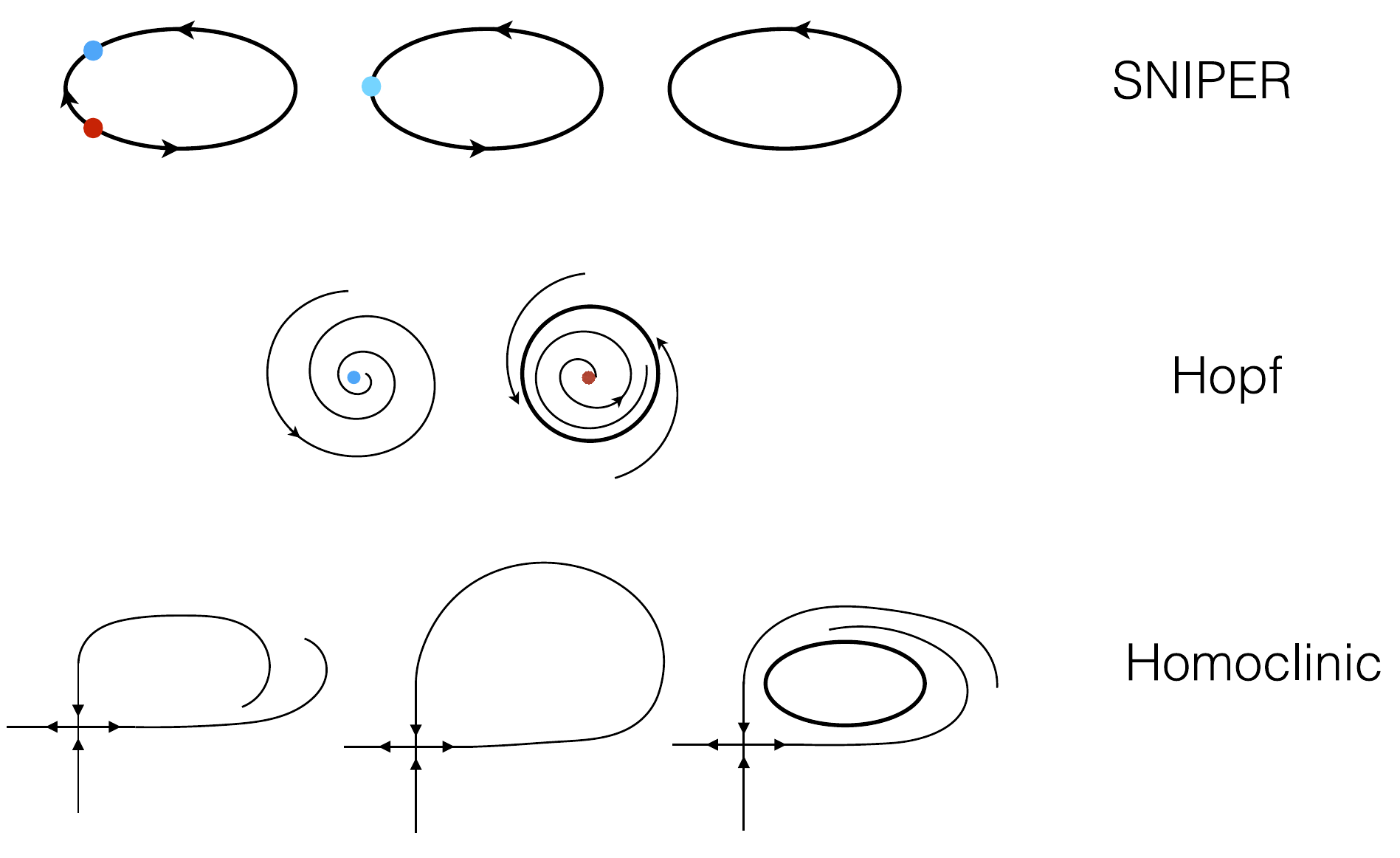,width=9.0cm}}}
\caption{(color online)
Three typical bifurcations appearing as the result of changing a single parameter. As the parameter changes for the SNIPER bifurcation, two fixed points collapse to a saddle node on the circle, and then the system oscillates. In the Hopf bifurcation, a periodic orbit appears after destabilization of the fixed point. In the homoclinic bifurcation, the stable and unstable manifolds of the saddle point join to form a homoclinic orbit, as the parameter changes; with further parameter change, the homoclinic orbit is destroyed and a periodic orbit appears.
}
\label{Bif}
\end{figure}

The iPRC and PRC are closely related to the bifurcation that led to the oscillatory behaviour  \cite{Ermentrout:96,Brown:04}. $Z(\phi)$ is a vector and, in our example in Sec.\ \ref{SyncEF}, the norm of $Z$ was proportional to $\sin (\phi + \beta)$. This is typical of Hopf bifurcations. In Fig.\ \ref{Bif} we present typical bifurcations in neuronal models for which the iPRC and PRC are relevant.

\noindent

In neuron models, the coupling is in one single variable: the membrane potential $V$. So we only need to compute the derivative with respect to $V$. Thus, $Z(\phi) = \partial \phi / \partial V$ \footnote{We are abusing the notation by using $Z$ to represent both the full gradient and the derivative with respect to a single variable. In theoretical neuroscience this convention is standard.}.  \citet{Izhikevich:00}  derived a phase model for weakly coupled relaxation oscillators and burster neurons \cite{Izhikevich:07}. \citet{Brown:04} obtained the phase sensitivity $Z$ for other interesting cases, including homoclinic oscillators. Neurons with stable repetitive firing (corresponding to a stable orbit) can be  classified as having PRC type I dynamics corresponding to a SNIPER bifurcation, or PRC type II dynamics corresponding to a Hopf bifurcation. The phase portraits for these two bifurcations are illustrated in Fig.\ \ref{Bif}. PRCs of type I are always positive whereas PRCs of type II have both negative and positive parts, as shown in Table \ref{TableQ}. The PRC type is indicative of the neuron's ability to synchronize: networks of neurons with  PRC type II  can synchronize via mutual excitatory coupling, but those of PRC type I cannot \cite{Ermentrout:96}.


\begin{table}[b]
\centering
\begin{tabular}{lr} 
\hline
Bifurcation    & $Z(\phi)$ \\
\hline
\hline
SNIPER & $1-\cos \phi$  \\
Hopf & $\sin(\phi - \beta)$  \\
Homoclinic & $\exp(-\lambda \phi)$  \\
Integrate and Fire & $2\pi$  \\
Leaky Integrate and Fire & $\exp(g \phi)$ \\ \hline
   \end{tabular}
   \caption{The phase sensitiveness for various models and their bifurcation after  \citet{Brown:04}.  \citet{Izhikevich:00,Izhikevich:07} obtained the phase sensitivity $Z$ for relaxation oscillators. They are discontinuous and are not shown here.}
   \label{TableQ}
\end{table}

\subsection{Globally coupled oscillators} \label{EnO}

Now suppose that we have $N$ coupled oscillators
\begin{equation}\label{expert}
 \dot{x_i}= f_i(x_i) + \varepsilon \sum_{j=1}^N H_{ij}(x_i, x_j).
\end{equation}
We assume that, when they are uncoupled $\varepsilon=0$, each system has an exponentially attracting periodic orbit. So the dynamics of the uncoupled system occurs on a torus $\mathbb{T}^N$ that is exponentially attracting. Moreover, we also assume that $f_i$ is close to $f$. As we turn the coupling on, the dynamics changes. The theory of normally hyperbolic invariant manifolds guarantees that the dynamics of system with small coupling will also take place on a torus \cite{Eldering:13}.
So the amplitudes remain roughly the same.  But the dynamics on the torus, that is, the phases can change a lot \cite{Turaev:15}.

Again if we know the isochrons for the phases we can use the same arguments to describe the system in terms of the phases. The corresponding phase model
$$
\dot \phi = \omega_i + \varepsilon \sum_{j} q_{ij}(\phi_i, \phi_j),
$$
where each oscillator has its own period $T_i$, and $q_{ij}$ is the coupling function describing the influence of the $j$-th oscillator on the $i$-th oscillator.
Here
$$
q_{ij}(\phi_i, \phi_j) = Q(\phi_i) \cdot  H_{ij}(\gamma_i(\phi_i), \gamma_j(\phi_j)).
$$
Note that $Z$ is independent of the index $i$ because we assumed that $f_i$'s are all close to $f$.
Again, we can average over the fast variables to obtain equations in terms of the phase difference \cite{Daido:96}.

\vspace{0.2cm}

In many cases, the isolated oscillators are close to a Hopf bifurcation. As discussed above, the coupling function after averaging takes the form
$$
q_{ij}(\phi_i , \phi_j) = \sin (\phi_i - \phi_j + \beta).
$$
This is by far the best-studied coupling function, and it has offered deep insights into collective properties for both globally coupled oscillators \cite{Acebron:05} and complex networks \cite{Arenas:08,Rodrigues:16}.

First, we consider $\beta=0$. The  model is then written as
$$
\dot \phi_i = \omega_i + \frac{\varepsilon}{N} \sum_{j=1}^N \sin(\phi_j -  \phi_i).
$$
If the oscillators are identical $\omega_i = \omega$ then any small coupling $\varepsilon >0$ leads to synchronization (the phases will converge to the same value).  If the distribution $g$ of natural frequencies $\omega_i$ is broad, then  at a critical coupling $\varepsilon_c$ a large cluster of synchronized oscillators appears and, with further increase of coupling, additional oscillators join the cluster.

The main idea of the analysis is to introduce an order parameter
$$
z= r e^{i \psi} = \frac 1 N \sum_{j=1}^N e^{i \phi_j}
$$
and to rewrite the equations in terms of the parameters $r$ and $\psi$ (which are now mean-field parameters)
$$
\phi_i = \omega_i + \varepsilon r \sin(\psi - \phi_i).
$$
Taking the limit $N\rightarrow \infty$, we can write the model in terms of self-consistent equations \cite{Kuramoto:84}.

When the distribution $g$ of the natural frequencies is an even, unimodal, and non-increasing function, and the coupling is weak, the incoherent state is neutral \cite{Strogatz:91}, but the order parameter $r$ vanishes (at a polynomial rate) if $g$ is smooth \cite{Fernandez:14}.
Moreover, on increasing the coupling, the incoherent solution $r=0$ bifurcates for
$$
\varepsilon>\varepsilon_c=\frac{2}{g(0)}.
$$
A systematic review of the critical coupling, including bi-modal distributions (keeping the coupling function purely harmonic) has been given by \citet{Acebron:05}.

\subsubsection{Coupling functions leading to multistability}

As we have seen, if the oscillators are close to a Hopf bifurcation as is typical, then the corresponding phase sensitivity $Z(\phi) \propto \sin(\phi + \beta)$. This means that the coupling function will have a phase shift
$$
q(\phi_i, \phi_j) = \sin (\phi_j - \phi_i + \beta), \mbox{   with }  |\alpha| < \pi/2.
$$
This coupling function is called the Sakaguchi-Kuramoto coupling \cite{Sakaguchi:86}.
The slight modification can lead to non-monotonic behaviour of synchronization \cite{Omel:12}. For certain unimodal frequency distributions $g$, the order parameter can decay as the coupling increases above the critical coupling,  and the incoherent state can regain stability. Likewise multistability between partially synchronized states and/or the incoherent state can also appear.

Although the dynamics of the model with this slight modification can be intricate, it is still possible to treat a more general version of the Sakaguchi-Kuramoto coupling
$$
q_{ij}(\phi_i,\phi_j) = B_j \sin(\phi_j + \beta_j - \phi_i - \alpha_i).
$$
This coupling function generalizes the standard  Sakaguchi-Kuramoto model as it allows for different contributions of oscillators to the mean  field, on account of the phase shifts $\alpha_i$ and $\beta_j$ and coupling factors $B_j$. In turn, the mean field acts on each oscillator differently. This scenario is tractable in terms of the self-consistency equations for the amplitude and frequency of the mean field \cite{Vlasov:14}. Also in this setting, solutions of the coupled phase oscillators approximate solutions of phase oscillators with an inertial term \cite{Dorfler:12} which plays a major role in power-grids.

{\bf Higher harmonics.} In the previous discussion, the coupling function $q$ contained one harmonic. $q(\phi) = \sin(\phi + \beta)$.  Depending on the underlying bifurcation we must now include further, higher-order, Fourier components,
$$
q(\phi) = a_1 \sin(\phi) + a_2 \sin(2\phi + b_2) + ... + a_n \sin(n\phi + b_n),
$$
where $a_i$ and $b_i$ are parameters. For example, synchronization  of weakly-coupled Hodgkin-Huxley neurons can be replicated using coupling functions consisting of the first four Fourier components \cite{Hansel:93}.

Moreover, considering the coupling function $q$ to be a biharmonic coupling function \cite{Hansel:93a}, there is a multiplicity of such states, which differ microscopically in the distributions of locked phases \cite{Komarov:13}. Higher harmonics in the coupling function can also lead to  the onset of chaotic fluctuations in the order parameter \cite{Bick:11}. Indeed, the coupling function alone can generate chaos, that is, even keeping the frequencies identical and having no amplitude variations.

\subsubsection{Designing coupling functions for cluster states and chimeras}

Ashwin and co-workers tailored the coupling function to obtain clusters states in identically and globally coupled phase oscillators \cite{Ismail:15,Orosz:09}. In this situation, because the oscillators are identical, the behaviour of the system is determined by the number of oscillators $N$ and the coupling function $q$. By carefully choosing the Fourier coefficients of the coupling function we obtain two major results: (a) any clustering can appear and be stable, and (b) open sets of coupling functions can generate heteroclinic network attractors. Heteroclinic networks are not confined to globally-coupled oscillators and they can appear robustly in complex networks \cite{Aguiar:11,Field:15}.

\red{In networks of identical oscillators a chimera state is defined as a spatio-temporal pattern in which synchronous and asynchronous oscillations coexist \cite{Abrams:04,Hagerstrom:12}.} Chimera states among phase oscillators appear only when a spatial (long-range) coupling is included. The approach described above, in which the coupling is tailored to obtain clusters and complicated attractors in identical and globally-coupled oscillators, can be extrapolated to construct chimeras. They can be obtained either by consideration of higher harmonics \cite{Ashwin:15}, or by perturbation of the coupling function in specific ways. 
\red{A partially coherent inhomogeneous pattern called chimera death, which combines the features of chimera
states and oscillation death, has been also established \cite{Zakharova:14}.}

\subsubsection{Coupling functions with delay}

A natural generalization of the coupling function is to introduce delay. A common case is the inclusion of transmission delays
$$
q_{\tau}(\phi_i(t), \phi_j(t)) = \sin (\phi_j(t-\tau) - \phi_i - \beta).
$$
The addition of delays makes the model infinite-dimensional and leads to a series of new phenomena such as bistability between synchronized and incoherent states, and unsteady solutions with time-dependent order parameters \cite{Yeung:99}. Multistability is very common in the presence of delays. In particular, it can also be observed in small variations of the previous coupling function
$$
q_{\tau}(\phi_i(t), \phi_j(t)) = b \sin \phi_i(t) +  \varepsilon \sin (\phi_j(t-\tau) - \phi_i - \beta),
$$
and multistablity can also be observed \cite{Kim:97}.

Carefully chosen  communication delays can also be used to encode patterns in the temporal coding by spikes. These patterns can be obtained by a modulation of the multiple, coexisting, stable, in-phase synchronized states, or traveling waves propagating along or against the direction of coupling \cite{Popovych:11}. 
\red{Coupling functions with delay can also be used for controlling the state of oscillation death \cite{Zakharova:13}.} Two limiting cases of delay can be treated. First, for very small delays, the theory of an invariant manifold can be applied. Secondly, in the case of large delays, developments due to \citet{Lichtner:11} and \citet{Flunkert:10}  can be used to determine the collective properties of ensembles of oscillators.

\subsubsection{Low dimensional dynamics}

A particularly striking observation is the low-dimensional dynamics of identical globally-coupled phase oscillators under the Sakaguchi-Kuramoto coupling function. Note that, in this case, writing the sinusoidal coupling in exponential form, we can express the coupled equations as
$$
\dot \phi_j = f e^{i \phi_j}  + g + \bar{f} e^{-i \phi_j},
$$
where $f$ and $g$ are smooth function of the phases which can also depend on time.

These identically coupled oscillators evolve under the action of the Moebius symmetry group $M$ (actually a Moebius subgroup). So, ensembles of identical, globally-coupled oscillators have $N-3$ constants of motion and their dynamics is three-dimensional \cite{Watanabe:93,Marvel:09}. That is,  all phases evolve according to the action of the same Moebius transformation
$$
e^{i \phi_j} = M_{\alpha,\psi} (e^{i \theta_j}),
$$
where the $\theta_j$ are constants and $\alpha \in \mathbb{C}$ and $\psi \in \mathbb{S}^1$ are the parameters of the Moebius group.

This approach can do more. In the limit of large $N$ it is possible to obtain nonlinear equations for the order parameter.
Choosing  uniformly-distributed constants of motion $\theta_j$, the complex order parameter $z$ follows a Riccati equation
$$
\dot z = i (f z^2 + g z + \bar{f}).
$$
This reduction was applied to study a number of nonlinear problems in arrays of Josephson junctions \cite{Vlasov:13,Marvel:09a}, discontinuous transitions in explosive synchronization \cite{Vlasov:15a} and to classify the attractors in the ensemble of oscillators. Indeed, the only attractors are fixed points or limit cycles where all but one oscillator are synchronized \cite{Engelbrecht:14}.

The above reduction follows from the group symmetry of the equations and it is valid only for identical frequencies. \citet{Ott:08} put forward a scheme allowing dimensional reduction for nonidentical frequencies. In the limit $N \rightarrow \infty$, the state of the oscillator system is described by a distribution
$$
f(\omega,\phi,t) = \frac{g(\omega)}{2\pi} \left( \sum_{n=1}^{\infty} f_n(\omega, t) e^{in\phi} + c.c\right),
$$
where c.c. stands for complex conjugate. Next we assume that
\begin{equation}\label{Ott}
f_n (\omega, t) = \alpha(\omega,t)^n,
\end{equation}
that is, the whole distribution $f$ is determined by only one function with $|\alpha|<1$. It is possible to show that the evolution of the system preserves this form of $f$. For various classes of distribution $g$ it is possible to obtain equations for $\alpha$ and thereby for the order parameter.

So the scheme will give low dimensional equations for the order parameter. This ansatz of Eq.\ (\ref{Ott}) has been successfully applied to understand the dynamics of globally coupled oscillators and the second order Kuramoto model \cite{Rodrigues:16}, and to  understand the formation of clusters when higher order harmonics are included in the coupling function \cite{Skardal:11}. The approach can also be used to study nonautonomous globally coupled ensembles of phase oscillators \cite{Petkoski:12a}.

\citet{Pikovsky:11} made a generalization to heterogeneous ensembles of phase oscillators, and connected the  Watanabe and Strogatz reduction to the Ott and Antonsen ansatz. In the limit of infinitely many oscillators, the Kuramoto order parameter $z$ can be written as an integral over the stationary distribution of phases $\nu$. Clearly $z$ does not characterize the distribution $\nu$, so one may consider the generalized order parameters
$$
z_m  = \int_0^{2\pi} \nu(\phi) e^{im\phi} d\phi,
$$
which are the Fourier coefficients of the distribution $\rho$. Clearly, $z_1 = z$ is the standard order parameter. If the distribution of the constants of motion is uniform, then
$$
z_m = z^m
$$
and, for this particular case, the order parameter $z$ completely determines the distribution. The Ott and Antonsen ansatz corresponds to the special case where the generalized order parameters are expressed via the powers of order parameter.

{\bf Sensitivity to the coupling function.} \red{This approach to finding low-dimensional dynamics is dependent on the coupling function being sinusoidal in shape.} 
The Watanabe and Strogatz reduction for $N$ globally coupled oscillators gives $N-3$ constants of motion. Therefore, the dynamics of the ensemble is neutral. The dynamics on these subspaces evolves under the identity map. Recent results show that, by perturbing the identity map, we can generate {\it any dynamics} \cite{Turaev:15}. So, small perturbations in the coupling function can lead to abrupt changes in the dynamics of the ensemble.

\subsubsection{Noise and nonautonomous effects}

If the oscillators are subject to noise, the phase reduction scheme can still be applied
but with some minor modifications \cite{Balanov:08,Ermentrout:06}. Even in the absence of coupling, the oscillators can synchronize if driven by a common noise. This is a general result by \citet{Jan:87} who showed that, when phase oscillators are driven by noise, the trajectories converge to a random fixed point (corresponding to the two oscillators going to the same trajectory). This result is well appreciated in the physics community as are also the differing effects of common and independent noises \cite{Lindner:04}.

In the context of interacting oscillators we can analyze the contributions of the coupling function and common noise in driving the oscillators towards synchronization \cite{GarciaAlvarez:09}.
Consider the following model of two coupled (or uncoupled) phase oscillators with common and independent noises:
\begin{eqnarray}
\dot \phi_1 &=& \omega_1 + \varepsilon \sin \psi + A_1 \xi(t) \sin(\phi_1) + B_1 \xi_1 \sin \phi_1 \nonumber \\
\dot \phi_2 &=& \omega_2 + \varepsilon_2 \sin \psi + A_1 \xi(t) \sin(\phi_1) + B_2 \xi_2 \sin \phi_1 \nonumber
\end{eqnarray}
where $\xi,\xi_1$ and $\xi_2$ are Gaussian noises of unit variance and, again, $\psi = \phi_1 - \phi_2$.
It is possible to obtain a stochastic differential equation for the generalised phase difference $\psi$. This equation is nonautonomous. An analytical approach is to write a Fokker-Planck equation for the probability density of $\psi$.

The probability density is almost independent of the fast variables $\phi_1$ and $\phi_2$, so a good  approximation is to integrate over these variables to obtain a proxy for a stationary probability distribution. This approach reveals three important effects: $(i)$ independent noises $\xi_1,\xi_2$ hinder synchronisation; $(ii)$ coupling-induced synchronisation takes place for low noise intensity and large coupling strengths; and $(iii)$ common-noise-induced synchronisation occurs for large common-noise intensities and small coupling strengths.

{\it Nonautonomous effects.} If the frequencies of the oscillators are nonautonomous but the oscillators are identical the reduction techniques can still be applied \cite{Watanabe:93,Marvel:09,Petkoski:12a}, so that phenomena such as synchronization can be studied. A new class of systems described by nonautonomous differential equations are {\it chronotaxic systems}. \red{These are defined as dissipative dynamical systems with internal sources of energy.} In such cases the coupling function is nonautonomous and the systems retain stable (time-dependent) amplitude and phase under external perturbation \cite{Suprunenko:13}.

\subsection{Networks of oscillators} \label{sec47:Thr_Net}

In this section, we generalize the discussion to networks of interacting systems with pairwise interaction. That is, we consider
$$
\dot x_i = f_i(x_i) + \varepsilon \sum_{j=1}^N W_{ij} H_{ij}(x_i,x_j),
$$
where $W_{ij}$ is the matrix encoding the strength of interaction between $j$ and $i$, and $H$ is the coupling function. Note that we allow each isolated vector field to be distinct. To be able to draw conclusions about the overall dynamics from the microscopic data for $f$, $W$ and $H$, we will consider a subclass of vector fields and coupling functions.

\subsubsection{Reduction to phase oscillators}

Assume that for $\varepsilon=0$, each isolated system has an exponentially-attracting periodic orbit. A typical assumption is that $H_{ij} = H$ (i.e.\ all coupling functions are identical). Then proceeding in the same way as in Sec. \ref{EnO} for globally coupled oscillators in the limit of small coupling strengths, we can reduce the dynamics to the phases
$$
\dot{\varphi}_i = \omega_i + \varepsilon \sum_{j=1}^N W_{ij} q(\varphi_i,\varphi_j).
$$
This model describes the dynamics of the phase oscillators in terms of complex networks of interactions. Most results relate to the sinusoidal coupling function $q(\varphi_i,\varphi_j) = \sin (\varphi_j - \varphi_i)$. The main questions lie in the realm of collective dynamics and transitions from incoherent to coherent states \cite{Rodrigues:16}. The situation here is less-well-understood. For instance, it is unclear how to generalize the low-dimensional reduction approach.

\subsubsection{Networks of chaotic oscillators}

A subclass of this model offers insight. Consider $W_{ij} = A_{ij}$, where $A_{ij}=1$ if $i$ receives a connection from $j$, and $A_{ij}=0$ otherwise. Moreover, consider the diffusive coupling functions $H_{ij}(x,y) = H(x-y)$. Suppose also that $f_i = f$, that is, all isolated nodes are identical. We also assume that the isolated systems are chaotic. This assumption is not necessary but {\ae}sthetically pleasant, because in this situation the only possible source for collective dynamics is through the coupling. This model corresponds to identical oscillators interacting diffusively, and it can then be rewritten as
$$
\dot x_i = f(x_i) + \varepsilon \sum_{j=1}^N A_{ij} H(x_j - x_i).
$$
The role of the coupling function is to attempt to bring the system towards synchronization $x_1=x_2 = \cdots = x_N$. The main questions now are: (a) when will the coupling function $H$ bring the system towards synchronization; and (b) how will the interaction structure $A_{ij}$ influence the system? We should analyse the growth of small perturbations $x_i = s + \xi_i$, where $\dot s  = f(s)$. But we face the challenges of having too many equations and, moreover, of all the $\xi_i$ being coupled.

{\bf Global Results:} Here we want to find the conditions on the coupling function guaranteeing that the network dynamics will converge to synchronization, regardless the initial conditions. The challenge is to construct a Lyapunov function, whose existence is a sufficient condition for a globally stable synchronous state.

\citet{Pogromsky:01} used control techniques and concepts of passive systems to obtain global synchronization results for arrays of interacting systems. Assuming that the coupling function is positive definite, they were able to construct a Lyapunov function for the array and to express its construction in terms of the spectrum of the network. They showed thereby that all solutions of the coupled equation are bounded and that, if the coupling is large enough, the network synchronizes. They also showed how diffusion-driven instabilities can appear in such arrays. This approach to passive systems was subsequently applied to neuron models to study their synchronization properties \cite{Steur:09}.

In a similar spirit to constructing Lyapunov functions, \citet{Belykh:04} developed the {\it connection graph stability method}. At its very heart, the method requires the existence of a Lyapunov function for the nonlinear equations of the perturbations $\xi_i$. The existence of this function is unclear from the beginning, however, and it may depend on the vector field $f$. The method relates the critical coupling necessary to attain synchronization to the total length of all paths passing through an edge on the network connection graph.

Another approach to studying synchronization is to tackle the equations for the perturbations $\xi_i$ using the theories of contraction \cite{Russo:09} and exponential dichotomy \cite{Pereira:14}. Here we use the coupling function to construct differential inequalities. At their cores, these approaches are equivalent to the construction of local Lyapunov functions. However, if the coupling function is such that the contraction theory can be applied (for example the coupling function is positive definite) then much information on the synchronization can be extracted. In particular, even if the network structure is time-varying the network may synchronize \cite{Lu:05}.

{\bf Local Results:} In the section above, we took account of the nonlinear behaviour of the perturbation $\xi_i$. In the local approach we consider only the linear terms in $\xi$.
\citet{Pecora:98} had the idea of block-diagonalizing the perturbations $\xi_i$ via a change of coordinates where $\xi_i$ goes to $\zeta_i$. In the new variables $\zeta_i$ the perturbations decouple and they all have the same form
$$
\dot z = [Df(s(t)) + \alpha \Gamma] z,
$$
where $\Gamma = DH(0)$. To recover the equation for $\zeta_i$ we only need to set $\alpha = \varepsilon \lambda_i$ where $\lambda_i$ is the $i$th eigenvalue of the Laplacian of the network. So the problem reduces to the case of two coupled oscillators. Obviously, there are additional challenges in understanding the graph structure via $\lambda_i$, but the main idea now boils down to the case of two oscillators. We classify the stability of the variational equation for $z$. The commonest criterion used for stability is the Lyapunov exponent, which gives rise to the master stability function $\Lambda$ (just as in the two oscillator case). This approach showed that the topology of the networks can exert systematic influences on the synchronization \cite{Barahona:02} and can be used to predict the onset of synchronization clusters \cite{Pecora:14,Williams:13}. In the last two decades this approach has been popular and it has been applied to a variety of network structures \cite{Arenas:08} and to problems of pinning control \cite{Sorrentino:07}.

The MSF approach was also extended to the case where the coupling function has time delays \cite{Li:04}. Moreover, in the limit of large delays it is possible to understand the behaviour of the level sets of the master stability function $\Lambda$. Indeed, the level sets tend to be circles whose radii increase monotonically in the complex plane \cite{Flunkert:10}. Some networks also have two types of coupling function. Typical examples are neural networks where electrical and chemical synapses  coexist. If the underlying matrices defining the chemical and electrical coupling commute, then the MSF can be used to understand the net effects of the coupling function on the synchronization \cite{Baptista:10}.

{\bf Generalizations:} So far, we have discussed networks of identical oscillators. If the network is composed of slightly nonidentical nodes, the MSF approach can still be applied, via a perturbation analysis \cite{Sun:09}. In general, to understand the effect of the network, combined with the effects of nonidentical nodes, further information about the coupling function is necessary. If the coupling function $\Gamma$ (linearized about the synchronized manifold) has a spectrum with a positive real part, then we can extract a great deal of information it. Adding random links to a network of nonidentical oscillators can substantially improve the coherence \cite{Pereira:13}. For directed networks, depending on the coupling function, improvements in the network topology such as link addition can destabilise synchronization \cite{Pade:15}. This phenomenon can also be observed in experiments with lasers when the coupling function has a time-delay \cite{Hart:15}.

\begin{figure*}
{\caption{(color online). Schematic illustration of the procedure for the inference of coupling functions. From left to right: measurement data $\mathcal M$; pre-estimation procedure where the phase or amplitude $\tilde{\mathcal M}$ are estimated from those data; the inference of a dynamical model from the $\tilde{\mathcal M}$ data; and the coupling function emerging as the end result of the procedure.} \label{fig:infr}}
{\includegraphics[width=0.85\textwidth,angle=0]{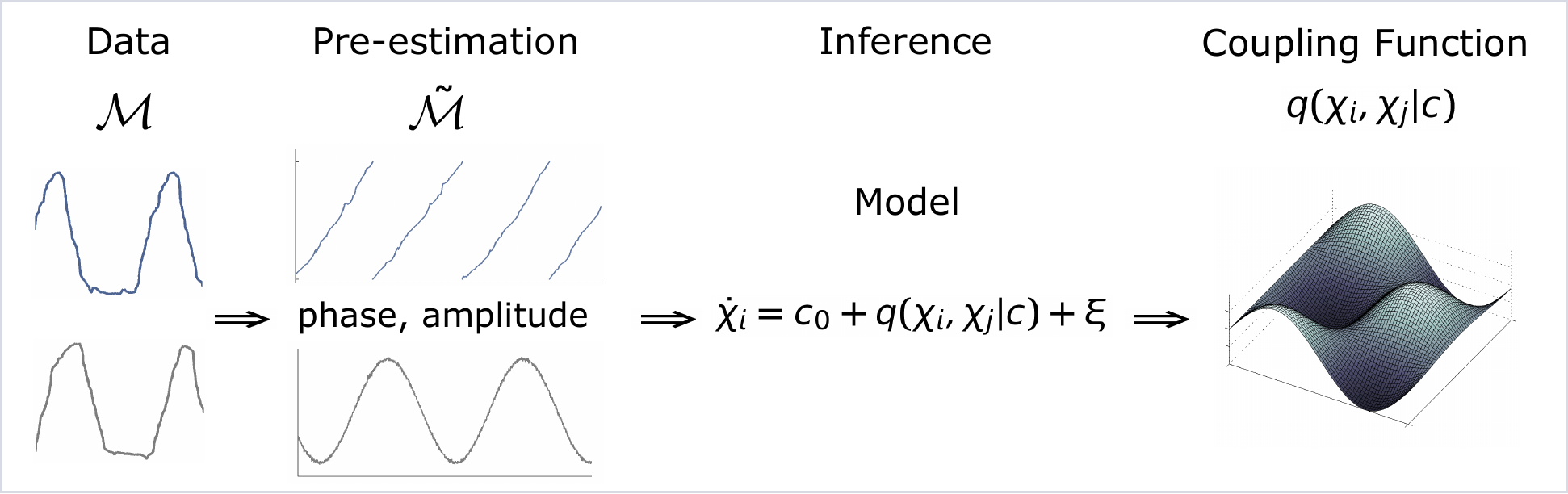}}
\end{figure*}

Moreover, if one adds a small perturbation on the nodes of systems interacting in a fully connected network, the collective dynamics will lead to smaller fluctuations than those expected if the oscillators were uncoupled and one applies the central limit theorem
\cite{Masuda:10}. That is, the central limit theorem would predict fluctuations of order $O(N^{-1/2})$, whereas the collective dynamics gives $O(N^{-1})$. One can classify the regular networks such that the improvement is significant \cite{Pereira:14}. For this class of coupling function, one can also gain insight into the speed of synchronization, that is, how fast the system converges towards synchronization. The speed is well known to depend on the network structure \cite{Timme:04,Grabow:10}. For the  above class of coupling functions, it is possible to show how the properties of the coupling function and network structure combine to determine the speed of convergence. For example, if the spectrum of the coupling function is real, then the speed towards synchronization is dictated by the real part of the Laplacian spectral gap.

So far efforts have been concentrated on the study of nonidentical nodes, while keeping the coupling function identical. Because the coupling function couples with the network structure of the equations, perturbations in the coupling function can have a drastic impact on the collective dynamics. For instance, if the network has a heterogeneous degree distribution, no perturbations in the coupling function are tolerated. Any perturbation in the large network limit will destabilize the synchronous motion \cite{Maia:15}.

\section{Methods}\label{sec5:Methods}

\subsection{ Inferring coupling functions}\label{sec61:Inference}


Before discussing methods for the inference of coupling functions, we mention earlier discussions and techniques \cite{Schiff:96,Cenys:91,Schreiber:00a,Stefanovska:99a,Arnhold:99} that paved the way for the subsequent introduction of coupling functions. Although the mathematical and computational facilities of the time did not allow for the full inference of coupling functions, this goal was nonetheless closely approached by different measures that detected the existence of a coupling relationship and characterized its nature.

In a study of this kind, \citet{Stefanovska:99a} investigated the coupled oscillators of the cardiovascular system from human blood flow signals. Among other methods, wavelet time-frequency analysis was used to detect the instantaneous frequency of the heart, through which the coupling from respiration was assessed. Similarly, many information-based measures were used for assessment of the coupling strength and directionality \cite{Schreiber:00a,Palus:03a,Barrett:13}. Even though these methods are very useful in detecting the net coupling effects, they are essentially directed functional connectivity measures and they are not designed to infer mechanisms.

\blue{
Another traditional approach for studying interactions is via transfer function analysis \cite{Boreman:01,Sanathanan:63,Saul:91,Cooke:98}. The transfer function is a mathematical representation that describes the linear relationships between the inputs and outputs of a system considered as a black-box model. Although this approach suffers from some limitations, it has nonetheless been used in the past for understanding interactions, and still is.
}

The inference of coupling functions involves the analysis of data to \emph{reconstruct a dynamical model} describing the interactions. The main pillar of the procedure is a method for dynamical inference, often referred to as {\it dynamical modelling} or {\it dynamical filtering} \cite{Toussaint:11,Kalman:60,Arulampalam:02,Voss:04}. The latter has been used historically as means of advanced ``filtering'', when one selects and detects the features of interest described by the model -- a celebrated example being the Kalman filter \cite{Kalman:60}. Fig.\ \ref{fig:infr} presents the main steps in obtaining the coupling function. In short, starting with the data $\mathcal M$ from two (or more) interacting dynamical systems, first the appropriate observable data $\tilde{\mathcal M}$, like the phase or amplitude, are estimated from the initial  data $\mathcal M$, so that they can be used by a method that infers a dynamical model from which one can extract the coupling functions.

The data $\mathcal M$, are usually represented by amplitude state signals measured dynamically i.e.\ they describe the time evolution of the system.
Very often the raw measurements require pre-processing and pre-estimation procedures. 
If the systems are of an oscillatory nature, the phase of the periodic signal is extracted; similarly the amplitude can be extracted from the signals. 
There can be further pre-processing, including filtering within desired intervals, removal of artifacts, noise suppression, removal of common source disturbances, filtering of power supply frequency, etc. 
The pre-processed signals then act as input for the inference methods.

The inference process aims to reconstruct a model to describe the interacting dynamical systems. It is given by a set of ordinary differential equations (ODEs) or, if there is dynamical noise, by stochastic differential equations (SDEs). The model in Fig.\ \ref{fig:infr} is given in terms of general variable $\chi$, while usually the model uses either the phase or amplitude domain. The dynamics of the system is modeled with a set of base functions, which are usually linearly independent. For example, the set of base functions can be a Fourier series of sine and cosine functions. 
Base functions can be either linear or nonlinear, and are specified by a set of parameters $c$ that usually act as scaling parameters. 
When appropriately parameterized by $c$, the base functions then combine to give the all-important coupling function.

The base functions are a part of the model that is assumed to be known beforehand, so \emph{the main task of the inference method is to determine the parameters $\cc$ from the data $\tilde{\mathcal M}$, given the (SDE or ODE) model $\dot \chi$.} The choice of the right model can be rather difficult, especially if the dynamical system does not posses some general characteristics. Nevertheless, a number of methods exist for optimal model selection \cite{Ljung:98,Burnham:02,Berger:96}. Given a model and a set of data, one can use different methods to perform the inference. These methods may differ considerably in their characteristics and performance, and we present below some examples of those that are most widely used.

\subsection{Methods for coupling function reconstruction}\label{sec62:MethodsRecons}
\subsubsection{Modeling by least-squares fitting }\label{sec621:LSF}

As mentioned above, one of the first works on the reconstruction of coupling functions from data was that developed by \citet{Rosenblum:01}. Their inference of the interaction is based on a least-squares fitting procedure applied to the phase dynamics of the interacting oscillators. The main goal of the method is the detection of coupling and directionality. Nevertheless, part of the results are functions that closely resemble the form of coupling functions.

The technique provides for experimental detection of the directionality of weak coupling between two self-sustained oscillators, from bivariate data. The approach makes use of the well-known fact that weak coupling predominantly affects the phases of the oscillators, not their amplitudes. The principal idea is to investigate and quantify whether the phase dynamics of one oscillator is influenced by the phase of the other. To achieve this, the model of the phase equations (Eq.\ \ref{eq:phs}) is fitted to the phase data. From the inferred model and its parameters, one can then quantify the coupling in one or the other direction.

First, for each point in time of the phase time-series, the increments $\Delta_{1,2}(k)=\phi_{1,2}(t_k+\tau)-\phi_{1,2}(t_k)$ are computed, where $\tau$ is a free parameter. These increments $\Delta_{1,2}(k)$ are considered as being generated by some unknown two-dimensional map
$$\Delta_{1,2}(k)=\mathcal F_{1,2}[\phi_{1,2}(k),\phi_{2,1}(k)].$$
The functions $\mathcal F_{1,2}[\phi_{1,2}(k),\phi_{2,1}(k)]$ are decomposed into Fourier series, and their dependences $\Delta_{1,2}(k)$ on $\phi_1$ and $\phi_2$ are modeled with the least-square fitting procedure. As base functions for the fitting, the Fourier series:
$$
\mathcal F_{1,2}=\sum_{m,l} A_{m,l} e^{im\phi_1+il\phi_2},
$$
with $|l|\leq3$ for $m=0$, $|m|\leq3$ for $l=0$ and $|l|=|m|=1$ were considered.

It is worth pointing that this notion is close to, though not exactly identical to, the dynamical inference of ODEs: the increments $\Delta_{1,2}(k)$ are close to the Euler method for first order differentiation which would have been $\Delta_{Euler;1,2}(k)=[\phi_{1,2}(t_{k+1})-\phi_{1,2}(t_k)]/h$, where $h$ is the sampling (differentiation) step. Therefore, the functions $\mathcal F_{1,2}[\phi_{1,2}(k),\phi_{2,1}(k)]$ are similar to the coupling functions $q_{1,2}(\phi_1,\phi_2)$, i.e.\ they are close to the form of the genuine coupling functions with close relative but not absolute coupling strength. Despite the differences, these were probably the first extracted functions of oscillatory interactions, and they were of great importance for the work that followed.

\begin{figure}
{\caption{The reconstructed functions of the phase interactions. (a) The function  $\mathcal F_{1}(\phi_{1},\phi_{2})$ for the influence of the second on the first oscillator, and (b) the function  $\mathcal F_{2}(\phi_{1},\phi_{2})$ for the influence of the first on the second oscillator. From \citet{Rosenblum:01}. } \label{fig:LSF}}
{\includegraphics[width=0.83\textwidth,angle=0]{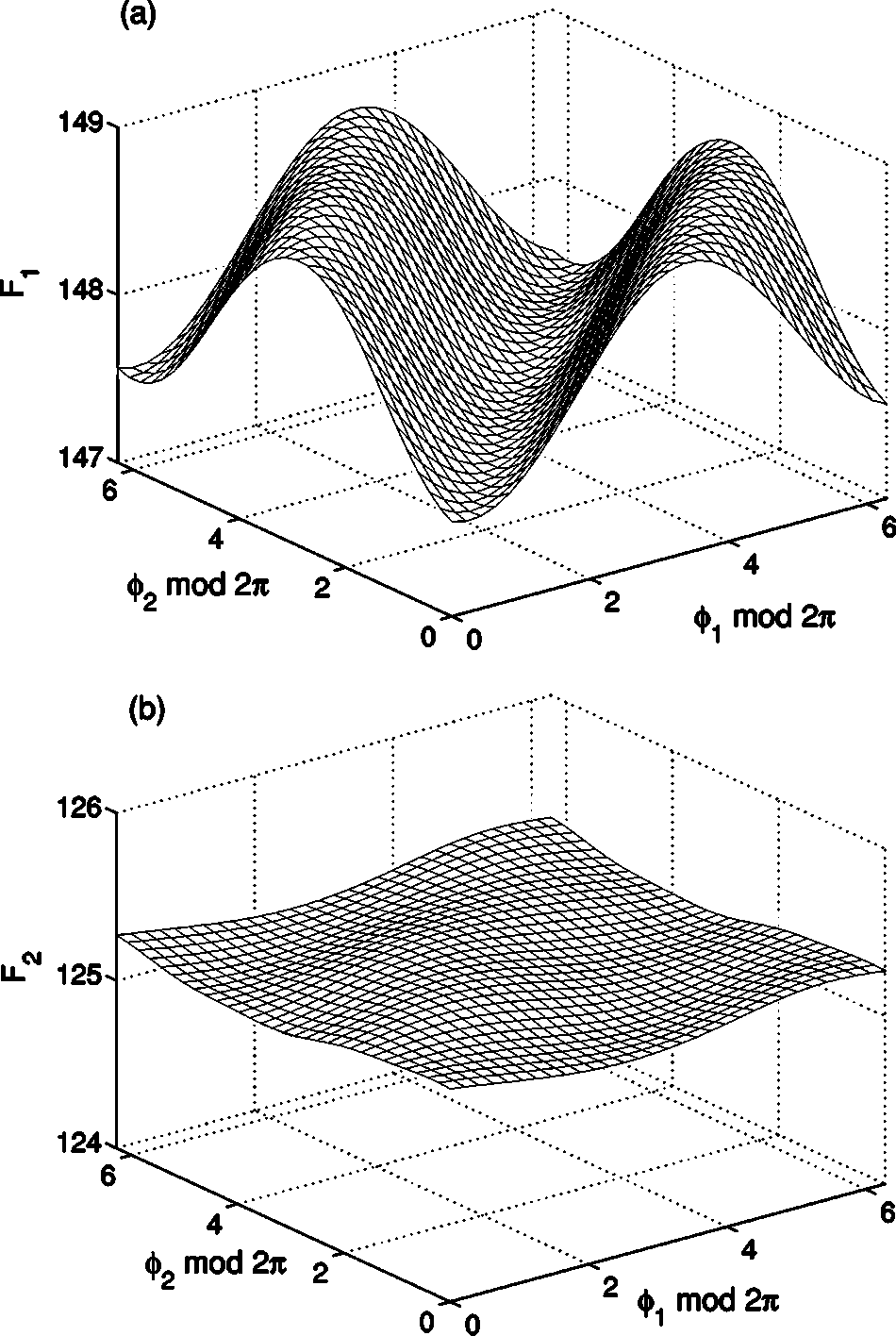}}
\end{figure}

The inference itself was performed by least-squares fitting, a widely-used method for finding the best-fitting curve to a given set of points \cite{Leon:80,Lawson:74}. The main objective of the fitting consists of adjusting the parameters of a model function to best fit a data set. The result of the fitting process is an estimate of parameters, given the model and the base functions. To obtain the parameter estimates, the least-squares method minimizes the summed square of residuals (often also called offsets). The residual $\wp(k)$ for the $k$-th data point is defined as the difference between the observed response value $\Delta(k)$ and the fitted response value $\tilde \Delta(k)$, and is identified as the error associated with the data. The summed square of residuals is then given as:
$$\mathcal O=\sum_k \wp^2(k)=\sum_k [\Delta(k)-\tilde \Delta(k)]^2.$$
The main estimation, aiming to minimize $\mathcal O$, involves partial differentiation with respect to each parameter, and setting the result equal to zero \cite{Leon:80,Lawson:74}. Such schemes can use linear, weighted or nonlinear fitting. More recent methods for coupling function reconstruction with fitting procedures often involve a kernel density estimation \cite{Kralemann:08}. The finally inferred parameters applied to the model base functions, provide explicit definitions of the functions $\mathcal F_{1,2}[\phi_{1,2}(k),\phi_{2,1}(k)]$.

To demonstrate the method,  a simple example of two coupled phase oscillators subject to white noise was considered:
\begin{equation}\label{eq:exmpLSF}
\begin{split} \nonumber
\dot \phi_1 &=\omega_1+\varepsilon_1 \sin(\phi_2-\phi_1)+\xi_1(t)\\
\dot \phi_2 &=\omega_2+\varepsilon_2 \sin(\phi_1-\phi_2)+\xi_2(t),
\end{split}
\end{equation}
where the coupling functions are sines of the phase difference (c.f.\ Kuramoto Eq.\ (\ref{eq:Kuramoto})), with frequencies $\omega_{1,2}=1\pm0.1$, and couplings $ \varepsilon_1=0.1$ and $ \varepsilon_2=0.02$ (i.e.\ weaker than $\varepsilon_1$). The noise is assumed to be white Gaussian with $\langle \xi_{1,2}(t) \xi_{1,2}(t')\rangle = \delta(t-t') 2D_{1,2}$, with $D_{1,2}=0.2$. One should note that the least-squares fitting only infers the deterministic part of the ODEs, and the noise here is used to introduce imprecisions only i.e.\ there is no inference of SDEs. Fig.\ \ref{fig:LSF} presents the two functions as reconstructed using least-squares fitting. Note from Fig.\ \ref{fig:LSF}(a) that the form of the reconstructed function $\mathcal F_{1}$ rightly resembles that of the genuine coupling function $\sin(\phi_2-\phi_1)$, i.e.\ a diagonal form of a wave determined by the phase difference $\phi_2-\phi_1$; and (b) the strength, or amplitude of $\mathcal F_{2}$  is much lower due to the weaker coupling strength.

\subsubsection{ Dynamical Bayesian inference}\label{sec622:DBI}

The recently introduced method for the dynamical Bayesian inference of coupling functions \cite{Stankovski:12b} relies on a Bayesian framework \cite{Toussaint:11,Smelyanskiy:05a,Friston:02,Bayes:63} and is applied to a stochastic differential model where the deterministic part is allowed to be time-varying.

The method attempts to reconstruct the coupling functions by inferring a model consisting of two weakly-interacting dynamical systems subject to noise. The model to be inferred is described by the stochastic differential equation
\begin{equation}
\dot \cci_i = \ff(\cci_i,\cci_j|\cc)+ \sqrt{\DD}\xi_i,
\label{eq:model}
\end{equation}
where $i\neq j={1,2}$, and $f(\cci_i,\cci_j|\cc)$ are base functions describing the deterministic part of the internal and the interacting dynamics. The parameter vector $\cc$ provides scaling coefficients for the base functions. The noise is assumed to be white, Gaussian, and parameterized by a noise diffusion matrix $\DD$. At this point we speak of $\cci_i$ in general, but later we will refer separately to the phase or amplitude coupling functions depending on the domain of the application.

Given the $2 \times M$ time-series ${\mathcal X} = \lbrace {\bf \cci}_{n} \equiv \cci(t_{n}) \rbrace$ ($t_n=nh$) provided, and assuming that the model base functions are known, the main task for dynamical inference is to infer the unknown model parameters and the noise diffusion matrix ${\mathcal P}=\{ \cc,\DD \}$. The problem eventually reduces to maximization of the conditional probability of observing the parameters ${\mathcal P}$, given the data ${\mathcal X}$. For this Bayes' theorem \cite{Bayes:63} is applied, exploiting the \emph{prior} density $p_{\mbox{\scriptsize prior}}(\mathcal P)$ of the parameters and the \emph{likelihood} function $\ell ( \mathcal X | \mathcal P )$ of observing $\mathcal X$ given the choice $\mathcal P$, in order to determine the \emph{posterior} density $p_{{\mathcal X}}({\mathcal P}|{\mathcal X})$ of the unknown parameters
${\mathcal P}$ conditioned on the observations ${\mathcal X}$:
\begin{equation}
p_{{\mathcal X}}(\mathcal P | \mathcal X) = \frac{\ell ( \mathcal X | \mathcal P ) \,
p_{\mbox{\scriptsize prior}}(\mathcal P) }{ \int{\ell ( \mathcal X | \mathcal P
) \, p_{\mbox{\scriptsize prior}}(\mathcal P) d \mathcal P}}. \nonumber
\end{equation}
The next task is to determine the likelihood functions in order to infer the final posterior result. From the time-series the midpoint approximation ${\cci}_{n}^{\ast} = (\cci_{n} + \cci_{n+1})/2$ is constructed, followed by the Euler differentiation $\dot \cci_{n}=(\cci_{n+1}-\cci_{n})/h$. Use of the stochastic integral for noise that is white and independent leads to the likelihood function, which is given
by a product over $n$ of the probabilities of observing $\cci_{n+1}$ at each
time \cite{Smelyanskiy:05a}. The negative log-likelihood function is then $\mathcal S=-\ln \ell({\mathcal X}|{\mathcal
P})$ given as:
\begin{equation}
\begin{split}
    \mathcal S &=   \frac{N}{2}\ln |{\DD}| + \frac{h}{2}\, \sum_{n=0}^{N-1}\Big(
     \cc_k \frac{\partial \ff_{k}(\cci_{\cdot,n}) }{\partial \cci}+\\
     &+ [\dot{\cci}_{n} - \cc_k {\ff}_{k}({\cci}_{\cdot,n}^{\ast})]^T {({\DD}^{-1})}  [\dot{\cci}_{n} - \cc_k {\ff}_{k}({\cci}_{\cdot,n}^{\ast})] \Big ),
\end{split}
    \label{eq:likelihood}
\end{equation}
with implicit summation over the repeated index $k$. The likelihood (\ref{eq:likelihood}) is of quadratic form. Thus if the prior is a multivariate normal distribution, so also will be the posterior. Given such a distribution as a prior for the parameters $\cc$, with mean $\bar {\cc}$, and covariance matrix ${ {\bf \Sigma_{\mbox{\scriptsize prior}} \equiv \Xi}^{-1}}_{\mbox{\scriptsize prior}}$, the final stationary point of $\mathcal S$ is calculated recursively from:
\begin{equation}
\begin{split}
    \label{eq:cD}
     \DD  &= \frac{h}{N} \left(
 \dot{\cci}_{n} - \cc_k {\ff}_{k}({\cci}_{\cdot,n}^{\ast}) \right)^T \left(\dot{\cci}_{n} - \cc_k {\ff}_{k}({\cci}_{\cdot,n}^{\ast}) \right) , \\
     \cc_k &= ({\bf \Xi}^{-1})_{kw} \,  {\bf u}_{w} , \\
    {\bf u}_{w}  & = ({\bf \Xi}_\text{prior})_{kw} \,  {\cc}_{w} +
      h \, {\ff}_{k}({\cci}_{\cdot,n}^{\ast}) \,
(\DD^{-1}) \, \dot{{\cci}}_{n} +\\
 &- \frac{h}{2} \frac{\partial \ff_{k}(\cci_{\cdot,n}) }{\partial \cci}, \\
{\bf \Xi}_{kw}  &= ({{\bf \Xi}_{\text{prior}}})_{kw}   + h \, {\ff}_{k}({\cci}_{\cdot,n}^{\ast}) \,
{(\DD^{-1})} \,   {\ff}_{w}({\cci}_{\cdot,n}^{\ast}),
\end{split}
\end{equation}
where summation over $n=1,\ldots,N$ is assumed and the summation over repeated indices $k$, and $w$ is again implicit. The initial prior can be set to be the non-informative flat normal distribution, ${{\bf\Xi}}_{\text{prior}}=0$ and $\bar \cc_{\mbox{\scriptsize prior}}=0$. These four equations (\ref{eq:cD}) are the only ones needed for implementing the method.
They are applied to a single block of data ${\mathcal X}$ and the resultant posterior multivariate probability ${\mathcal N}_{\mathcal X}(c|\bar{c},\Xi)$ explicitly defines the probability density of each parameter set of the model (\ref{eq:model}).

In dynamical Bayesian inference each new prior distribution depends on and uses the previously inferred posterior distribution. In this framework, however, the information propagation is amended in order to allow the method to follow the time-variability of the parameters \cite{Stankovski:12b}. The new prior covariance matrix becomes $\Sigma_{\text{prior}}^{n+1} = \Sigma_{\text{post}}^n + \Sigma_{\text{diff}}^n$, where $\Sigma_{\text{diff}}^n$ describes how much some part of the dynamics can change with time.

\red{Given the use of Bayesian inference with informative priors, the method is not prone to the overfitting of parameters, and it does not require much data within the windows because, in each new block of data, it only updates the parameters \cite{Duggento:12}. For analyses of dynamical oscillators, one can use data windows containing 6 to 10 cycles of the slowest oscillation; care is needed to ensure that the windows are long enough in cases where there is modulation that is slow relative to the eigenfrequencies \cite{Clemson:16}. The confidence of the fit is given by the resultant covariance matrix $\Sigma_{\text{post}}$.}

The description above is for two interacting oscillators. Nonetheless, the theory also holds for a larger number of oscillators and the dynamical Bayesian inference has been generalized to infer networks of systems with multivariate coupling functions  \cite{Stankovski:15a}.

\begin{figure}[t!]
{\caption{(color online). Application of dynamical Bayesian inference to cardiorespiratory interactions when the (paced) respiration is time-varying. (a) The inferred time-varying respiration frequency. (b) The coupling directionality between the heart and respiration (on this figure denoted as h and r, respectively). (c),(d),(e) The cardiorespiratory coupling function evaluated for the three time-windows whose positions are indicated by the gray arrows. From \citet{Stankovski:12b}. } \label{fig:DBI}}
{\includegraphics[width=0.99\textwidth,angle=0]{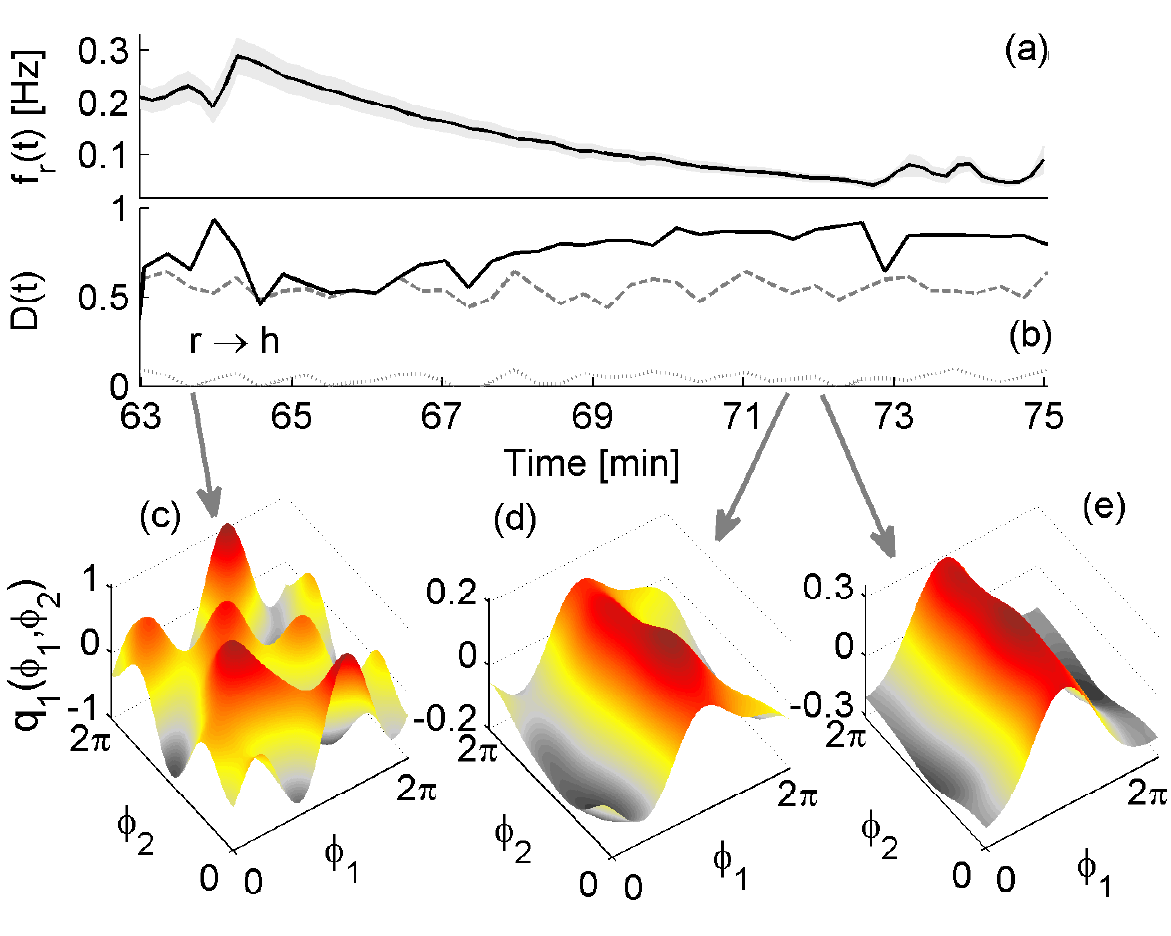}}
\end{figure}

Fig.\ \ref{fig:DBI} shows an application of dynamical Bayesian inference to cardiorespiratory interactions from a resting human subject whose paced respiration was ramped down with decreasing frequency. The inference of the dynamics and the coupling functions were reconstructed from the cardiorespiratory phase dynamics \cite{Stankovski:12b}. Fig.\ \ref{fig:DBI}(a) indicates the reconstructed respiration frequency, showing the linearly-decreasing trend. The inferred coupling directionality, defined as $Dirc=(\varepsilon_r-\varepsilon_h)/(\varepsilon_r+\varepsilon_h)$, is also time-varying, with predominant direction of influence from the respiration to the heart.
The reconstructed cardiorespiratory coupling functions Fig.\ \ref{fig:DBI}(c)-(e), are described by complex functions whose form changes qualitatively over time -- cf.\ Fig.\ \ref{fig:DBI}(c) with Fig.\ \ref{fig:DBI}(d),(e). This implies that, in contrast to many systems with time-invariant coupling functions, the functional relations for the interactions of an open (biological) system can themselves be time-varying processes. By analyzing consecutive time windows, one can follow the time evolution of the coupling functions. 

The time-variability of biological systems and the ability of the method to reconstruct it has implications for the detection of chronotaxic systems, which are a class of nonautonomous self-sustained oscillators able to generate time-varying complex dynamics \cite{Suprunenko:13}. Such systems have drive-response subsystems which are inherently connected with appropriate coupling functions. Dynamical Bayesian inference has been applied to reconstruct such chronotaxic systems for cases where the model was known or could be closely approximated \cite{Clemson:14}.

\subsubsection{ Maximum likelihood estimation: multiple-shooting}\label{sec623:MLE}

The reconstruction of coupling functions has been performed using techniques for maximum likelihood estimation, an approach that was employed for reconstruction of the coupling functions of electrochemical interactions \cite{Tokuda:07,Tokuda:13}. For this a multiple-shooting method, as one type of maximum likelihood estimation, was used.

The maximum likelihood estimation \cite{Aldrich:97,Myung:03} is a statistical method of seeking that probability distribution which makes the observed data most likely, which means that one needs to find the value of the parameter vector that maximizes the likelihood function. The procedure of maximization, intuitively describes the ``agreement'' of the selected model with the observed data, and for discrete random variables it maximizes the probability of the observed data under the resulting distribution. Maximum likelihood estimation gives a unified approach to estimation, which is well-defined in the case of the normal distribution and many other problems. It is of fundamental importance in the theory of inference and provides the basis for many inferential techniques in statistics.

Maximum likelihood estimation is in general different from least-squares fitting (Sec.\ \ref{sec621:LSF}), as the former seeks the most likely parameters, while the latter is a descriptive tool that seeks the parameters that provide the most accurate description of the data. There is a situation, however, in which the two methods intersect and the same parameters are inferred. This is when observations are independent of one another and are normally distributed with a constant variance \cite{Myung:03}.

The multiple-shooting method used for coupling function reconstruction is based on maximum likelihood estimation \cite{Baake:92,Tokuda:13,Voss:04}. A known approach for inference of the trajectories and the parameters is the so-called initial value approach, where initial guesses for the states $x(t_1)$ and parameters $\cc$ are chosen and the dynamical equations are solved numerically. However, a problem can appear in such approaches -- the inferred trajectory may converge only to a local maximum.

The multiple shooting algorithm provides a possible solution to the problem. In this approach, initial conditions are estimated at several points along the time series, so that the shooting nodes, and thus the estimated trajectory, can be made to stay closer to the true values for a longer time. This task is considered as a multi-point boundary value problem. The  interval for fitting $(t_1, t_N)$ is partitioned into $m$ subintervals:
$$t_1=\tau_1<\tau_2<\ldots<\tau_{m+1}=t_N.$$

Local initial values $x_j = x(\tau_j)$ are introduced as additional parameters for each subinterval $(\tau_j, \tau_{j+1})$. In the case of independent Gaussian noise, maximization of the likelihood amounts to minimization of the cost function $\zeta^2(x_1,x_2 \ldots, x_m,\cc)$, which is the sum of the squared residuals between the data and the model trajectory, weighted by the inverse variance of the noise:
$$ \zeta^2(x_1, \ldots, x_m,\cc)=\sum_{i=1}^N \frac{(y_i-G(x_i(x_1, \ldots, x_m,\cc),\cc))^2}{\sigma_i},$$
where $x$ and $y$ are the state and the observed data, respectively, $G$ is a function for the dynamics, and $\sigma_i$ gives the noise variance. Thus in the multiple-shooting method the dynamical equations are integrated piecewise and the cost function is evaluated and minimized on the multiple samples from each subinterval.

Assuming that the dynamical parameters $\cc$ are constant over the entire interval, the local initial values are optimized separately in each subinterval. The latter leads to an initially discontinuous trajectory and the final step is to linearize them so as to provide continuous states. This task, called condensation, is often achieved by use of the generalized Gauss-Newton method.

\begin{figure}[t!]
\begin{center}
{\caption{  (a) The estimated natural frequencies (vertical axis) of 32 electrochemical oscillators versus their measured natural frequencies (horizontal axis). (b) The coupling function estimated by the multiple-shooting method (dotted line), compared with that estimated by application of the perturbation to a single isolated electrochemical oscillator (full curves). From \citet{Tokuda:07}. } \label{fig:MLE}}
{\includegraphics[width=0.97\textwidth,angle=0]{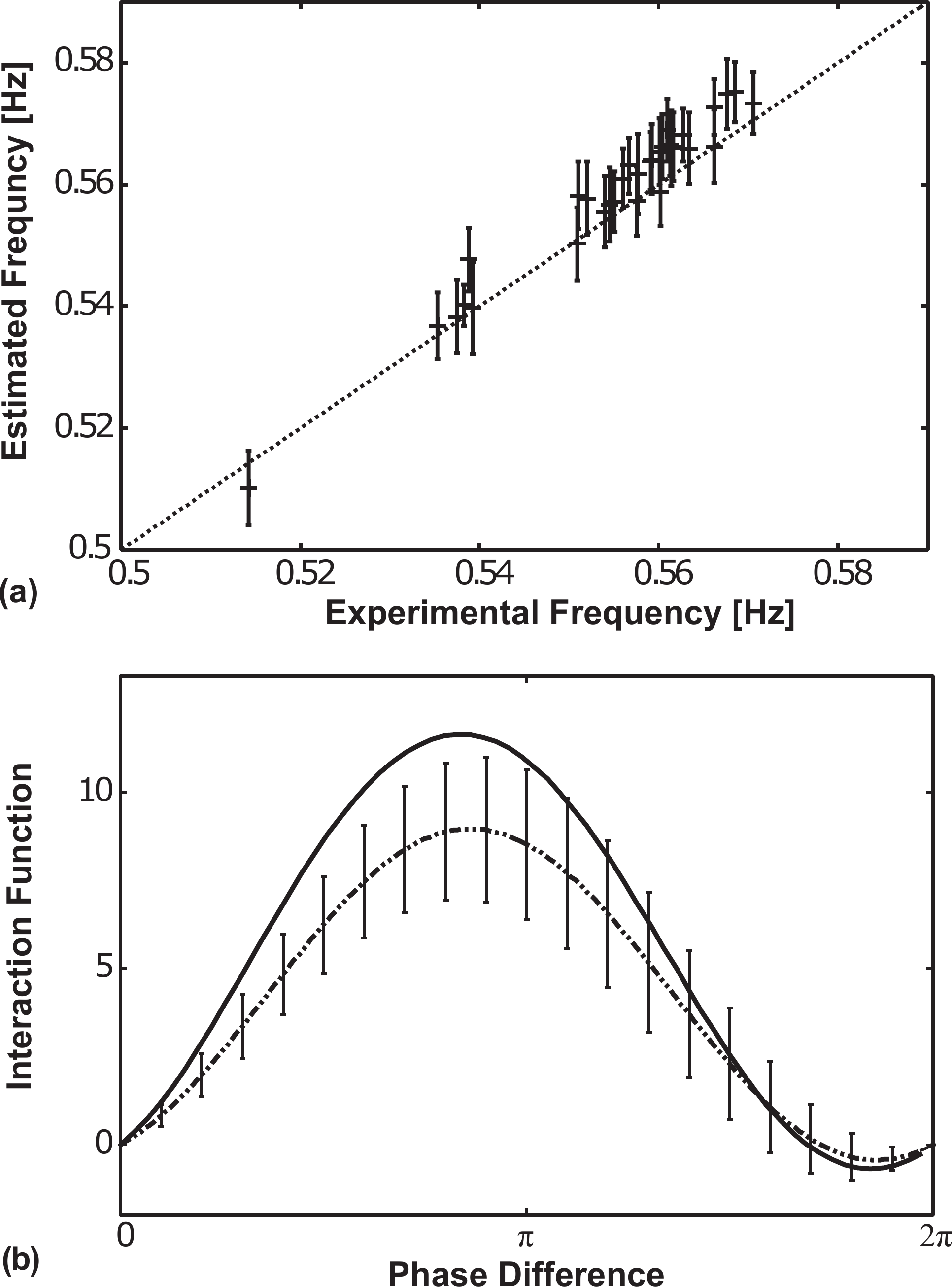}}
\end{center}
\end{figure}

The multiple-shooting method has been used to model the phase dynamics of interacting electrochemical oscillators in order to reconstruct their coupling functions \cite{Tokuda:07}, using Fourier series as base functions. The particular application used an electrochemical oscillatory system in which the coupling function had previously been calculated \cite{Kiss:05} (see also Sec.\ \ref{sec35:SyncPrediction}) by applying the perturbation method to a single oscillator, and thus a direct comparison could be made between the two approaches.

Fig.\ \ref{fig:MLE} shows the inference of a network of 32 electrochemical oscillators. The effective natural frequencies are well estimated, with slightly higher values than those obtained with for completely uncoupled systems Fig.\ \ref{fig:MLE}(a). The form of the estimated coupling function is in reasonable agreement with that found by applying the perturbation method to a single isolated electrochemical oscillator Fig.\ \ref{fig:MLE}(b), with a difference in amplitude of 23.7\% between the two. The coupling function is said to be of a form consistent with theoretical predictions for Stuart-Landau oscillators close to a Hopf bifurcation. In a similar way, the technical dependences and conditions, including dependence on the observational noise, network size, number of defects, and data length have also been examined \cite{Tokuda:13}.

\subsubsection{Random phase resetting method}\label{sec624:resettingMethod}

The method of random phase resetting can be used for the dynamical inference of interacting systems, and also for the reconstruction of their coupling functions \cite{Levnajic:11}. Initially, the method was designed for the reconstruction of network topology i.e.\ the inference of coupling strengths; nevertheless, the framework employed dynamical inference and the inferred model allows for the coupling functions to be reconstructed as well.

The main idea relies on repeatedly reinitializing the network dynamics (e.g., by performing random phase resets), in order to produce an ensemble of the initial dynamical data. The quantities obtained by averaging this ensemble reveal the desired details of the network structure and the coupling functions.

\begin{figure}[t!]
{\caption{ Inference of interacting dynamics by application of the random phase-resetting method. (a) Four-node network of interacting phase oscillators. (b) Reconstruction of the four-node network. Circles are the actual parameter values; crosses are the inferred values; left $a^{(1)}_{ij}$, right $b^{(1)}_{ij}$, for each pair $i\rightarrow j$. (c) Coupling function in respect of the phase difference $\psi_{42}=\phi_2-\phi_4$ from the reconstructed parameters $a^{(1)}_{42}$ and $b^{(1)}_{42}$. From \citet{Levnajic:11}. } \label{fig:Reset}}
{\includegraphics[width=0.98\textwidth,angle=0]{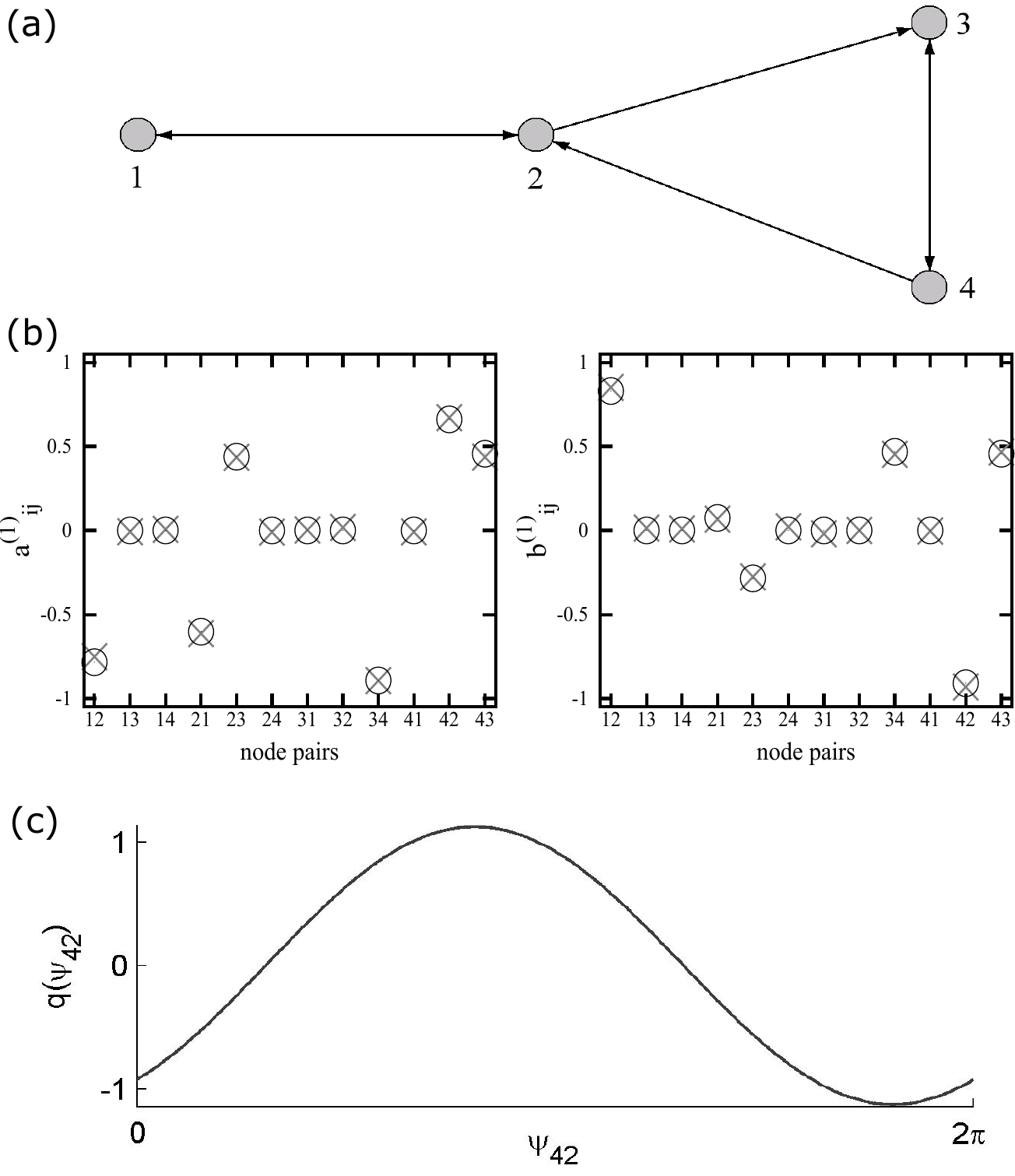}}
\end{figure}

The method considers a dynamical model of interacting phase oscillators such as that in Eq.\ \ref{eq:phs}, with coupling functions in terms of the phase difference $q_i(\phi_j-\phi_i)$.  The dynamics starts from a set of initial phases  which are denoted as $\boldsymbol{\phi}=(\phi_1,\ldots,\phi_N)(t=0)$, chosen from a distribution $\hbar(\boldsymbol{\phi}) > 0$ normalized to $(2\pi)^N$. The method is based on two assumptions: (i) that one is able to arbitrarily reinitialize the network dynamics $\mathcal K$ times, by independently resetting the phases of all nodes to a new state $\boldsymbol{\phi}$, and (ii) that one is able to measure all the values $\boldsymbol{ \phi}_l$, and all initial instantaneous frequencies $ \dot{\boldsymbol{ \phi}}_l$, each time the dynamics is reinitialized (for $l=1,\ldots,\mathcal K$). A $2\pi$-periodic test function $\mathcal T=\mathcal T(\phi_i-\phi_j)$ with zero mean is given as input and the coupling functions are taken to be represented by Fourier series, to obtain an expression for the index $\Upsilon_{ij}$:

\begin{equation}\label{eq:reset1}
\begin{split}\nonumber
\Upsilon_{ij}[\mathcal T]=& (2\pi)^{-N}\sum_{k=1}^N\sum_{n=1}^\infty \int_{[0,2\pi]^N}d \boldsymbol{\phi} \mathcal T(\phi_i-\phi_j)\\
&\times[a_{kj}^{(n)}\sin(n\phi_k-n\phi_j)+b_{kj}^{(n)}\cos(n\phi_k-n\phi_j)].
\end{split}
\end{equation}

The dynamical network described by the phase equations can be reconstructed by computing $\Upsilon_{ij}$ for a suitable $ \mathcal T$ function, e.g.\ $ \mathcal T(\phi)=e^{i n \phi}$. The practical implementation in terms of the data involves the representation of $ \dot{\boldsymbol{ \phi}}_l$ with a kernel smoother \cite{Wand:94,Kralemann:08}, and appropriate averaging, to get $\Upsilon_{ij}[\mathcal T]$:
$$\Upsilon_{ij}[\mathcal T]=\Bigg \langle \frac{\dot \phi_j \mathcal T(\phi_i-\phi_j)}{\hbar(\boldsymbol{ \phi})} \Bigg \rangle=\frac{1}{\mathcal K}\sum_{l=1}^{\mathcal K}\frac{\dot \phi_j (\boldsymbol{ \phi}_l) \mathcal T(\phi_i-\phi_j)}{\hbar(\boldsymbol{ \phi}_l)}.$$

To mimic an experimentally feasible situation, $\mathcal K$ random phase resets of the
network dynamics separated by the time interval $\tau$ are performed. A network of four phase oscillators, coupled as shown in Fig.\ \ref{fig:Reset}(a), is considered. \red{After applying the inference method, the reconstructed pairwise connections are shown to be in reasonable agreement with the actual coupled values, and especially in identifying the non-couplings, as shown in} Fig.\ \ref{fig:Reset}(b). Once the parameters of the phase model $a^{(1)}_{ij}$ and $b^{(1)}_{ij}$ have been inferred, one can also reconstruct the form of the coupling function. Fig.\ \ref{fig:Reset}(c) presents an example of the coupling function showing the influence that the fourth oscillator is exerting on the second oscillator.

The approach is related to the methods for reconstruction of phase response curves (see Sec.\ \ref{sec42:PRC}). Here, however, phase resetting is used somewhat differently i.e.\ the focus is on the network's internal interactions, rather than on its response to stimuli. 
The power of this method lies in a framework that yields both the topology and the coupling functions. Its downside is that it is \textit{invasive} -- requiring one to interfere with the on-going system dynamics (via phase-resets or otherwise), which is often experimentally difficult and sometimes not even feasible.

\subsubsection{Stochastic modeling of effective coupling functions}\label{sec625:StochMethod}

An important feature of the interacting dynamics in real systems is the presence of noise. Explicit consideration of the stochastic nature of the dynamics can provide a better means of describing the coupling functions, and how they are affected. In Sec.\ \ref{sec622:DBI} we discussed the dynamical Bayesian method which treats stochastic dynamics, and is able to infer the deterministic part of the coupling function separately from the random noise. Often when considering noise-induced oscillations, however, one may wish to determine the effective coupling functions including the effect of noise \cite{Schwabedal:10}. When performed on the effective phase dynamics with an invariant phase defined in a nonperturbative way, the  phase will depend on the noise intensity, and so will all the corresponding characteristics such as the coupling function.

The authors consider an effective phase model describing periodically-driven, noise-induced oscillations:
\begin{equation}\label{eq:effect_model}
\begin{split}
\dot \theta=h(\theta)+g(\theta)\xi(t)+f(\vartheta(t),\theta),
\end{split}
\end{equation}
where $\vartheta=\Omega t$ is a $2\pi$-periodic driving phase. The aim is to describe the effective phase dynamics $\mathcal H(\theta, \vartheta)$ and the corresponding effective coupling functions. One can express the effective dynamics (\ref{eq:effect_model}) as: $\dot \theta=\mathcal H(\theta,\vartheta)=\mathcal H_m(\theta)+\mathcal F(\theta, \vartheta)$, where $\mathcal H_m(\theta)$ is $\vartheta$-independent marginal effective velocity and $\mathcal F(\theta, \vartheta)$ is the effective coupling function. By integrating Eq.\ (\ref{eq:effect_model}) over $\vartheta$, the marginal effective velocity $\mathcal H_m(\theta)$ can be determined. Hence, using $\mathcal F(\theta, \vartheta)=\mathcal H(\theta,\vartheta)-\mathcal H_m(\theta)$, one can determine the effective coupling function:
$$
\mathcal F(\theta, \vartheta)=f-\int_0^{2\pi} f\frac{P}{P_m}d\vartheta - g^2\partial_\theta \ln\frac{P}{P_m},
$$
where $P=P(\theta, \vartheta)$ and $P_m=P_M(\theta)$ are the probability densities of the full and the marginal dynamics, respectively. The variable $\theta$ can be considered as a protophase and can be further transformed by $\phi=\mathcal C(\theta)=2\pi\int_0^\theta P(\eta)d\eta$ to yield an invariant effective phase dynamics:
\begin{equation}\label{eq:effect_model2}
\dot \phi=\omega+2\pi P_m [\mathcal C^{-1}(\phi)]\mathcal F[\vartheta,\mathcal C^{-1}(\phi)]=\omega+q(\vartheta,\phi).
\end{equation}
Equation (\ref{eq:effect_model2}) provides the effective phase dynamics of the periodically-driven noise-induced oscillations with an effective coupling function $q$ that depends on the noise intensity.

This theoretical description can be illustrated on a noise-driven FitzHugh-Nagumo model as an example of an excitable system:
\begin{equation}\label{eq:FGH}
\begin{split}\nonumber
\epsilon \dot x &= x-\frac{x^3}{3}-y,\\
\dot y &=x+a+\sigma \xi(t) +b\cos(\Omega t),
\end{split}
\end{equation}
where $a=1.1$, $\epsilon=0.05$ are the parameters of the system, $\xi(t)$ is an additive noise which induces oscillations, and the cosine external function provides the interactions in the system. After estimation of the protophase time-series $\theta$ and its transformation to the phase $\phi$, the effective coupling function $q(\Omega t,\phi)$ can be determined. For this a double Fourier series decomposition was used with least-squares fitting of the model to the data. In this sense, the core of the inference is the same as the least-squares fit discussed in Sec.\ \ref{sec621:LSF}, even though the difference here is that one reconstructs a stochastic model.

\begin{figure}[t!]
{\caption{(color online). Coupling functions for noise-induced oscillations in the FitzHugh-Nagumo model with $b= 0.1$ and two different values of the noise intensity $D$: for $D= 0.08$ (left panel), the mean frequency is $\omega\approx0.62$); and for $D= 0.11$ (right panel), $\omega\approx0.95$. From \citet{Schwabedal:10}. } \label{fig:Stochs}}
{\includegraphics[width=0.98\textwidth,angle=0]{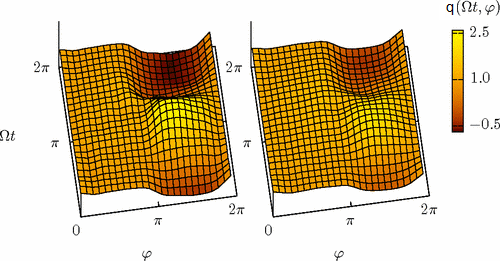}}
\end{figure}

The results of the analysis indicated an increase in the effective coupling for vanishing noise, and masking of the coupling for driven noise-induced oscillations of the FitzHugh-Nagumo model. Fig.\ \ref{fig:Stochs} presents an unusual case with implications for the  interpretation of effective coupling functions. Namely, the effective coupling function was computed with two noise intensities for the same coupling strength. By comparing the two plots in Fig.\ \ref{fig:Stochs} one can see that the amplitude of $q$ decreases with increasing noise intensity. The change in amplitude may have been related to a more pronounced masking of the coupling induced by the frequency shift, or due to the generic decrease in effective coupling for stronger noise because of flattening of the marginal probability.

\subsubsection{ Comparison and overview of the methods}\label{sec626:CompareMethods}

The methods discussed for reconstruction of coupling functions possess some characteristics that are in common, as well other features that differentiate them. The latter eventually lead to different choices of method for use, depending on the circumstances and conditions for the dynamics and the coupling functions to be inferred.

Table \ref{tab:methods} summarizes the difference and performance of the methods discussed in the previous sections \ref{sec621:LSF}-\ref{sec625:StochMethod}. Inference of stochastic dynamics which treated SDEs and the influence of dynamical noise that can cause noise-induced qualitative changes (e.g.\ phase slips), can be performed with dynamical Bayesian inference and stochastic modelling of the effective coupling functions. The other methods treat ODEs with possible measurement noise that can affect the statistics and precision of the inference.

Often the model for inference has more parameters than the real system. In such cases \emph{overfitting of parameters} can occur and some methods can infer random error or noise instead of the underlying dynamical relationship. A model that has been overfitted will generally be a poor representation of the real system, as it can exaggerate minor fluctuations in the data. Bayesian inference uses informative prior probabilities and can avoid the problem of overfitting parameters.

The speed of calculation varies between methods, especially as some methods perform additional steps and therefore take longer. Dynamical Bayesian inference has a recursive loop, evaluated for each time-point within a data window until a certain precision is reached; the multiple shooting method requires additional initial conditions, the shooting nodes, to be estimated at several points along the time series; the random phase resetting method uses a large number of additional random initial resetting points. These additional processing steps relative to the initial handling of the time-series, inevitably require more computing power and thus reduce the calculation speed.

The coupling functions are usually evaluated for a sequence of time-series each defined by a certain window-length whose choice determines how well a method is able to follow the time-evolution of the coupling functions. Dynamical Bayesian inference updates the new probabilities within a window of data, based on prior knowledge; the multiple-shooting method exploits the initial shooting nodes; and the random-resetting method also uses resetting points, which are said to require shorter data windows and, in turn, to provide good time-resolution of the inference.

\begin{table}[]
\centering
\caption{Comparison of methods for the inference of coupling functions in terms of four characteristics (columns) including, respectively: stochastic treatment; absence of parameter overfitting; calculation speed; and the size of data windows. The methods (rows) are as described in the previous sections: least square fitting (LSF); dynamical Bayesian inference (DBI); maximum likelihood estimation with multiple shooting (MLE-MS); random phase resetting (RPR); and the stochastic modelling of effective coupling functions (SMECF). The symbol $\checkmark$ indicates if a method possesses that characteristic, and a $\times$ if not.    }
\label{tab:methods}
\begin{tabular}{l|c|c|c|c}
  \hline
$\text{ }$   & Stochastic & No-overfitting & Calc.\ speed & Data Size \\
  \hline \hline
LSF    & $\times$   & $\times$       & $\checkmark$          & $\times$      \\
DBI    & $\checkmark$   & $\checkmark$       & $\times$          & $\checkmark$      \\
MLE-MS & $\times$   & $\times$       & $\times$          & $\checkmark$      \\
RPR    & $\times$   & $\times$       & $\times$          & $\checkmark$      \\
SMECF  & $\checkmark$   & $\times$       & $\checkmark$          & $\times$ \\
\hline
\end{tabular}
\end{table}

A difficulty in common for all the methods is the reconstruction of coupling functions (and coupling strength in general) when the systems are highly synchronized and coherent in the 1:1 frequency ratio \cite{Rosenblum:01,Kiss:05}. Namely, in the 1:1 phase synchronous state there is a definite strong relationship between the phases and the trajectory on a $(\phi_1,\phi_2)$ torus which is one line; hence  $\phi_1$ and $\phi_2$ are not independent, and the coupling functions of the two variables cannot be separately estimated i.e.\ one is not able to separate the effect of interaction from the internal dynamics of autonomous systems. In order to obtain information about the coupling one needs to observe deviations from synchrony, either due to dynamical noise or due to the onset of quasiperiodic dynamics outside the synchronization region. Synchronous states with larger $n$:$m$ frequency ratios are favorable, because many revolutions cover the surface of the torus, and the inference is then more successful.

A similar situation applies for the inference of coupling functions of dynamical systems in amplitude states. In such cases, the systems are multidimensional, e.g.\ two coupled Lorenz chaotic systems, and the inference of the coupling functions is more plausible in a 1:1 generalized synchronization sense \cite{Stankovski:14a}. Complete synchronization and  very strong coupling will again constrict the available space for inference, leading to difficulties in reconstructing the amplitude coupling functions.

\subsection{ Towards coupling function analysis}\label{sec63:CFAnalysis}

Often one needs to draw conclusions based on comparison and quantitative measures of the coupling functions, after they have been inferred. Such situations occur in experimental studies of some real system interactions, e.g.\ in biomedicine or chemistry. For example, the biomedical expert would like to have a quantitative measure of the coupling functions to be able to describe or compare different states or behaviours in health and disease.

One needs to quantify some characteristics that describe the coupling functions, and in particular features that are unique to the coupling functions and cannot be obtained from other measures. The form of the function can describe the mechanism of the interactions, so being able to quantify it is of obvious interest.

Quantifying a function is not a trivial task, in general. However, many coupling functions can be decomposed, or are inferred through decomposition into functional components, like for example when the phase coupling functions are decomposed into Fourier series. Therefore, the problem of quantification of the coupling function can be reduced to quantification of its components, and in particular the parameters obtained for the components. In this way, one is left to work with a one-dimensional vector of parameters.

\begin{figure}[t!]
{\caption{(color online). Boxplots illustrating the similarity of cardiorespiratory coupling functions. (a) The correlation coefficient $\rho$ and (b) the difference measure $\eta$, for all available pairs of functions (high similarity corresponds to large $\rho$  and small $\eta$). ES: similarity between the respiration-ECG coupling functions  of the same subject, obtained from two trials. EG: same relation similarity between different subjects in the group demonstrates low interpersonal variability. PS and PG: intra- and interpersonal similarities, respectively, for the respiration-arterial pulse coupling function. EPS and EPG: intra- and interpersonal similarities, respectively, between between the two types of coupling function. From \citet{Kralemann:13b}. } \label{fig:CFsiml}}
{\includegraphics[width=1\textwidth,angle=0]{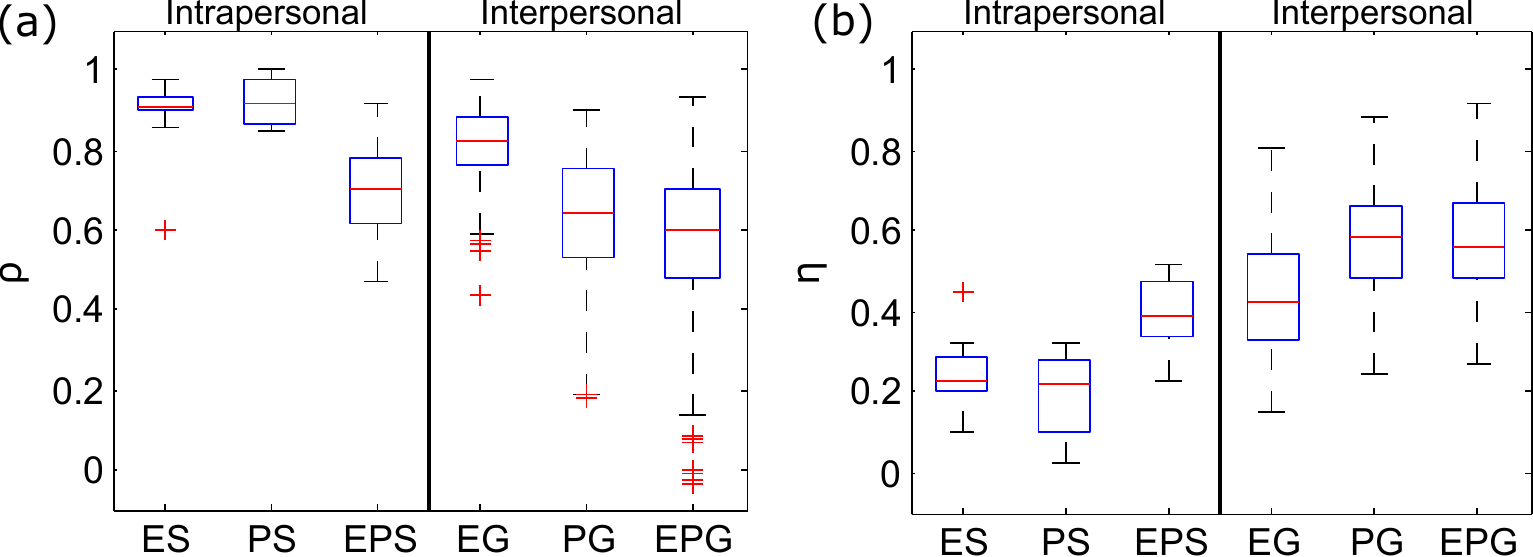}}
\end{figure}

One way to perform such a quantification is through the correlation coefficient and the difference measure evaluated from the inferred  coupling functions \cite{Kralemann:13b}. The first index $\rho$ measures the similarity of two coupling functions $q_1$ and $q_2$, irrespectively of their coupling strengths. It is calculated as the correlation coefficient:
$$\rho=\frac{\langle \tilde q_1 \tilde q_2 \rangle}{||\tilde q_1||\text{ }||\tilde q_2||},$$
where $\langle \circ \rangle$ denotes spatial averaging over the 2D domain $0\leq\phi_1,\phi_2\leq2\pi$, $\tilde q=q-\langle q \rangle$ and $||q||=\langle qq \rangle^{1/2}$. The similarity index $\rho$ is of great interest as it relates only to the form of the function and is a unique measure of the coupling functions. In a similar way, the difference measure is defined as:
$$\eta=\frac{|| \tilde q_1- \tilde q_2 ||}{||\tilde q_1||+||\tilde q_2||}.$$
The difference measure is of less interest than $\rho$ as it relates to the coupling strengths, which can be assessed in different ways through other measures.

Two measures were used to analyze the similarity and difference of cardiorespiratory coupling functions, as shown in Fig.\ \ref{fig:CFsiml}.  It was found  that the functions have a well-pronounced characteristic shape for each of the subjects and that the correlations between the coupling functions obtained in different trials with the same subjects were very high Fig.\ \ref{fig:CFsiml}(a). Naturally, the correlation between the functions of different subjects was lower, reflecting the interpersonal variability; however, it is high enough to demonstrate the high similarity of the interactions in the group of subjects. The similarity of the coupling functions, obtained from different observables such as the ECG and the arterial pulse for the cardiac oscillations, support the validity of the use of invariant phase. The similarity index $\rho$ has been also used for quantifying the form of the brain coupling functions \cite{Stankovski:15a}, quantifying significant differences in the form of the coupling functions when altered by the use of different anaesthetics \cite{Stankovski:16}.

\subsection{ Connections to other methodological concepts}\label{sec6:Other}

\subsubsection{Phase reconstruction procedures}\label{sec41:PhReduction}

When analysing data one needs first to reconstruct the phase, before attempting to detect the underlying phase coupling functions. Various methods exist for phase reconstruction from data, including the marked events method (the marking a particular time event, e.g.\ a maximum or a zero-crossing, within a cycle of oscillation), the Hilbert transform, and wavelet transform based methods \cite{Quiroga:02,Daubechies:11,Iatsenko:14b,Pikovsky:01,Gabor:46}. The effect of the method used can have a direct impact on the form of the reconstructed coupling function. It is therefore important to choose a method to reconstruct a phase that is as genuine as possible. For example, the marked events method reduces the inter-cycle resolution and, despite its limited usefulness in synchronization analysis, it is not appropriate for coupling function analysis. \citet{Kralemann:08} proposed a protophase-to-phase transform that obtains an invariant phase in terms of the genuine, observable-independent phases. This technique can be very useful in checking consistency with the phase estimated by use of the Hilbert transform. Some other phase estimates have also been discussed, noting that use of the synchrosqueezed  wavelet transform means that one does not need explicit protophase-phase preprocessing, as it estimates directly the genuine phase \cite{Daubechies:11,Iatsenko:13a}.  Recently,  \citet{Schwabedal:16} introduced a method that facilitates a phase description of collective, irregular-oscillatory dynamics from multichannel recordings and they demonstrated it on EEG recordings. Such phase estimates have a potential for the reconstruction of coupling functions from collective dynamics. In any case, one should be very careful when estimating phases for coupling functions, in particular from experiments, as otherwise this can lead to spurious descriptions of the coupling functions.

\begin{figure}
{\caption{Experimental estimation of neuronal phase response curves (PRC). a) Raw estimation of the PRC (dots) and smoothing over a $2\pi/3$ interval (gray line) compared with the estimated PRC (black line) from the approach in \cite{Galan:05}. Both curves match, which indicates that the raw data are consistent with a phase model. b) Same as a) but after shuffling the raw data. The PRC is roughly flat and yields inconsistent results with the smoothing, implying that the shuffled data cannot be described by a phase model. From \citet{Galan:05}. }\label{fig:PRC1}}
{\includegraphics[width=0.99\textwidth,angle=0]{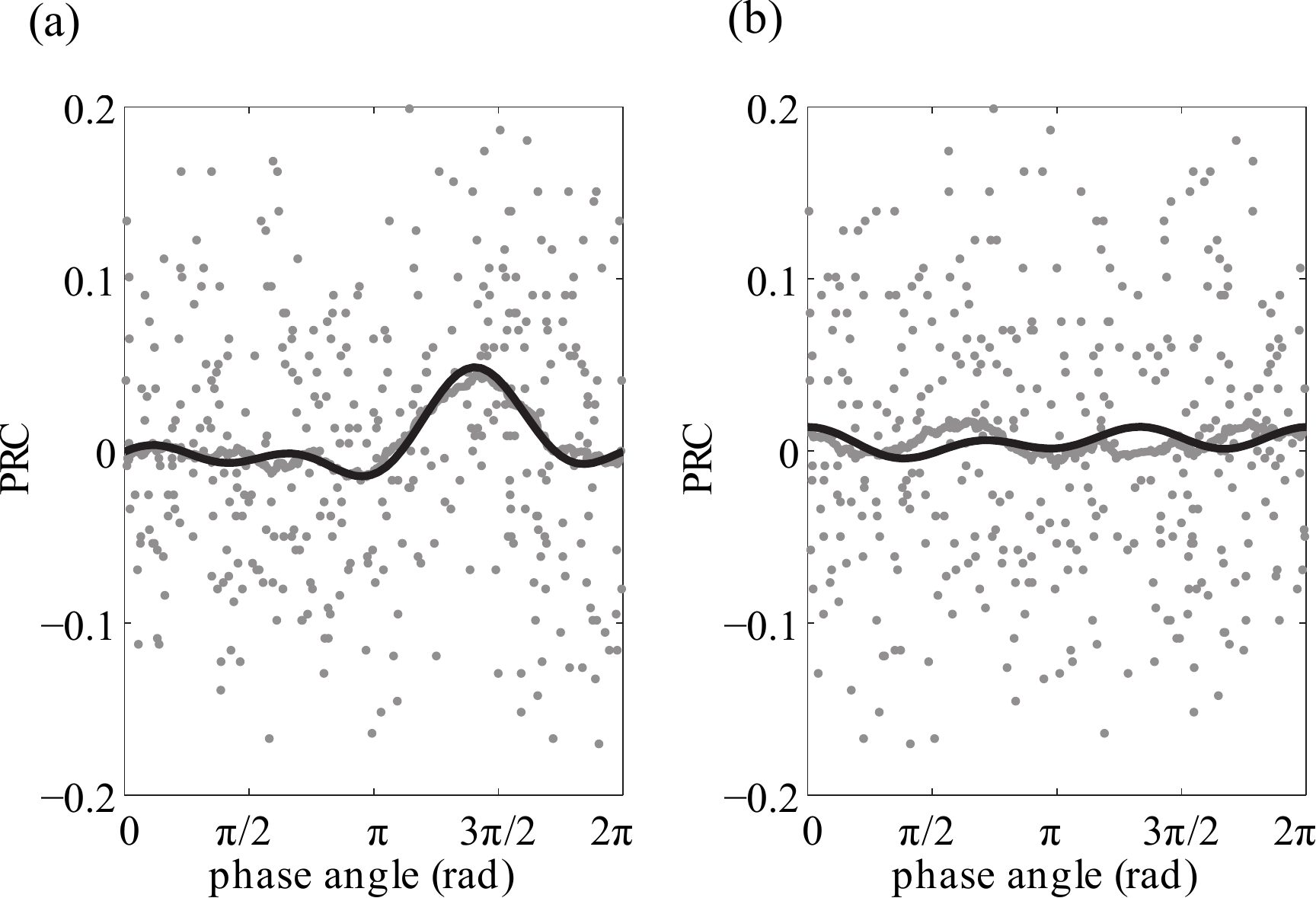}}
\end{figure}

\begin{figure*}
{\caption{(color online). Phase response curve and effective forcing for the cardiorespiratory interactions. (a) Individual PRCs $Z$ and (b) effective forcing $I$ for all ECG-based coupling functions (grey curves). In both \blue{main panels the thick (blue)} lines show the average over all the individual (grey) curves. The \blue{thick (red)} lines are obtained by decomposition of the averaged coupling function. The small panel on the top in (a) shows for comparison the average ECG cycle as a function of its phase.  The small panel on the top in (b) shows the average respiratory cycle as a function of its phase, with the epochs of inspiration and expiration marked (approximately). From \citet{Kralemann:13b}. }\label{fig:PRC2}}
{\includegraphics[width=0.85\textwidth,angle=0]{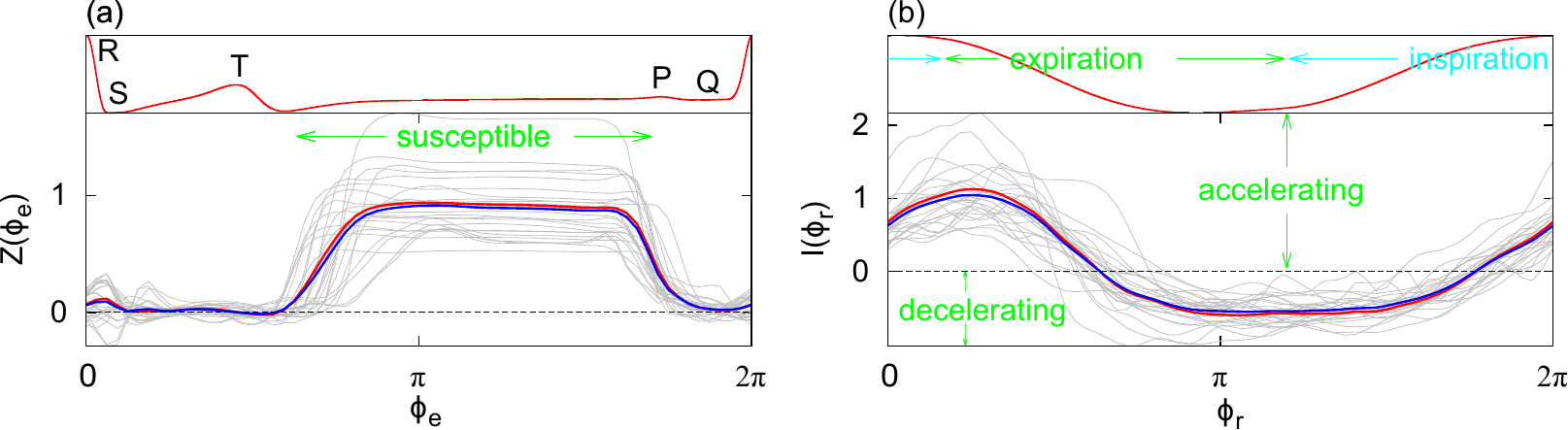}}
\end{figure*}

\subsubsection{Relation to phase response curve in experiments}\label{sec42:PRC}

The phase response curve (PRC), describes how an oscillator responds to an external perturbation \cite{Winfree:80,Kuramoto:84,Tass:99}. The response of the affected oscillator is manifested as shift of its phase. It has been used in various fields, especially in biological oscillations including the heartbeat, circadian rhythms and neuronal activity \cite{Tass:99,Oprisan:04,Ermentrout:96,Czeisler:89,Preyer:05,Ko:09,Hannay:15}.

The phase response curve is a function expressed in terms of one phase variable from the affected oscillator (for detailed theoretical description see Sec.\ \ref{sec:PRC_theory}). In this way, the phase response curve is a similar concept to a coupling function, with that difference that the latter describes the interactions on the whole (two-dimensional) space i.e.\ depending on the two phase variables. In fact, \emph{the phase response curve is a functional component of the coupling function}. In terms of the general theory of phase dynamics \cite{Kuramoto:84, Winfree:80}, the coupling function $q_1(\phi_1,\phi_2)$ can be expressed as the product of two functions:
\begin{equation}\label{equ:prc}
q_1(\phi_1,\phi_2)=Z_1(\phi_1)I_1(\phi_2),
\end{equation}
where $Z_1(\phi_1)$ is the phase response curve, while $I_1(\phi_2)$ is the perturbation function.

The reconstruction of functional curves from data has been used widely, elucidating the mechanisms underlying the oscillations found in nature \cite{Tass:99,Czeisler:89,Batista:12}. This approach is widely used in neuroscience \cite{Schultheiss:11,Tateno:07,Gouwens:10,Ermentrout:96,Galan:05}. For example, the phase response curve has been estimated with electrophysiological experiments on real neurons from the mouse olfactory bulb \cite{Galan:05}. A constant current was injected into the neuron to make it fire at a constant frequency within the beta/gamma frequency band. By following the responses of the neurons to the injected current stimulation, the phase response curve was reconstructed. Fig.\ \ref{fig:PRC1}(a) shows the experimental dots and the fitted phase response curve, which matches well the one from the phase model of the study. The surrogate estimation (shuffled dots) in Fig.\ \ref{fig:PRC1}(b) validates this result.  Thus, the method allowed for a simplification of the complex dynamics from a single neuron to a phase model. This study also demonstrates the relationship to the coupling function, which was reconstructed from the convolution of the phase response curve and the perturbation function -- an approach used in chemistry as well \cite{Kiss:05}.

Going in the opposite direction, the \emph{phase response curve can be estimated by decomposition of the  coupling function }\cite{Kralemann:13b}. This can be done by decomposing the reconstructed coupling function into a product of two functions Eq.\ (\ref{equ:prc}) and searching for a minimum in the decomposition error by means of an iterative scheme. In this way, the interactions coming from the second oscillator are used as the perturbation to the first oscillator under consideration, whence there is no need for additional external stimulation -- a procedure referred by the authors as \emph{in vivo} estimation of the phase response curve.  This method was applied to the reconstruction of the cardiac phase response curve as perturbed by the respiratory oscillations, as shown in Fig.\ \ref{fig:PRC2}. One can clearly see the interval where the phase response curve is non-zero, so that the cardiac system is susceptible to the respiratory perturbation.  Intervals of positive (negative) effective forcing are the intervals where respiration is accelerating (decelerating) the heart rate.

Interesting and relevant parallels could be drawn between coupling functions and amplitude response curves, or phase-amplitude response curves \cite{Huguet:13,Castejon:13}. The latter are similar to phase response curves, with the difference that there is also a response to the amplitude on increasing or decreasing the strength of the oscillations.

\subsubsection{   General effective connectivity modeling}\label{sec43:Effective}

Quite generally, methods of modeling dynamical systems from data often contain coupling functions \cite{Toussaint:11,Smelyanskiy:05a,Friston:11,Voss:04}. The extent to which these coupling functions resemble the same concept as that discussed in this review can vary, depending on the design of the method and the model itself. For example, there can be a model of one larger system which is different from the interaction of two or many systems, but there can be functions within the model that are coupling certain variables or dimensions.

Similar implications hold for the general description of methods for effective connectivity modeling which exploit a model of differential equations and allow for dynamical mechanisms of connectivity to be inferred from data. Such effective connectivity has particularly wide use in neuroscience, where the methods infer the links on different scales of connectivity and spatially distributed regions within the heavily connected brain network.  Although such methods have much in common with coupling function inference methods they do not, however, consider the coupling function as an entity, and nor do they assess or analyze the coupling functions as such.

\section{ Applications and Experiments}\label{sec64:Applc}

In this section we review a number of important applications of the methods for reconstruction of coupling functions and their use for the study and manipulation of the interactions, in various fields.

\subsection{ Chemistry}\label{sec641:MetChem}

The interactions of chemical oscillations have been studied extensively, including in connection with coupling functions \cite{Kiss:05,Kiss:07,Kiss:02,Blaha:11,Tokuda:07,Tokuda:13,Miyazaki:06,Kori:14}. Experiments on chemical, or electrochemical, oscillations  provide a convenient way of studying and manipulating interactions and coupling functions under controlled laboratory conditions.

One of the more prominent coupling function applications to chemical oscillators is for engineering complex dynamical structures \cite{Kiss:07}. The work exploits the simplicity and analytical tractability of phase models and, in particular, their reconstructed coupling functions in order to \emph{design} optimal global, delayed, nonlinear feedback for obtaining and tuning the desired behavior. It uses a feedback design methodology capable of creating a large class of structures describable by phase models for general self-organized rhythmic patterns in weakly interacting systems with small heterogeneities. The electrochemical oscillations were achieved with electrode potentials during the electrodissolution of nickel wires in sulfuric acid.

\begin{figure}[t!]
{\caption{(color online). Engineering a system of four non-identical oscillators using a specific coupling function to generate sequential cluster patterns. (a) The target (solid line, $H(\Delta \phi) = \sin(\Delta \phi – 1.32) – 0.25\sin(2\Delta \phi)$) and optimized coupling function with feedback (dashed line). (b) Theoretical and experimentally observed  heteroclinic orbits and their associated unstable cluster states. (c) Time series of the order parameter ($R1=\sum_{j=1}^N \exp(i\phi_j)$) along with some cluster
configurations. (d),(e) Trajectories in state space during slow switching. The black lines represent calculated heteroclinic connections between cluster states (fixed points). The \blue{(red)} surface in (e) is the set of trajectories traced out by a heterogeneous phase model. $H(\Delta\phi)$ on the plots is equivalent to the $q(\psi)$ notation used in the current review. From \citet{Kiss:07}. } \label{fig:ChemApp1}}
{\includegraphics[width=1\textwidth,angle=0]{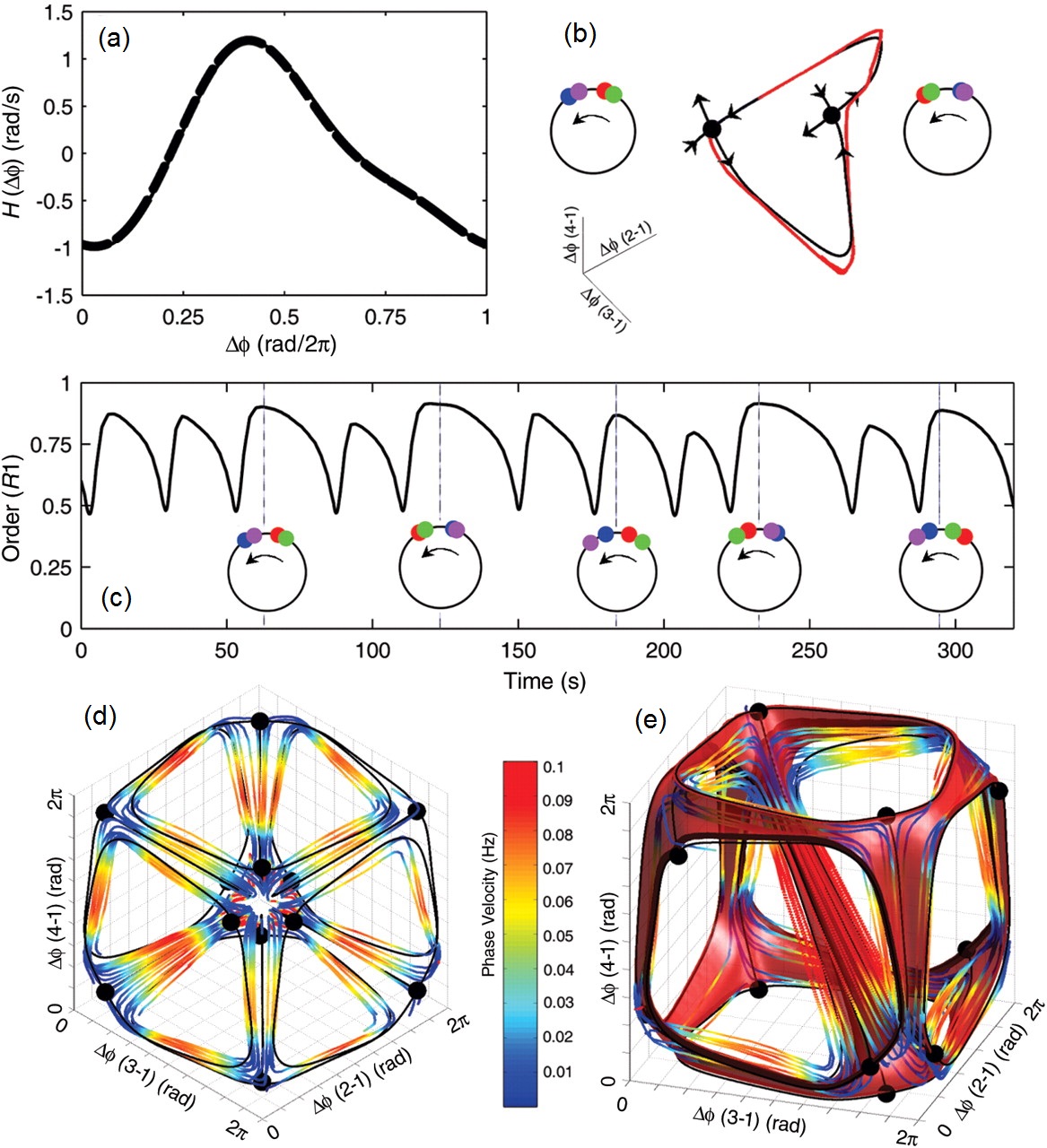}}
\end{figure}

The engineering of the interactions to the desired behavior is achieved in a population of $N$ oscillators through the imposition of nonlinear, time-delayed feedback in the amplitude state. This reduces to a phase model of a population of oscillators with weak, global (all-to-all) coupling described by the Kuramoto model \cite{Kuramoto:84} with a general diffusive coupling function $q(\phi_j-\phi_i)$ (i.e.\ notation $H(\phi_j-\phi_i)$ in this work). In this way, one can also derive the phase response function $Z(\phi_i)$ in connection to the feedback function. Given such a feedback function and response function $Z(\phi_i)$, one can in principle obtain the coupling function $q(\phi_j-\phi_i)$ for use in the phase model. However, in the work discussed, \citet{Kiss:07} proceeded in the opposite manner: they chose a coupling function to produce the desired states, and then designed a feedback loop with optimized feedback parameters to give the desired $q(\phi_j-\phi_i)$.

The method is demonstrated with three interesting and important experiments: (i) tuning the desired arbitrary phase differences between two dissimilar oscillators (see also Fig.\ \ref{fig:app}); (ii) generation of complex patterns that include self-organized switching between unstable dynamical states and clusters; and (iii) the physiologically important problem of desynchronization of oscillators. Below, we devote particular attention to case (ii) involving the generation of sequential states and clusters \cite{Ashwin:05a}.

\begin{figure}
{\caption{The mechanisms of synchronization for Belousov-Zhabotinsky chemical oscillations, as determined by their coupling functions. The full curve shows $Q(\psi)=q(\psi)-q(-\psi)$ estimated from the coupling functions $q(\psi)$. $q(\psi)$ and $q(-\psi)$ are presented as dashed and dot-dashed lines, respectively. Stable and unstable solutions of Eq.\ (\ref{eq:chem_ph}) are shown as solid and open circles, respectively. $\Delta\omega$ on the plots is equivalent to the $\Delta_\omega$ notation used in the present review. From \citet{Miyazaki:06}. } \label{fig:ChemApp2}}
{\includegraphics[width=0.75\textwidth,angle=0]{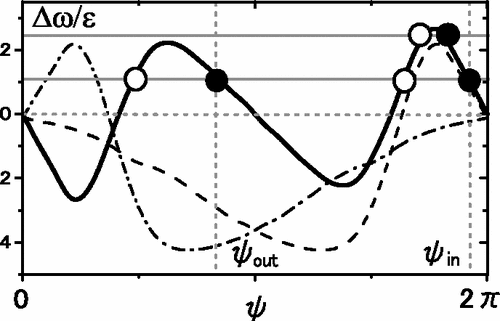}}
\end{figure}

Quadratic feedback to a population of four oscillators is used to reproduce a coupling function proposed for slow switching, Fig.\ \ref{fig:ChemApp1}(a). The experimental system with feedback that sequentially visits (unstable) two-cluster states with two oscillators in each cluster shows two (saddle type) cluster states in state space, Fig.\ \ref{fig:ChemApp1}(b). In agreement with the experiments, the phase model predicted a switch between these states due to the existence of heteroclinic orbits connecting them. These switches can be seen as a fluctuation of the system order, as shown in Fig.\ \ref{fig:ChemApp1} (c).  The engineered feedback produces configurations of two clusters, each containing two elements, connected by heteroclinic orbits. Two types of transitions have been  observed: intra-cluster and inter-cluster transitions as presented by the trajectories of the experimental system, and illustrated as phase space plots in Fig.\ \ref{fig:ChemApp1}(d) and (e).

In a similar way, \citet{Kiss:05} developed a method for reconstruction of coupling functions from electrochemical oscillations, which are then used to predict synchronization, as also discussed in Sec.\ \ref{sec35:SyncPrediction}. Similarly, \citet{Tokuda:07,Tokuda:13} used a different technique for inference of the coupling function of multivariate electrochemical oscillations: see Sec.\ \ref{sec623:MLE}. Also, to capture the whole nature of the interaction of electrochemical oscillations (and not only the synchronization-related ones) the coupling functions were reconstructed in the full two-dimensional ($\phi_1,\phi_2$) space i.e.\ not only for the one-dimensional diffusive coupling difference $\Delta \phi=\phi_2-\phi_1$ \cite{Blaha:11}.

Of particular interest is a coupling function method that \citet{Miyazaki:06} applied for studying the interactions of Belousov-Zhabotinsky chemical oscillations. This class of reactions serves as a classical example of non-equilibrium thermodynamics, resulting in the establishment of a nonlinear chemical oscillator \cite{Strogatz:01b}. The method infers the phase dynamics with diffusive coupling functions from the experimental phase time-series. The coupling function was already discussed in Sec.\ \ref{sec34:CFMechanism}, as shown in Fig.\ \ref{fig:BZ_chemical}. Here we further review the interpretation and use of such coupling functions.

\begin{figure}
{\caption{Experiments on a three-cluster state close to a Hopf bifurcation with negative global coupling of 64 electrochemical oscillators. (a) Current time series and the three cluster configuration. Solid, dashed, and dotted curves represent the currents from the three clusters. (b) Cluster configuration. White, black, and gray circles represent the three clusters. (c) Response function and waveform (inset) of the electrode potential from a current of single oscillator. (d) Phase coupling function. $\Gamma(\Delta\phi)$ on the plot (d) is equivalent to the $q(\psi)$ notation used in the current review. From \citet{Kori:14}. } \label{fig:ChemApp3}}
{\includegraphics[width=1\textwidth,angle=0]{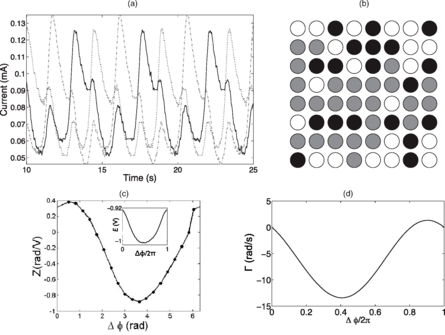}}
\end{figure}

\begin{figure*}
\floatbox[{\capbeside\thisfloatsetup{capbesideposition={right,top},capbesidewidth=3.6cm}}]{figure}[\FBwidth]
{\caption{(color online). Cardiorespiratory coupling functions for the study of human ageing. Typical time-averaged coupling functions for (a,c) a young subject aged 21 years and (b,d) an old subject aged 71 years. (a,b) are from the cardiac, while (c,d) are from the respiration phase dynamics.  From \citet{Iatsenko:13a}. }\label{fig:CRCF}}
{\includegraphics[width=0.74\textwidth,angle=0]{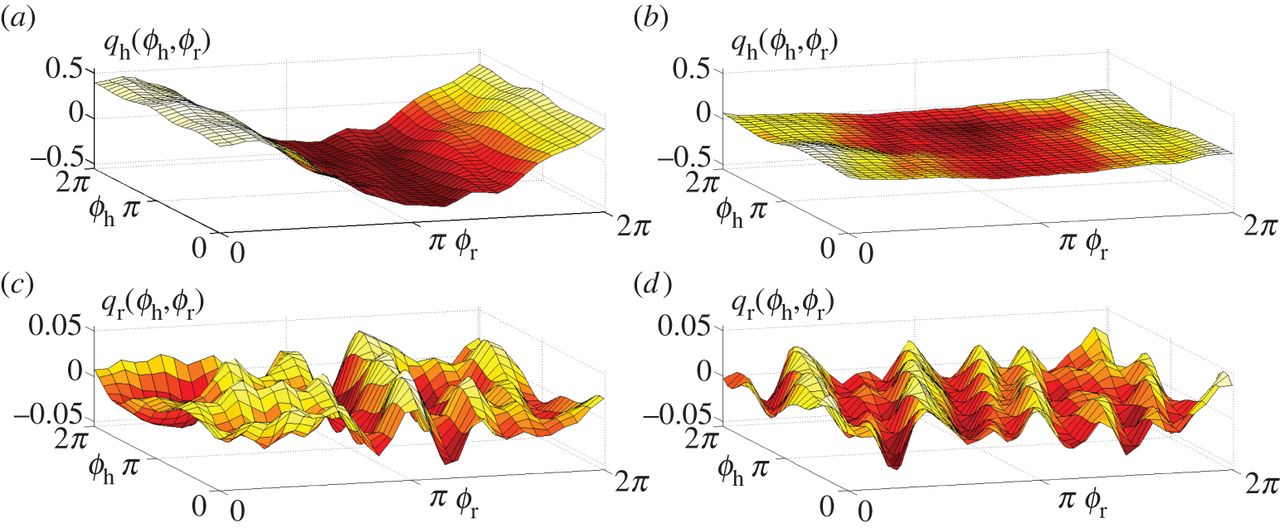}}
\end{figure*}

The inferred coupling function $q(\psi)$ is used to describe the mechanisms of the various synchronous states in two mutually coupled Belousov-Zhabotinsky reactors. The dynamics of the phase difference $\psi$ can be expressed as:
\begin{equation}\label{eq:chem_ph}
  \dot \psi=-\Delta_\omega + \varepsilon Q(\psi),
\end{equation}
where $Q(\psi)=q(\psi)-q(-\psi)$, $\Delta_\omega=\omega_2-\omega_1$ and $\varepsilon=\varepsilon_{12}-\varepsilon_{21}$. Then, by varying the frequency $\Delta_\omega$ mismatch and coupling strength $\varepsilon$, one can study and predict the occurrence of synchronization. Fig.\ \ref{fig:ChemApp2} shows that with increase of $\varepsilon$, a pair of stationary solutions of Eq.\ (\ref{eq:chem_ph}) are obtained as the intersection points of $Q(\psi)$ and $\Delta_\omega/\varepsilon$ (the first horizontal line from the top in Fig.\ \ref{fig:ChemApp2}). There is one stable solution (solid circle) and in-phase synchronization is realized. If one increases $\varepsilon$ further, a new stable solution appears slightly below $\pi$ in addition to that corresponding to in-phase synchronization -- the intersection with the second horizontal line from the top in Fig.\ \ref{fig:ChemApp2}.  This corresponds to out-of-phase synchronization. Thus, a bistability between out-of-phase and in-phase synchronization can appear.

\citet{Kori:14} have performed a comprehensive theoretical analysis and experimental verification of phenomena in electrochemical oscillators, investigating the general occurrence of phase clusters in weakly, globally-coupled oscillators close to a Hopf bifurcation. The amplitude equation with a higher-order correction term, valid near a Hopf bifurcation point, is derived and it is used to calculate analytically the phase coupling function from given limit-cycle oscillator models. Such phase coupling functions, allowed the stability of phase clusters to be analyzed, as demonstrated on the Brusselator model.

Experiments on electrochemical oscillators have demonstrated the existence of three-cluster states near the Hopf bifurcation with negative coupling. Electric potentials were used to control the nature of the oscillations, and they were chosen initially such that the oscillators exhibited smooth oscillations near the Hopf bifurcation. Fig.\ \ref{fig:ChemApp3}(a) shows the current from one oscillator of each of the three clusters. The nearly balanced three-cluster state with configuration
(25:20:19) is shown on a grid of 8 $\times$ 8 circles in Fig.\ \ref{fig:ChemApp3}(b).  Phase response curves (Fig.\ \ref{fig:ChemApp3}(c)) and coupling functions (Fig.\ \ref{fig:ChemApp3}(d)) for these oscillators were found experimentally by introducing slight perturbations to the oscillations. The stability of the cluster states was determined, and it was found that the three-cluster state is the most stable, while  four- and five-cluster states were observed also at higher potentials.  Further increase in the potential resulted in complete desynchronization of the 64 oscillators.

\subsection{ Cardiorespiratory interactions}\label{sec642:CR}

The heart and the lungs have physiological functions of great importance for human health and their disfunction may correspond to severe cardiovascular disease. Both organs are characterised by a pronounced oscillatory dynamics, and the cardiorespiratory interactions have been studied intensively using the theory and methods from the nonlinear coupled-oscillators approach \cite{Stefanovska:00a,Stefanovska:99a,Schaefer:98,Kenner:76}.

The cardiorespiratory coupling functions are therefore a subject of great interest i.e.\ the mechanisms through which respiration influences the cardiac period and, in particular, how this relates to different states and diseases. The cardiorespiratory analysis performed with dynamical Bayesian inference \cite{Stankovski:12b}, as discussed in Sec.\ \ref{sec622:DBI} and Fig.\ \ref{fig:DBI}, revealed the form of the coupling functions in detail. The use of a changing respiration frequency in a linear (ramped) way showed that the form of the reconstructed coupling functions is in itself time-varying. Recently, the method was applied to the study of the effects of general anaesthesia on the cardiorespiratory coupling functions \cite{Stankovski:16}. A similar form  of the function was reconstructed  for the awake measurements as in the previous studies, while its form was more varying and less deterministic for the state of general anaesthesia.

\begin{figure*}
{\caption{(color online).  Coupling functions for the human cardiorespiratory system. The reconstructed functions specify the dependence of the instantaneous cardiac frequency, measured in radians per second, on the cardiac and respiratory phases. The functions $Q_p(\phi_r,\phi_p)$ are computed from the arterial pulse and respiration.  Results from the subject who had the lowest levels of determinism and similarity to the coupling functions obtained from ECG phases $Q_e$ are shown in (a), and those for the subject with the highest determinism and similarity in (b). Panel (c) presents the averaged coupling function, over all measurements for all subjects.  From \citet{Kralemann:13b}. }\label{fig:CRCF2}}
{\includegraphics[width=0.95\textwidth,angle=0]{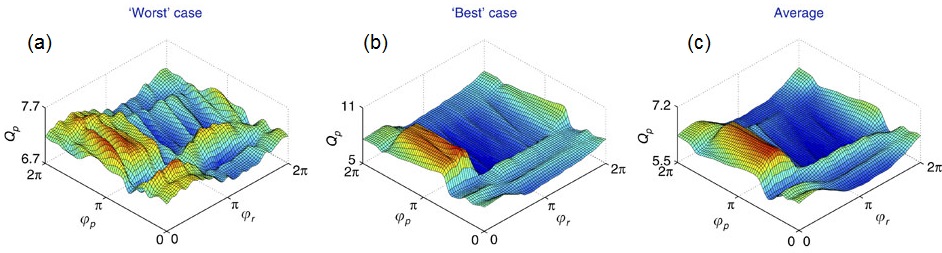}}
\end{figure*}

Dynamical Bayesian inference was used to study the effect of ageing on the cardiorespiratory interactions \cite{Iatsenko:13a}. Analyses were performed on cardiac and respiratory time series recorded from 189 subjects of both genders, aged from 16 to 90 years. By application of the synchrosqueezed wavelet transform,  the respiratory and cardiac frequencies and phases were pre-estimated. By applying dynamical Bayesian inference to the phase time-series, measures such as synchronization, coupling directionality and the relative contributions of different mechanisms were then estimated.

The cardiorespiratory coupling function was thus reconstructed, and its time evolution and age dependence were assessed. Fig.\ \ref{fig:CRCF} shows the time-averaged versions of the cardiorespiratory coupling functions typical of a younger and an older subject. Figs.\ \ref{fig:CRCF}(a) and (b) show the coupling functions of the heart dynamics $q_h(\phi_h,\phi_r)$. The form of the functions (especially noticeable in Fig.\ \ref{fig:CRCF}(a)) is changing mostly along the respiration phase $\phi_r$ axis, while it is nearly constant along the $\phi_h$ axis, indicating that this coupling is predominantly defined by the direct influence of respiration on the heart. In physiology, this modulation is known as respiratory sinus arrythmia (RSA). By comparing the coupling functions for the young and old subjects one can see a clear decrease of the RSA amplitude with age. It can also be noted that RSA remains the main stable contribution to the $q_h(\phi_h,\phi_r)$ coupling function, irrespective of age and that it survives after time-averaging. The coupling function from respiratory dynamics $q_r(\phi_h,\phi_r)$ shown in Fig.\ \ref{fig:CRCF}(c) and (d) was very low and seemed to be quite irregular and not age-dependent. From analysis of the time-variability of the form of the coupling functions it was observed that, in older people, the heart coupling function $q_h(\phi_h,\phi_r)$ becomes less stable in time, dominated by the highly time-variable indirect contributions. At the same time, the dynamics of the respiratory coupling function  $q_r(\phi_h,\phi_r)$ did not seem to change with age, being irregular and unstable.

\citet{Kralemann:13b} looked at the cardiorespiratory coupling functions, as an intermediate result, in order to obtain the phase response curve of the heart and the perturbation inserted by respiration. They studied the respiratory and cardiac oscillations of 17 healthy humans while resting in an unperturbed state. The cardiac oscillation was assessed through two different observables -- the electrocardiograph (ECG) and the arterial pulse signal. The idea of using two different observables is to demonstrate that an invariant phase can be obtained from each of them, describing a common inherent interaction between respiratory and cardiac oscillations.

By analysing the phase dynamics, first by estimating the protophases and transforming them into genuine phases, the cardiorespiratory phase coupling functions were reconstructed. Fig.\ \ref{fig:app}(b) shows the reconstructed cardiorespiratory coupling functions $Q_e$ using the cardiac phase extracted from the ECG signal. The coupling functions estimated from the phases of the arterial pulse signal are shown in Fig.\ \ref{fig:CRCF2}. The forms of the functions reveal the detailed mechanism through which respiration influences the cardiac oscillations i.e.\ the regions with high values of the function mean higher frequencies (acceleration), whereas low regions correspond to lower frequencies (deceleration) of the cardiac oscillations due to the respiratory influence. The existence of such cardiorespiratory coupling functions was tested statistically in respect of inter-subject and intra-subject surrogates.

The high similarity of the cardiorespiratory coupling functions obtained from the phases of the ECG observables Fig.\ \ref{fig:app}(b) and those from the arterial pulse phases Fig.\ \ref{fig:CRCF2}(b) demonstrates that the proposed method correctly identified the underlying interaction mechanism. This was achieved because the method was able to transform protophases from different observables into invariant phase dynamics from which a common form of the coupling functions was obtained. The minor differences in the form of the functions, as compared to the previous one with ageing (Fig.\ \ref{fig:CRCF2}), are related to possible inter-subject variations and the different phase estimation approaches.

The similarity of the cardiorespiratory coupling functions among different subjects and between the two observable phases was further quantified with the similarity indices \cite{Kralemann:13b}, as discussed in Sec.\ \ref{sec63:CFAnalysis} and shown in Fig.\ \ref{fig:CFsiml}. The similarity index that quantifies the correlation between the form of the functions, also proved very useful in assessment of the state of general anaesthesia \cite{Stankovski:16}.
It was found that the intersubject correlation similarity of the cardiorespiratory coupling functions, in comparison to the awake state, decreased with the onset of propofol-induced general anaesthesia, and to an even greater extent when sevoflurane was used.

\subsection{ Neural coupling functions}\label{sec643:MetNeuro}

\begin{figure*}
{\caption{(color online).  Neural cross-frequency coupling functions between $\delta$ and $\alpha$ oscillations in general anaesthesia.  (a)-(c) The average coupling functions from all subjects within the group. Note that, for realistic comparison, the vertical
scale of coupling amplitude is the same in each case. Here,  \emph{Awake} refers to
the state when the subject is conscious and resting; and \emph{Propofol} and \emph{Sevoflurane} to states when the subject is anaesthetized with propofol
or sevoflurane, respectively.  From \citet{Stankovski:16}. }\label{fig:Neuro2}}
{\includegraphics[width=0.88\textwidth,angle=0]{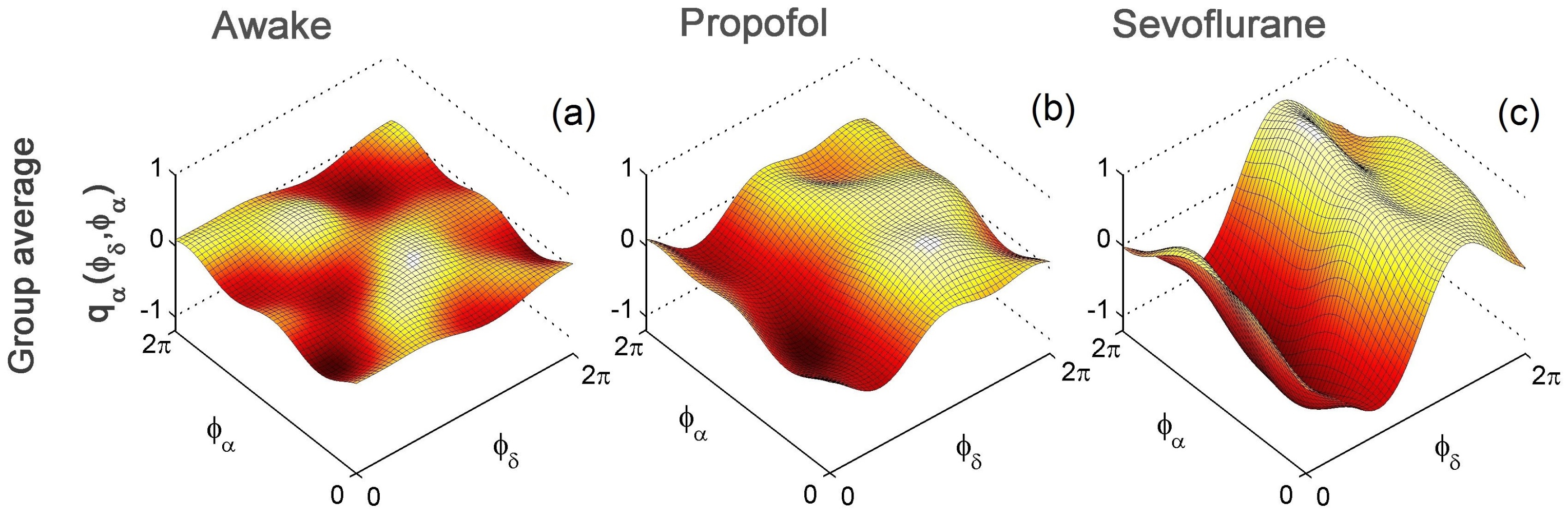}}
\end{figure*}

Neural states often manifest themselves as changes in brain electrophysiological activity,
which emanates from the dynamics of large-scale cell ensembles oscillating synchronously within characteristic frequency intervals. Individual ensembles communicate to integrate their local information flows into a common brain network. One way to describe such an integration or communication is through \emph{cross-frequency coupling}, an approach that has led to numerous studies elucidating the respective roles of cognition, attention, memory and anaesthesia \cite{Canolty:06,Jirsa:13,Lakatos:08,Stefanovska:07}. Unlike these cross-frequency coupling strength approaches, the methods discussed here can assess the neural states through the computation of the coupling functions, hence describing the functional forms and mechanisms of individual cross-frequency interactions. In this way, one infers \emph{neural cross-frequency coupling functions }\cite{Stankovski:17c}.

The methods for the reconstruction of coupling functions have been applied to electroencephalogram (EEG) recordings. The brainwave intervals, including the $\delta$ (0.8-4 Hz), $\theta$ (4-7.5 Hz), $\alpha$ (7.5-14 Hz), $\beta$ (14-22 Hz) and $\gamma$ (22-80 Hz), were first extracted from a single EEG channel recording. The phase was then extracted from each filtered time series, using for example the Hilbert transform or the synchrosqueezed wavelet transform. During this preprocessing procedure, particular care was taken to minimise overlap between the spectra of the intervals \cite{Lehnertz:14}: overlaps of consecutive frequency intervals would result in overestimation of the corresponding phase-to-phase coupling. Dynamical Bayesian inference was then used to reconstruct the coupling functions from the multivariate five-phase oscillators. In a similar manner, dynamical Bayesian inference was applied to a study of neural interactions during epileptic seizures \cite{Wilting:15}, though not for cross-frequency coupling.

\begin{figure}
{\caption{(color online). Examples of neural cross-frequency coupling functions. (a) Spatial distribution of the $\delta$-$\alpha$ coupling functions over the head, based on the different probe locations. (b) Average coupling function along all the probes for the $\delta$-$\alpha$ coupling relation. Each $\delta$-$\alpha$ coupling function $q_\alpha(\phi_\delta,\phi_\alpha)$ is evaluated from the $\alpha$-dynamics and depends on the bivariate $(\phi_\delta,\phi_\alpha)$ phases.  From \citet{Stankovski:15a}. }\label{fig:Neuro1}}
{\includegraphics[width=1\textwidth,angle=0]{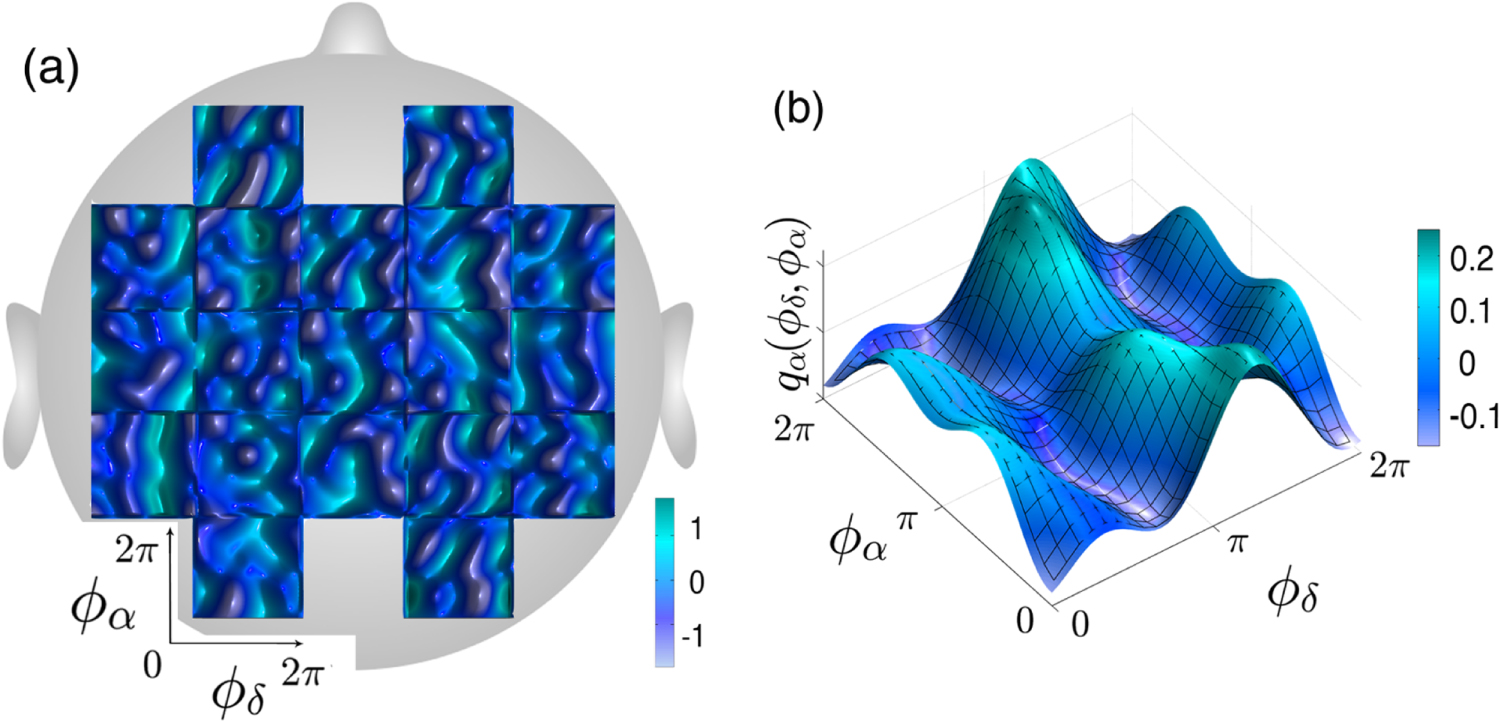}}
\end{figure}

As the brain is a highly complex system that can mediate a variety of functions from fixed structure \cite{Park:13}, the coupling relationships between the brainwaves can be different. One important coupling relation is the $\delta,\alpha\rightarrow\alpha$, as it has been found that the $\delta$-waves typical of deep sleep in adults can influence the $\alpha$-activity, which is related to the processing of information \cite{Feinberg:87,Jirsa:13}. Fig.\ \ref{fig:Neuro1}(a) shows how the form of the $\delta,\alpha\rightarrow\alpha$ coupling functions varies in relation to their spatial locations on the head. It can be seen that the tridimensional waves propagate mostly in the $\delta$ dimension. This tendency can be seen better in Fig.\ \ref{fig:Neuro1}(b), which shows the averaged coupling function. Its form depends predominantly on the direct delta oscillation, changing mostly along the $\phi_\delta$-axis. This reveals how and when within one cycle the $\delta$ oscillations accelerate and decelerate the $\alpha$ oscillations. Other coupling relationships could include for example the pairwise $\theta,\gamma\rightarrow\gamma$ and $\alpha,\gamma\rightarrow\gamma$, or the multivariate triplet $\theta,\alpha,\gamma\rightarrow\gamma$, as shown previously in Fig.\ \ref{fig:CF_multi2}.

\begin{figure*}
{\caption{(color online).  Coupling functions of the change in democracy $q_1(D,G)$, from the interaction between GDP and democracy. For (a) the two term model is used, while for (b) the five term model. The black line is the solution $\dot D=0$. The strength of the coupling functions are encoded by the colorbars shown on the side of each figure.  From \citet{Ranganathan:14}. }\label{fig:SocialCF}}
{\includegraphics[width=0.935\textwidth,angle=0]{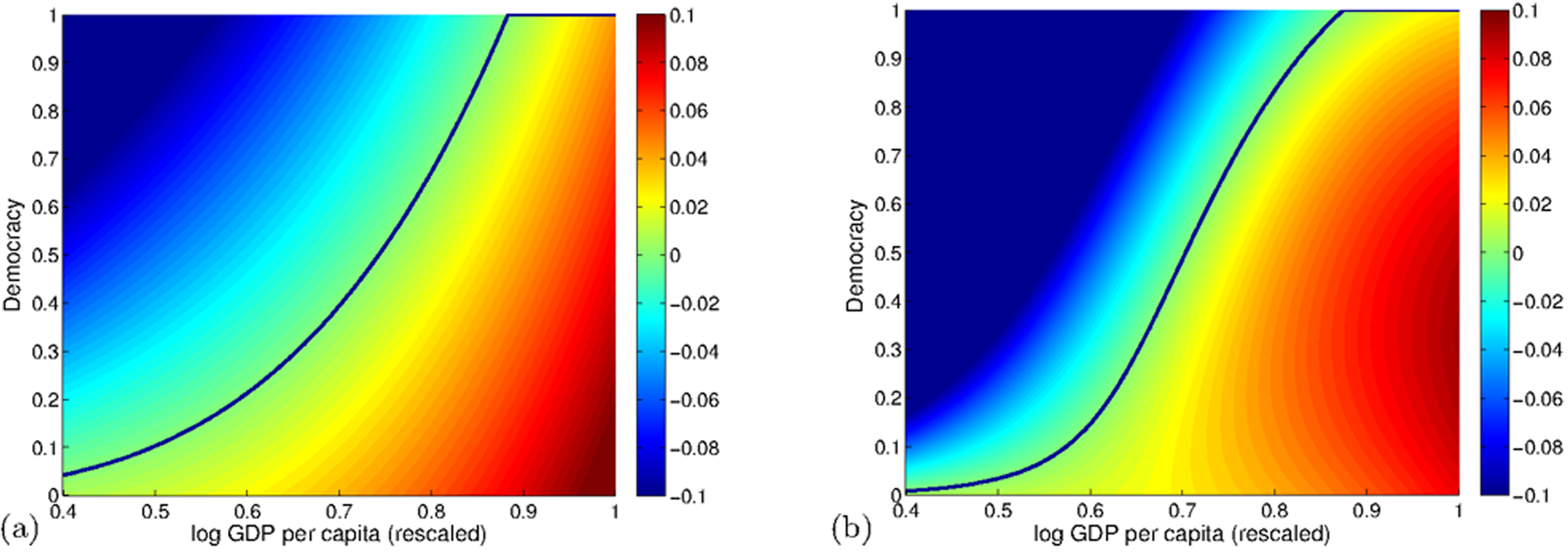}}
\end{figure*}

Neural cross-frequency coupling functions were used recently to elucidate the mechanisms of general anaesthesia \cite{Stankovski:16}. In fact, the analyses also included the cardiac and respiratory oscillations (in a sense integrating Secs.\ \ref{sec642:CR} and \ref{sec643:MetNeuro}). Here, we review an important finding based on the neural $\delta$-$\alpha$ coupling functions. The study included 25 awake and 29 anaesthetized healthy subjects, of which 14 subjects were anaesthetized with  the intravenous anaesthetic \emph{propofol} and 15 subjects with the inhalational anaesthetic \emph{sevoflurane}. The aim of the study was to determine if there are any differences in the interaction mechanisms in respect of the three states: awake; and anaesthetized with either propofol or sevoflurane.

Fig.\ \ref{fig:Neuro2} shows the group $\delta$-$\alpha$ coupling functions for the three states. The coupling functions for the awake resting, propofol, and sevoflurane states are evidently quite different from each other, both in the forms and strengths of the couplings. The $\delta$-$\alpha$ coupling function for the awake state has a relatively complex and varying form, and low amplitude. The coupling functions for propofol and sevoflurane are similar to each other and they look significantly different from those for the awake state. The sevoflurane coupling function has the largest coupling amplitude. Careful surrogate testing showed that 
the propofol and sevoflurane coupling functions are statistically significantly different from the corresponding surrogates. The qualitative form of the $\delta$-$\alpha$ coupling function has a sine-like form along the $\phi_\delta$-axis, while remaining nearly constant along the $\phi_\alpha$-axis. This implies that much of the $\delta$-$\alpha$ coupling comes from the direct contribution of the delta oscillation. The specific form of the function (e.g.\ Fig.\ \ref{fig:Neuro2}(c)) reveals the underlying coupling mechanism, i.e.\ it shows that, when the delta oscillations are between $\pi$ and $2\pi$, the sine-wave coupling function is higher and the delta activity accelerates the alpha oscillations; similarly, when the delta oscillations are between $0$ and $\pi$, the coupling function is decreased and delta decelerates the alpha oscillations.

The delta-alpha coupling has been linked to the coding mechanism of feedback valence information \cite{Cohen:09}, non-REM sleep \cite{Bashan:12} and the eyes-closed state \cite{Jirsa:13}. The findings with an{\ae}sthesia are consistent with, and have further extended, these findings. The form of the $\delta$-$\alpha$ coupling functions (e.g.\ Fig.\ \ref{fig:Neuro2} (c)) indicates that the influence is direct modulation from delta to alpha, where the couplings are significantly stronger in an{\ae}sthesia than when awake. This showed that, once the subject is anaesthetized, delta activity influences the alpha oscillations by contributing to the reduction of information processing and integration.


\subsection{ Social sciences}\label{sec6434:Social}

In a recent application in social sciences, \citet{Ranganathan:14,Spaiser:14} have identified the coupling functions that capture interactions between social variables, employing a Bayes factor to decide how many interaction terms should be included in the model.

The work presents an interesting study of the relationship between democracy and economic growth, identifying nonlinear relationships between them. Economic growth is assessed through the GDP per capita (from the World Bank), while the level of democracy is gauged from the democracy index (from Freedom House) \cite{Ranganathan:14}. It is well known that the GDP per capita and democracy are highly correlated: higher GDP implies more democracy. The linear Pearson correlation coefficient between the two variables is 0.571 ($p<0.01$). However, by use of coupling functions one can try to determine a more precise, causal relationship between the variables, revealing the underlying mechanism.

In its general form, the model considered is:
$$\dot{\mathcal{D}}=q_1(\mathcal{D},\mathcal{G}); \text{  } \text{  } \dot{\mathcal{G}}=q_2(\mathcal{D},\mathcal{G}),$$
where $\mathcal D$ denotes the democracy and $\mathcal G$ is GDP per capita. In this way the change of the variables $\mathcal D$ and $\mathcal G$ is represented with ordinary differential equations, even though the original data are discrete and one should really use difference rather than differential equations. Nevertheless, this approximation was used for mathematical simplicity. Further, the model can have some of the functions from a set of seventeen base functions of polynomial form, including reciprocal, quadratic and cubic terms. The main idea of the method is to select the optimal base functions thereby reducing the number of terms in the model.

The inference itself consists of two main steps. The first is an inferential fitting to obtain a model from the data, based on a maximum likelihood procedure, and involving multiple linear regression (similar to that discussed in Sec.\ \ref{sec623:MLE}). The second step uses a Bayesian \cite{Berger:96} model selection procedure\footnote{We note that the Bayes factor uses the Bayesian probability theory too; however, as it is purely statistical procedure, it differs from the DBI as discussed in Sec.\ \ref{sec622:DBI}.}. Here, the method decides how many interaction terms should be included in the model, i.e.\ it selects a subgroup of base functions of the seventeen polynomials available, after trying all possible combinations among them. Thus, the method punishes overly complicated models and identifies the models with the most explanatory power. Such procedures could benefit greatly if a surrogate testing procedure (see Sec.\ \ref{sec22:CplDirc}) were used to determine whether the finally selected model is genuinely reliable.

The method was applied to model the interaction of democracy and GDP per capita for the years 1981-2006 for 74 countries. The resulting coupling functions for two selected models are shown in Fig.\ \ref{fig:SocialCF}. The simplest model shown in Fig.\ \ref{fig:SocialCF}(a) includes a coupling function with two terms $\dot D=0.11G^3-0.067D/G$, while the best fit five-function model Fig.\ \ref{fig:SocialCF}(b) was given as $\dot D=0.77G^3+1.9D-0.85D/G-0.96DG-0.14D^2$. For the middle GDP, both of the coupling functions show dependences that closely relate to the linear dependence determined with the simple correlation coefficient. However, there were some nonlinear deviations from this, especially for very low and very high GDP. In particular, the threshold for very high GDP indicated that there is no significant improvement in democracy with further GDP growth.

\red{Similarly, the best model for $G$ was inferred to be $\dot G=0.014+0.0064DG-0.02G$, which shows primarily that the GDP is growing at a constant rate, but in addition demonstrates that it is positively affected by democracy interacting with GDP, and that the growth is self-limiting at high levels of GDP.}

Finally, we point out that the method was further applied to investigate the interactions of other social variables, such as the case of interactions between democracy, development and cultural values \cite{Spaiser:14}. These works could benefit significantly from further coupling function assessment and analysis.

\subsection{ Mechanical coupling functions}\label{sec644:Mechanic}

Mechanical clocks and oscillators provide an important cornerstone in the study of interactions and synchronization phenomena, starting from the earliest observations of the phenomenon in pendulum clocks by \citet{Huygens:7386}, up to the more comprehensive and detailed studies based on current methods \cite{Kapitaniak:12}.


\begin{figure}
{\caption{(color online).  Coupling functions for the two coupled mechanical metronomes. (top) Experimental apparatus with the two metronomes, placed on a rigid support. (a),(b) The coupling function in each direction from the case of coupling with one rubber band; (c),(d) with two rubber  bands. (e),(f) coupling functions for the ``uncoupled'' without any rubber bands. The vertical scales are the same so that one can clearly see the reduction of the coupling function in the uncoupled case. From \citet{Kralemann:08}. }\label{fig:Mech2}}
{\includegraphics[width=1\textwidth,angle=0]{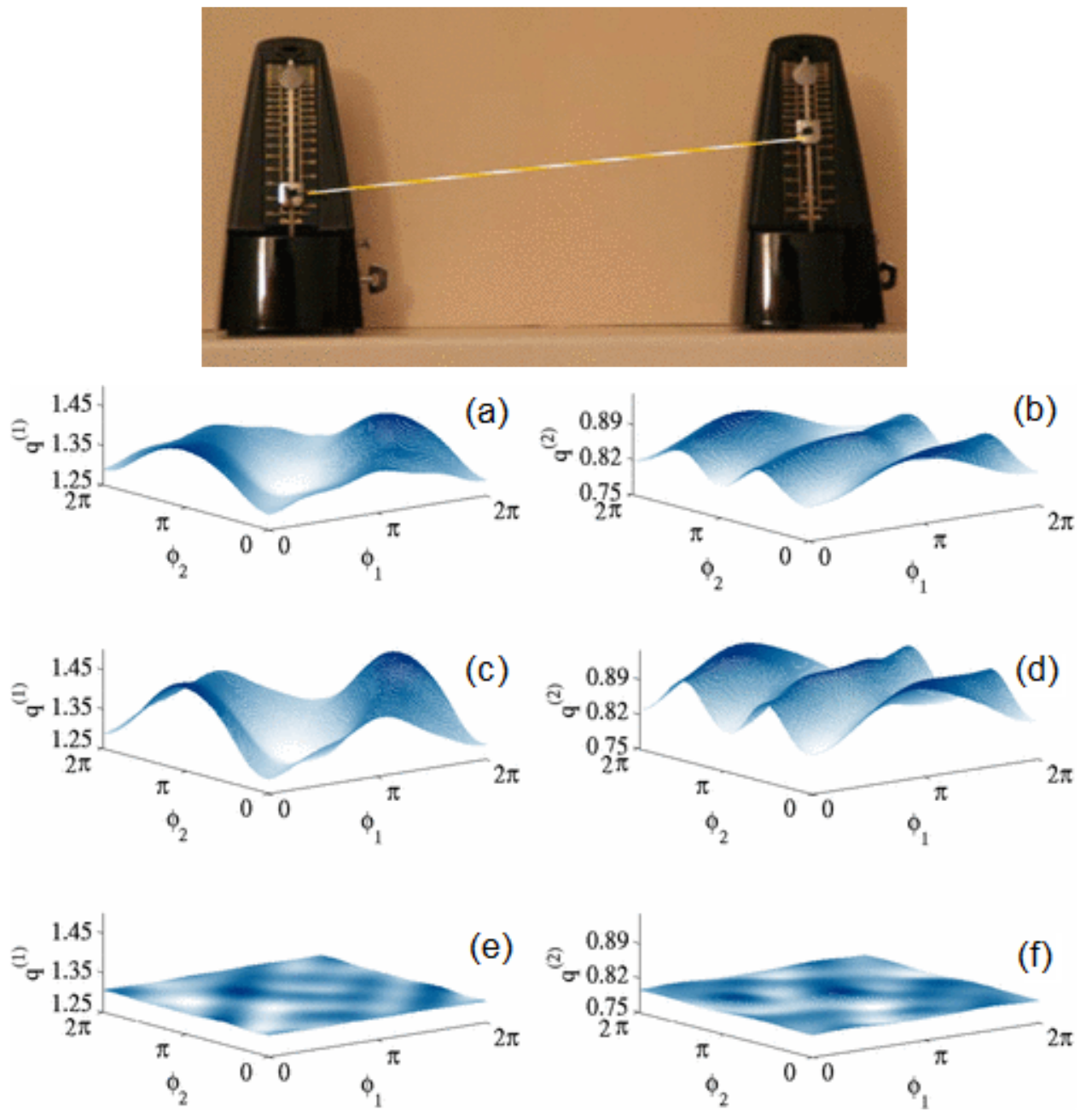}}
\end{figure}

\begin{figure*}
{\caption{(color online).   Schematic diagram showing the communication protocol based on coupling functions. Messages $s_1,\ldots,s_n$ are encrypted by modulation of the coupling functions connecting two dynamical systems at the transmitter. Only two signals are transmitted through the public domain.  The receiver consists of two systems of the same kind with the same coupling functions (forming the private key) and uses dynamical Bayesian inference to reconstruct $s_1,\ldots,s_n$. From \citet{Stankovski:14a}. }\label{fig:Com1}}
{\includegraphics[width=0.935\textwidth,angle=0]{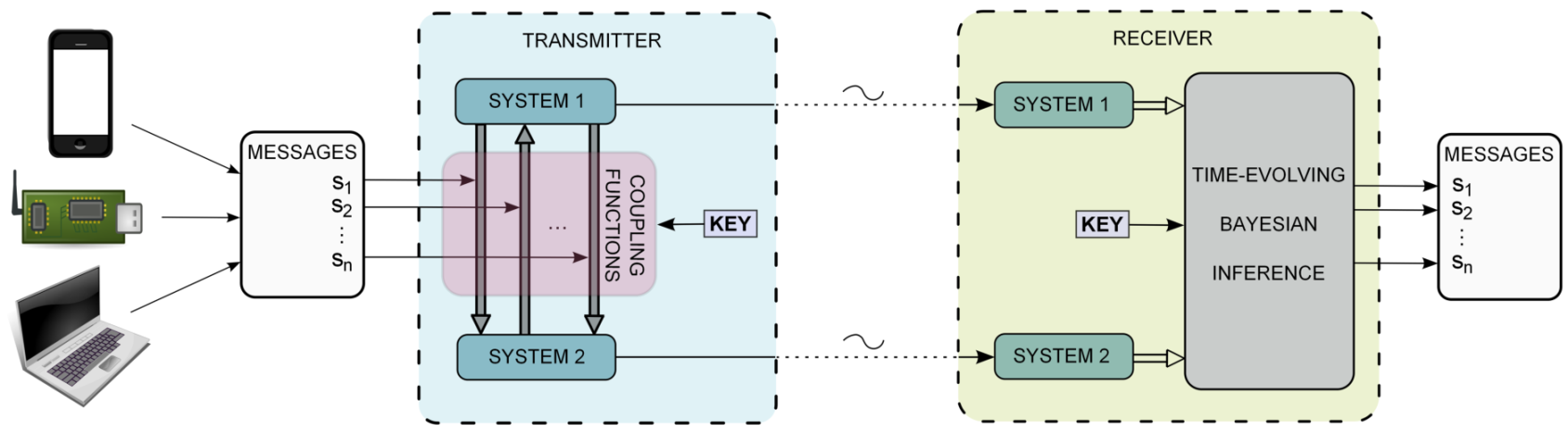}}
\end{figure*}

\citet{Kralemann:08} describe an experiment using two coupled mechanical metronomes for the analysis of coupling functions. The metronomes were placed on a rigid base and the coupling through which they interact and influence each other was achieved by connecting them with an elastic rubber band -- Fig.\ \ref{fig:Mech2}(top). A digital camera was used for acquiring the data, from which the oscillatory signals were extracted. Coupling functions were determined for three different experimental conditions, when: (i) the pendulums of the metronomes were linked by a rubber band; (ii) the pendulums were linked by two rubber bands; and (iii) the metronomes were uncoupled.

From the extracted signals, the Hilbert transform protophases were first estimated, and then transformed to genuine phases. The coupling functions were then reconstructed with a fitting procedure based on kernel smoothing. Fig.\ \ref{fig:Mech2} shows the results for the three cases. By comparison of the coupling functions in the case of one rubber coupling Fig.\ \ref{fig:Mech2}(a),(b) with the coupling of two rubber coupling Fig.\ \ref{fig:Mech2}(c),(d), one can see that the form is very similar, while the coupling strength is slightly higher for the case of two-band coupling. The coupling functions is of a complex form, changing along both axes, thus reflecting the bidirectional influence and contribution within the couplings. The methodology correctly detects extremely weak coupling functions Fig.\ \ref{fig:Mech2}(e),(f) for the case of no explicit coupling. \red{Future developments of this work could benefit from comparison of the extracted coupling functions with the actual mechanics of the coupled metronomes, as well as from validation of the weak coupling regime for better justification of the use of phases.}

\subsection{ Secure communications}\label{sec645:Comm}

The findings that the cardiorespiratory coupling function can be decomposed into a number of independent functions, and that the latter can have a time-varying nature \cite{Stankovski:12b} and Sec.\ \ref{sec642:CR}, inspired the creation of a new class of secure communications characterized by high efficiency and modularity \cite{Stankovski:14a}.

The protocol (Fig.\ \ref{fig:Com1}) starts with a number of information signals coming from different channels or communications devices (e.g.\ mobile phone, sensor networks, or wireless broadband) needing to be transmitted simultaneously. Each of the signals $s_i$ is encrypted in an amplitude coupling function; i.e.\ they serve as scaling parameters in the nonlinear coupling functions between two self-sustained systems in the transmitter. The coupling functions constitute the private key and, in principle, have an unbounded continuum of possible combinations. Two signals, one from each system, are transmitted through the public channel. At the receiving end, two similar systems are enslaved, i.e.\ completely synchronized, by the two transmitted signals. Finally, by applying time-evolving dynamical Bayesian inference (as discussed in Sec.\ \ref{sec622:DBI}) to the reconstructed systems, one can infer the model parameters and decrypt the information signals $s_i$.

This application is similar to that where amplitude coupling functions are reconstructed from data. The coupled systems are multidimensional and may be, e.g.\ chaotic Lorenz or R\"{o}ssler systems. The great advantage is that the model of the coupled systems is known exactly on the side of the receiver where the inference is performed. Thus the problem of not knowing the amplitude model and its dimensionality does not exist. The main task of the decryption lies in inferring the time-evolution of the parameters.

The protocol can encrypt multiple signals simultaneously as time-evolving parameters. Each of them scales one of the coupling functions, which are nonlinear and mutually linearly-independent. Thus the method inherently allows for multiplexing i.e.\ simultaneous transmission of multiple signals. Another property of the protocol is that it is highly resistent to channel noise. This is because the dynamical Bayesian inference is performed for stochastic dynamics, so that the method is able very effectively to separate the unwanted noise from the deterministic dynamics carrying the messages.

\begin{figure}
{\caption{(color online).  Transmission of ten pseudorandom binary signals encrypted in different coupling functions. The high values (binary `1') at the transmitter, are indicated by grey shading. The received signals, after decrypting, are shown by \blue{thick (red)} lines, each of which (a-j) represents one information signal $s_i(t)$. The particular amplitude coupling functions that were used for encrypting each signal are indicated on the ordinate axis. The bit words  are indicated by $m_1$-$m_4$ on the top of the figure.
From \citet{Stankovski:14a}. }\label{fig:Com2}}
{\includegraphics[width=1\textwidth,angle=0]{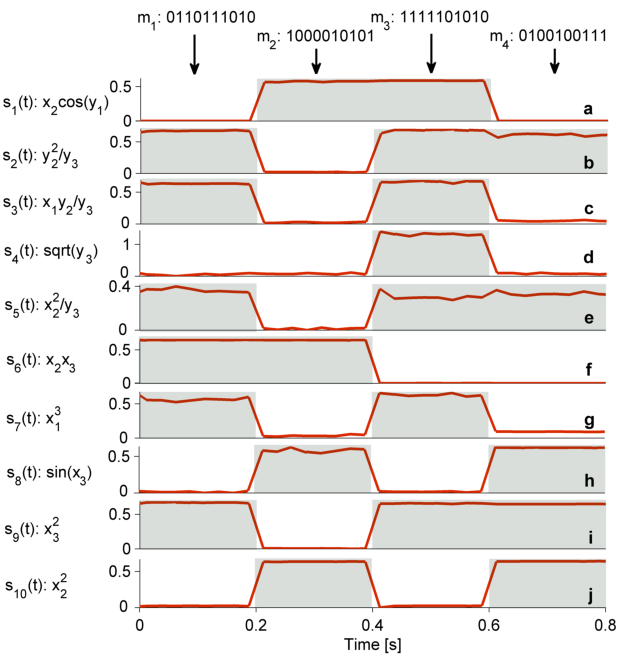}}
\end{figure}

The method is demonstrated on two bidirectionally coupled chaotic Lorenz systems. Ten information signals $s_1,\ldots,s_{10}$ are encrypted with ten coupling functions, as indicated on the ordinate axis in Fig.\ \ref{fig:Com2}. The choice of the particular forms of the coupling functions prescribe the private key. After the transmission, the systems are reconstructed on receiver side and, by application of Bayesian inference, the information signals are reconstructed. Fig.\ \ref{fig:Com2} shows good agreement of the original with the decrypted bits.

\section{ Outlook and Conclusion }\label{sec7:Future}

\subsection{ Future directions and open questions}\label{sec4:Other}

Coupling functions have been studied as early as some of the first theoretical works on interactions and they remain a very active field of research that is attracting increasing interest from the scientific community. They bring a certain complexity in understanding, but at the same time they also illuminate, and provide deeper insight into, the interaction mechanisms. As such, coupling functions pose many open questions and there still remain many related aspects that are not well understood. Below, we discuss some of the open questions and current possibilities for further developments related to coupling functions.

\subsubsection{  Theory}\label{sec4:Other}

The theoretical development of coupling functions will lead to a better understanding of the mechanisms responsible for the resulting overall dynamics, and they may help to incorporate seemingly different models into a general overall framework. Future theoretical studies need to \emph{identify the classes of coupling functions} that lead to particular physical effects. In doing so, one needs to determine if there are some classes of functions which demonstrate unique characteristics, and more importantly whether some particular functions lead to common effects.  As a consequence, these can then lead to classes of functions to be used in engineering, for controlling or predicting the outcome of the interactions.

The experimental results suggest two important directions for the theoretical development. First, \emph{coupling functions can be nonautonomous}. Secondly, \emph{coupling functions can lead to the coexistence of attractors}.

The theory of nonautonomous dynamical systems has gained recent interest, mainly in relation to finite-time bifurcations \cite{Kloeden:11}. These mathematical developments will play a major role in the theory of coupling functions. The theoretical studies should include systematic and comprehensive descriptions of the different classes of coupling function, including the nonautonomous case.

The coexistence of attractors has gained considerable attention. As experimental results show, coupling functions are important, especially in relation to network structure and the effect on the basin of attraction, e.g.\ the basin of attraction of synchronization. Recent results have shown that the roles of coupling function and network structure can be nontrivial \cite{Menck:13}. We still need new theoretical methods to tackle problems associated with the involvement of coupling functions in the coexistence of attractors. These questions are intimately related to the stability of the system and will play a role in important practical applications, such as to electrical power-grids.

\subsubsection{  Methods}\label{sec4:Other}

The future development of coupling function methods needs to take into account all the advantages and pitfalls of current methods, e.g.\ as outlined in the critical comparison Sec.\ \ref{sec626:CompareMethods}. So far, all the coupling functions methods have been applied to pairs of coupled systems, or small to medium size networks. New methods should allow \emph{applications to the more prevalent large-scale networks}. In line with this, they should aim to \emph{achieve faster calculations} so as to facilitate the ever-growing demand for extensive computation.

There is also a room for improvements of the amplitude coupling function methods. The search for more \emph{generally applicable amplitude models} remains open. These should be as general as possible, or at least general enough for specific sub-systems.

There is a need to overcome the problem of inferring coupling functions from systems that are \emph{highly synchronized and coherent} in 1:1 frequency ratio. Currently, this is a common deficiency in all of the available methods for coupling (function) reconstruction. A possible direction for solving this issue could lie in the use of perturbation, for example starting from different initial conditions or with some other form of temporary deviation from the highly coherent state.

To enhance the inference of more general interactions,  efforts are needed to develop and design robust methods for distinguishing direct from indirect couplings, better surrogates and null models for determining significant couplings and interpretation of the couplings in high-dimensional networks.

\subsubsection{ Analysis }\label{sec4:Other}

We also point to the need for \emph{further development in coupling function analysis}: although the basic coupling function theory and methods are relatively well developed, there is a pressing need for measures able to exploit the computed coupling functions to better effect. The current tools, including similarity analysis and coupling function decomposition, are very useful, but there is a clear need for development of even more analysis tools. The task here is to find better and more systematic ways of quantifying and describing the form of the coupling functions and the other functional characteristics unique to coupling functions.

The development of such methods for analysing and characterizing coupling functions could be linked to the \emph{mathematical theory of functional analysis}. To date, this theory including the main concepts in vector spaces, and measures of mappings between the functions, have not yet been fully exploited in relation to coupling functions.

\subsubsection{  Integration theory-applications}\label{sec4:Other}

We emphasize that further \emph{interplay between theory and applications} for the development of coupling functions is still needed. Although the applications usually take into account theoretical developments, recent experimental findings have not yet been properly addressed by the theory. For example, the theoretically most studied form of coupling functions is that for diffusive coupling, which includes the state or phase difference as an argument. The latter is mostly used because it provides convenient solvable solutions (see e.g.\ history, Sec.\ \ref{sec32:History}). On the other hand, the coupling decomposition experiments have shown that it is the direct coupling function that often predominates in reality, especially in biological oscillatory interactions. Hence, further theoretical studies are needed to establish the phenomena and the nature of interactions for direct coupling functions. Such theoretical investigations can usefully be performed numerically in cases where the relevant model cannot be solved analytically.

\subsubsection{Applications}

Coupling functions have universal implications for all interactions between (dynamical) systems. As such, they can describe mechanisms operating between systems that are seemingly of very different natures. We have reviewed a number of important applications, including for example chemical, biomedical, mechanical, social and secure communications; however, the unique features of coupling functions promise even further application in these and in other fields. We outline below some foreseeable directions for new applications, notwithstanding that many others are also possible.

Recently, there has been a significant interest and developments in the study of interactions and synchronization in \emph{power grids} \cite{Rubido:15,Rohden:12}. To ensure a reliable distribution of power, the network should be highly controllable and synchronized. It is therefore very important  that the state of synchronization should be highly stable (i.e.\ deep in the Arnold tongue), so as to ensure that small disruptions and glitches will not interrupt the function of the network. Coupling functions should be investigated in order to establish how to design and engineer a persistent \cite{Pereira:14} and very stable state of synchronization.

Similar problems occur with the control, synchronization and optimization in \emph{transport grids} \cite{Rodrigue:99,Albrecht:04}, for example in a rail network. In such cases, of vital importance are the dynamical and the time-varying events. The developed methods and theory for time-varying coupling functions could be of great use in these applications.

Increasing the scale of the networks often leads to higher-level organization, including \emph{networks of networks} and {\it multilayer networks} \cite{Stern:13,Kivela:14}. In such high-dimensional spaces, a variety of different physical effects can be observed, e.g.\ synchronization, chimeras, and clustering. The coupling functions of different levels and layers could provide deeper insight into the functions or subfunction integration of the networks.

Coupling functions have been found very useful in studying the interactions between macroscopic physiological systems, such as those between the cardiorespiratory and neural systems reviewed above. Further coupling function investigations will probably be developed between different oscillations in integrated \emph{network physiology} \cite{Bashan:12,Stefanovska:07}. In a similar way, coupling functions between microscopic physiological organizations could be developed. The latter could explore \emph{coupling functions between cells} including, for example, the oscillations of neurons or stem cells \cite{Murthy:96,Eytan:06,Mendez:08,Jackson:01}.

\subsection{ Conclusion }\label{sec4:Other}

In recent years, the investigation of coupling functions has developed into a very active and rapidly evolving field. Their study and use have brought huge progress in the understanding of the mechanisms underlying the diverse interactions seen in nature. The enterprise has now reached a critical mass, offering increased potential for new and important discoveries, and in this way the topic has attained a substance and unity justifying the present review. 

The concept of the {\it function} in the coupling functions is perhaps its most important characteristic. Yet, precisely because of being a function, it is inevitably harder to interpret, assess and compare than is the case for quantitative measures such as the coupling strength. In attempting to integrate and pull together existing knowledge about coupling functions, therefore, we have tried to organize, explain and, as far as possible, to standardize their description in the hope of making them more generally accessible and useful.


Interactions underlie many important phenomena and functions of the systems found in nature, and it is of great importance to be able to describe and understand the mechanisms through which the interactions occur. Coupling functions are opening up new perspectives on these interactions and we envisage that they will catalyze increased research activity on coupled dynamical systems and their interactions in the future.

\acknowledgements
We gratefully acknowledge valuable and continuing discussions with
Ralph Andrzejak,
Peter Ashwin,
Murilo Baptista,
Miroslav Barabash,
Chris Bick,
Philip Clemson,
Andreas Daffertshofer,
Andrea Duggento,
Jaap Eldering,
Deniz Eroglu,
Dmytro Iatsenko,
Viktor Jirsa,
Peter Kloeden,
Ljupco Kocarev,
\blue{Thomas Kreuz},
Juergen Kurths,
Jeroen Lamb,
Gemma Lancaster,
Klaus Lehnertz,
Zoran Levnaji\'{c},
Dmitry Luchinsky,
Hiroya Nakao,
Milan Palu\v{s},
Thomas Peron,
Spase Petkoski,
\blue{Arkady Pikovsky},
Antonio Politi,
Alberto Porta,
Martin Rasmussen,
Francisco Rodrigues,
Michael Rosenblum,
Rajarshi Roy,
Bj\"{o}rn Schelter,
Ivana Stankovska,
Sinisa Stojanovski,
Yevhen Suprunenko,
Peter Tass,
Valentina Ticcinelli,
and Dmitry Turaev.
The work was supported
by the Engineering and Physical Sciences Research Council UK (grant numbers EP/100999X1, EP/M015831/1 and EP/M006298/1),
by the ITN COSMOS programme (funded by the EU Horizon 2020 research and innovation programme under the Marie Sklodowska-Curie Grant Agreement No.\ 642563),
by Action Medical Research UK (grant number GN1963),
by Lancaster University Department of Physics,
by the Funda\c{c}\~{a}o de Amparo \`{a} Pesquisa  do Estado de S\~{a}o Paulo Brazil (grant number FAPESP 2013/07375-0),
by the Institute of Pathophysiology and Nuclear Medicine, Faculty of Medicine, UKIM Skopje, Macedonia
and by the Slovenian Research Agency (Program No. P20232).


\end{document}